\documentclass[preprint]{aastex}

\usepackage{color}
\usepackage{graphicx}
\usepackage{epsfig}
\usepackage{natbib}
\usepackage{rotating}
\usepackage{amsmath}

\begin{document}

\title{WISE Y Dwarfs As Probes of the Brown Dwarf-Exoplanet Connection}
\author{C. Beichman$^{1,2,3}$, Christopher R.\ Gelino$^{1,3}$, J.\ Davy Kirkpatrick$^1$, Michael C.\ Cushing$^4$, Sally Dodson-Robinson$^5$, Mark S. Marley$^{6}$,  Caroline V. Morley$^{7}$, and E.L. Wright$^8$}
\affil{1) Infrared Processing and Analysis Center, California Institute of Technology, Pasadena CA 91125}
\affil{2) Jet Propulsion Laboratory, California Institute of Technology, 4800 Oak Grove Dr., Pasadena CA 91107}
\affil{3) NASA Exoplanet Science Institute, California Institute of Technology,770 S. Wilson Ave., Pasadena, CA 91125}
\affil{4) Department of Physics and Astronomy, The University of Toledo,
2801 West Bancroft Street, Toledo, OH 43606}
\affil{5) Department of Astronomy,  University of Delaware, Newark, DE 19716}
\affil{6) NASA Ames Research Center, Mountain View, CA 94035}
\affil{7) Department of Astronomy, University of California at Santa Cruz, Santa Cruz, CA 95064}
\affil{8) Department of Astronomy, University of California Los Angeles, PO Box 951547, Los Angeles CA 90095}

\email{chas *at* ipac.caltech.edu}

\begin{abstract}

We have determined  astrometric positions for  15 WISE-discovered late-type brown dwarfs (6 T8-9 and 9 Y dwarfs) using the Keck II telescope, the {\it Spitzer Space Telescope}, and the {\it Hubble Space Telescope}. Combining data from 8 to 20 epochs we derive parallactic and proper motions for these objects which put the majority within 15 parsecs. For ages greater than a few Gyr, as suggested from kinematic considerations, we find masses of 10-30 M$_{Jup}$ based on standard models for the evolution of low mass objects with a range of mass estimates for individual objects depending on the model in question. Three of the  coolest objects have effective temperatures $\sim$ 350 K and inferred masses of 10-15 M$_{Jup}$. Our parallactic distances confirm earlier photometric estimates \citep{Kirkpatrick2012} and direct measurements \citep{Marsh2013, Beichman2013, Dupuy2013} and suggest that the number of objects with masses below about15 M$_{Jup}$ must be flat or declining relative to higher mass objects. The masses of the coldest Y dwarfs may be similar to those inferred for recently imaged planet-mass companions to nearby young stars.  Objects in this mass range, which appear to be rare in both the interstellar and proto-planetary environments, may both have formed via gravitational fragmentation: the brown dwarfs in interstellar clouds and companion objects in a protoplanetary disk. In both cases, however, the fact that objects in this mass range are relatively infrequent, suggests that this mechanism must be inefficient in both environments.

\end{abstract}


\section{Introduction}

% COMPILE WITH DVI FOR EPS TO WORK

Our  understanding of the gravitational collapse of interstellar gas clouds to form stars is one of the great success stories of modern astrophysics. The discovery of ``protostars'' in molecular clouds via infrared and millimeter observations started with high luminosity stars in giant molecular clouds, e.g. the Becklin-Neugebauer \citep{Becklin1967} and Kleinmann-Low \citep{Kleinmann1967} objects in the Orion Molecular Cloud \citep{Wilson1970} and progressed steadily through to the discovery of young stars of solar mass in clouds like Taurus \citep{Beichman1986} and to objects of still lower masses with Spitzer \citep{Dunham2013}. The theory of  star formation progressed hand in hand with observations, from initial discussions of the cloud collapse \citep{Larson1985} to detailed models incorporating disks and outflows \citep{Shu1987}. Long standing questions in star formation theory concern the distribution of stellar masses (the Initial Mass Function, IMF) produced by this process and the end-points of the process, i.e. the largest and smallest self-gravitating objects that can be formed via gravitational collapse. The discovery of substellar objects, ``brown dwarfs," orbiting nearby stars \citep{Nakajima1995} and in early sky surveys, e.g. The Two Micron All sky survey (2MASS; \citealt{Skrutskie2006}), DEep Near Infrared Southern Sky Survey (DENIS; \citealt{Epchtein97}) and Sloan Digital Sky Survey (SDSS; \citealt{York2000}) pushed the low mass limit of the IMF well below the 0.07 M$_\odot$ stellar limit \citep{Kirkpatrick1999,davy2005}. Data from the UKIRT Infrared Deep Sky Survey (UKIDSS) added many new L and T dwarfs and improved our knowledge of their space densities \citep{Burningham2013}. Most recently, the launch of the Wide-field Infrared Survey Explorer (WISE; \citealt{Wright2010}) has led to the identification of over 250 brown dwarfs with extremely low effective temperatures, T$_{eff}$, including the first Y dwarfs with T$_{eff}\sim$ 250-500 K \citep{Kirkpatrick2011, Kirkpatrick2012, Cushing2011}. 

In the early 1980's, before the advent of theories of non-baryonic dark matter, it was thought that sharply increasing low mass stellar and brown dwarf mass functions could account for the local missing mass inferred from galaxy rotation curves \citep{Bahcall1985}. This conjecture was ultimately ruled out as the shape of the low mass IMF was determined with results from the Hubble Space Telescope  (HST) \citep{Flynn1996},  SDSS and 2MASS, as well as by the incidence of  microlensing events determined by the MACHO project \citep{Alcock1996}. Thus, while the low mass shape of the IMF is no longer of cosmological importance, it remains an important question for star formation theory and role of gravitational instability in the origin of the IMF. A related question about gravitational instability arises due to the existence of planetary mass companions  on extremely wide orbits, e.g. HR 8799 and Fomalhaut \citep{Marois2008, Kalas2004} which is difficult to reconcile with models of planet formation via core accretion \citep{Dodson2009}. The formation of low mass objects via gravitational instability appears to be important in the proto-planetary environment as well. 

Thus, we investigate Y dwarfs found with WISE as probes of the low mass IMF and as analogs to the massive planets  orbiting nearby stars. Our long term goals are to understand better the physical properties of these objects and to assess how they might form, in either the interstellar or proto-planetary environments. A key step toward this goal is to determine the distances to the closest, lowest mass objects found by WISE. The first paper in this program reported a parallax for one of the coldest WISE Y dwarfs, WISE1828+2650, classified as a $\ge$Y2 object with a temperature of $\sim$300-500 K and a mass of $\sim$ 5 M$_{Jup}$ for an assumed age of $\sim$5 Gyr \citep{Beichman2013}. We report here on parallax determinations of 15 WISE objects with spectral types of T8 or later made using imaging with the {Hubble Space Telescope} ({\it HST}), the {\it Spitzer Space Telescope}, and the Keck-II telescope. In what follows, we define the sample ($\S$\ref{sample}), describe the observations ($\S$\ref{obs}) and derive the kinematic parameters ($\S$\ref{astromsection}). In $\S$\ref{discussion} we use the spectral energy distribution and absolute magnitudes to estimate the masses of the Y dwarfs, address the possible ages of the sample objects on the basis of their  kinematic properties, and discuss the apparent cutoffs in the distributions of brown dwarf and planetary companions in the range of $<$15 M$_{Jup}$.  

\section{The Sample\label{sample}}

One of the key goals of the WISE mission was the detection of ultra-cool T and Y brown dwarfs with the properties of the instrument tailored such that the W2 filter at 4.6 $\mu$m was positioned to sit at the peak of the cool brown dwarf spectral energy distribution while the shorter wavelength W1 filter at 3.5 $\mu$m sits in a region of methane absorption \citep{Burrows1997}. Thus, the prominent red W1-W2 color of brown dwarfs makes them relatively easy to identify among the millions of WISE sources so that the objects studied in this paper (Table~\ref{Objects}) are selected primarily for their extreme color, W1-W2$>$ 2.5 mag \citep{Kirkpatrick2011,Kirkpatrick2012, Cushing2011, Mace2013,Cushing2013}. Approximately 17 Y dwarfs are presently known, including field objects from WISE, a T dwarf companion \citep{Liu2012}, and a white dwarf companion\footnote{The Spitzer colors and absolute magnitude suggest a Y dwarf classification, but no confirming spectrum has yet been obtained.} \citep{Luhman2011}. In this paper we study nine WISE field Y dwarfs as well as six slightly warmer, late T dwarfs.

As discussed in \citet{Kirkpatrick2012} we suggest that the Y dwarf sample discussed here is relatively complete at the WISE W2 magnitude limits appropriate to low ecliptic latitudes.  While the $V/V_{max}$ value of 0.3 indicates that the late T and Y dwarf sample out to 10 pc are modestly  incomplete \citep{Kirkpatrick2012, Schmidt1968},  a number of investigations are underway to identify additional  Y dwarfs with WISE, including improved processing and more follow-up observations. The sample studied here of 9 Y dwarfs limited only by a declination limit of $\delta>-36^o$, represents a large fraction of the available Y dwarfs from WISE. An additional 6 objects, late T dwarfs, were   included in the sample to help to elucidate the transition between these two spectral types.

\section{Observations\label{obs}}

As described in \citet{Beichman2013} we piece together positional information with a variety of telescopes in the 1-5 $\mu$m range. In the near-infrared where the Y dwarfs are intrinsically faint, we have used the Keck-II telescope with laser guide star adaptive optics and, for nine objects, the Hubble Space Telescope (HST). In the 3-5 $\mu$m range where the sources are much brighter, we have the original WISE measurements which are of low positional accuracy as well as Spitzer observations which offer  higher resolution and higher signal to noise ratio (SNR). Individual positional uncertainties with the various telescopes range between 5-10 mas (Keck and HST), 60 mas (Spitzer) and 250-500 mas (WISE). We then have to tie together multiple astrometric reference frames which adds an additional layer of positional uncertainty. While this multiplicity of telescopes presents the challenge of  matching astrometric reference frames, we gain the advantage of a long temporal baseline and denser sampling of the Y dwarf motions that would be difficult to achieve with a single facility. Table~\ref{Objects} lists the WISE sources, their spectral types, and the number of observations with a particular facility. Table~\ref{AstrometryData} gives  the observing log for each facility as well the astrometric data at each epoch ($\S$\ref{astromsection}).

\subsection{WISE Observations}

The WISE mission had three distinct phases: the 4-band cryogenic period, the 3-band cryogenic period, and 2-band warm mission. Depending on position on the sky, especially ecliptic longitude, sources were observed in one or more of these phases. We determined positions and magnitudes for each period separately with a median date of observation spanning 1-2 days. Positions and associated uncertainties were obtained by averaging the source positions in the multi-band extractions from each individual orbit. The uncertainty in the WISE astrometric frame is approximately 80 mas based on the input 2MASS catalog used for WISE position reconstruction \citep{Cutri2011}. Typically, however, the positional uncertainties in the WISE detections are much larger than this, $\sim$250-500 mas, due to its large beamsize and detection at only one or at most two wavelengths, or to the effects of confusion with other nearby objects. 

In Table ~\ref{MagData} we report averages of the 4.5 \micron\ magnitudes (W2) for the various epochs and include 3.5 \micron\ (W1) when available. Upper limits in the two longer wavelength bands, 12 (W3) and 22 \micron\ (W4) are  high and do not significantly constrain the spectral energy distributions. We have converted the magnitudes to flux densities using the zero points from \citet{Wright2010}, but because of the unknown and extremely non-blackbody-like nature of brown dwarf spectral energy distributions (SED), we have not color-corrected these flux densities. Table~\ref{VarData} indicates that there is no evidence for variability in the [4.6] magnitudes at the 2-3 \% level for any of these objects. While not varying in the Spitzer bands,  WISE2220-3628 shows evidence for variability in the comparison of the ground-based J and HST/F125W photometry, with a nearly 1 mag difference between the two bands. Further monitoring of this object may be warranted.

\subsection{HST Observations}
Nine objects were imaged with HST's WFC3/IR in the F105W, F125W or F140W filters as precursor observations in support of subsequent grism measurements. The final images are quite heterogeneous, consisting of from one to four dithered exposures with exposure times ranging from 312 to 2412 seconds. In some cases multiple exposures were taken with small offsets to reduce the effects of cosmic rays and the undersampling of the individual frames. The Space Telescope Science Institute's (STScI) ``AstroDrizzle" mosaic pipeline was used to process these data to produce final mosaicked images. The pipeline corrects for the geometric distortion of the WFC3-IR camera to a level estimated to be $\sim$ 5 mas \citep{Kozhurina2009} which is of the same order or less than the extraction uncertainties of the faint target. Sources were extracted using the Gaussian-fitting IDL FIND routine to determine centroid positions and the APER routine\footnote{All the photometric measurements reported herein were made using this routine from the IDL ASTRO library, http://idlastro.gsfc.nasa.gov/contents.html. A number of other IDL routines are taken from this library as well.} with a 3 pixel radius for photometric measurements. The Full Width at Half Maximum (FWHM) for these undersampled data is $\sim$ 2 pixels or 0.26\arcsec\ consistent with STScI analyses \citep{Kozhurina2009}. Analysis of fields with multiple HST observations, e.g. WISE1541-2250, shows that after registration onto a common reference frame, the repeatability of individual source positions is $\sim$ 5 mas for bright objects located within 90\arcsec\ of the brown dwarf. The photometry was calibrated using the appropriate zero-points for Vega magnitudes\footnote{http://www.stsci.edu/hst/wfc3/phot\_zp\_lbn} from the WFC3 Handbook \citep{Rajan2010}.

\subsection{Spitzer Observations}

Observations with the {\it  Spitzer Space Telescope} were made using a variety of General Observer (GO) programs (PI D. Kirkpatrick) and some Director's Discretionary Time  (A. Mainzer and T. Dupuy). In all cases the observations were obtained during the Warm Mission phase using the IRAC camera \citep{Fazio2004} in its full array mode to make observations at 3.6 (Channel-1) and/or 4.5 $\mu$m (Channel-2). We analyzed post-BCD mosaics from the Spitzer Science Center (SSC) to make photometric and astrometric measurements, extracting sources using a 4 pixel radius  aperture, a 4-12 pixel annulus for sky subtraction, and normalizing the resultant counts using SSC-recommended aperture corrections\footnote{The post-BCD images have  0.6\arcsec\ pixels derived from the 1.22\arcsec\ native pixel data. The aperture correction is described in http://irsa.ipac.caltech.edu/data/SPITZER/docs/irac}.

\begin{figure*}
 \includegraphics[scale=0.6]{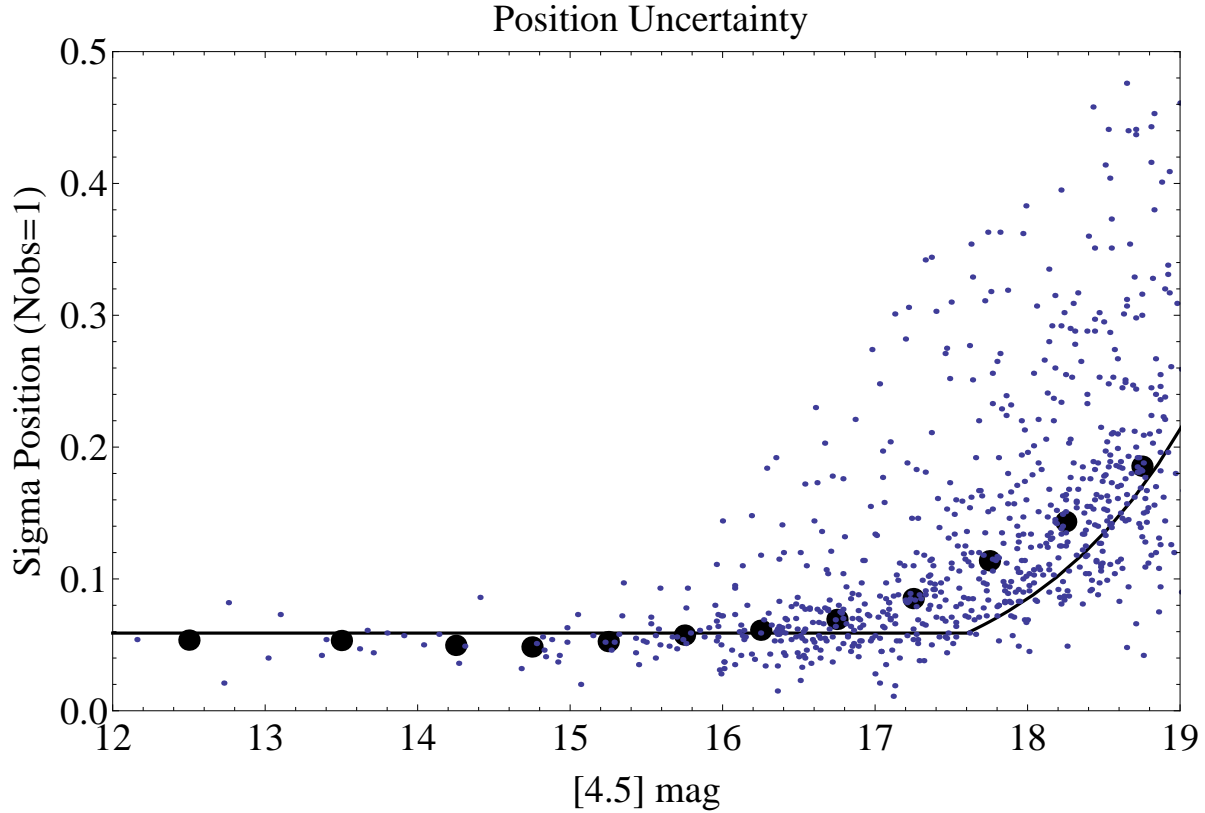}\caption{The dispersion in Spitzer positions from one epoch to the next is shown as a function of [4.6] Spitzer magnitude. Individual reference sources (small circles) were used to register the Spitzer frames and were drawn from a region within 60\arcsec\ - 90\arcsec\ of each brown dwarf target. The single axis uncertainties have been normalized to a single epoch according to N$_{obs}^{1/2}$ and are thus representative of the uncertainties for our single epoch brown dwarf measurements. The large filled circles represent the median uncertainty in 0.5 mag wide bins (1.0 mag bins for the 2 brightest bins). The solid line shows a model fitted to these values with a constant uncertainty of $\sigma_0=58\pm8$ mas for sources brighter than [4.6]$<$17.6$\pm0.2$ mag and an uncertainty increasing as SNR$^{-1}$ for fainter objects. Outliers in the distribution are typically due to confused or extended sources. Our brown dwarf targets are located in bright source portion of the positional uncertainty distribution. \label{SpitzerPointing}}
 \end{figure*}

For each target we put all epochs of Channels-1 and -2 onto a common reference frame by averaging  the positions of all bright sources within $\sim$60-90\arcsec\ of the target, typically 25-50 objects per frame, and calculating small offsets from one epoch to the next to register all frames to the average value. The largest offsets were of order 200 mas and typically much smaller, around 50 mas. We kept the size of the overlap region smaller than the overall size of the IRAC field of view to minimize the effects of optical distortion. The dispersion around the average bright source position is typically 60 mas in both right ascension and declination, or  1/20$^{th}$ of the native 1.2\arcsec\ pixel (Figure~\ref{SpitzerPointing}). These values are less than 100 mas distortions quoted by the SSC\footnote{http://irsa.ipac.caltech.edu/data/SPITZER/docs/irac/iracinstrumenthandbook/26/} in part because we have confined our observations to the small regions at the center of the IRAC arrays. Figure~\ref{SpitzerPointing} shows the positional uncertainty in multiple observations (N$_{obs}$=2 to 13) for 800 reference sources  from  all of our target fields as a function of IRAC [4.6] magnitude. These single axis uncertainties have been normalized to a single epoch according to N$_{obs}^{1/2}$ and are thus representative of the uncertainties for our single epoch brown dwarf measurements. The final positions for reference sources are improved relative to these values by N$_{obs}^{1/2}$. The solid line shows a simple model  to the positional uncertainty  with a constant value of 58$\pm$8 mas for sources brighter [4.6]=17.6$\pm$0.2 mag and a value which increases monotonically as SNR$^{-1}$ to fainter levels \citep{Monet2010}.  Our bright brown dwarf targets are always  in the flat part of the uncertainty distribution.

\subsection{Keck NIRC2 Observations}

Targets were observed in the H-band using using NIRC2 on the Keck-2 telescope  with the laser guide star Adaptive Optics (AO) system \citep{Wizinowich2006,vanDam2006} and tip-tilt stars located  10-50\arcsec\ away. The wide-field camera (40 mas/pixel scale; 40\arcsec\ field of view) was used to maximize the number of reference stars for astrometry. At each epoch, dithered sequences of images with offsets of 1.5\arcsec--3\arcsec\ in right ascension or  declination and total integration times of 1080 sec were obtained at airmass of 1.0-2.0. The majority of sources were observed at airmasses $<$ 1.5. The individual images were sky-subtracted with a sky frame created by the median of the science frames and flat-fielded with a dome flat using standard and custom IDL routines. Individual images were ``de-warped'' to account for optical distortion in the NIRC2 camera \citep{Beichman2013}. The reduced images were shifted to align stars onto a common, larger grid and the median average of overlapping pixels was computed to make the final mosaic. The source positions obtained from the Keck images were corrected for the effects of differential refraction relative to the center of the field using meteorological conditions available at the CFHT telescope weather archive\footnote{http://mkwc.ifa.hawaii.edu/archive/wx/cfht/} to determine the index of refraction corrected for wavelength, local temperature, atmospheric pressure and relative humidity \citep{Lang1983} and standard formulae \citep{Stone1996}. As discussed in \citet{Beichman2013}, for the small field of view of the NIRC2 images and the relatively low airmasses under consideration here, the first order differential corrections are small, $<$10 mas across the $\pm$20\arcsec\ field and proportionately less at smaller separations.

The effects of optical distortion in the wide-field NIRC2 camera were corrected using a distortion map derived by comparing Keck data of the globular cluster M15 \citep{Alibert2005}. Details of this distortion mapping are described in \citep{Beichman2013} but the correction amounts to $<1$ pixel (40 mas) across most of the array and up to 2 pixels at the edges of the array. After our correction procedure the residual distortion errors are less than 10 mas over the entire field.

\section{Astrometric Data Reduction \label{astromsection}}

The first step in determining the position of a target is to put all the available datasets onto a common reference frame. When HST observations were available, sources seen in common between HST and Spitzer were used to register the two fields onto a common frame with a typical accuracy of $<20$ mas, considerably less than the uncertainty in Spitzer positions themselves (50-60 mas). We used HST and Keck images to reject obviously extended objects from consideration as obtaining a good centroid position for these objects can be difficult, particularly in Keck images. However, whenever possible, objects with only slight extent ($<$0.2\arcsec) were included since these extragalactic sources help to anchor the positions to an absolute reference frame.

\begin{figure*}
 \includegraphics[scale=0.85]{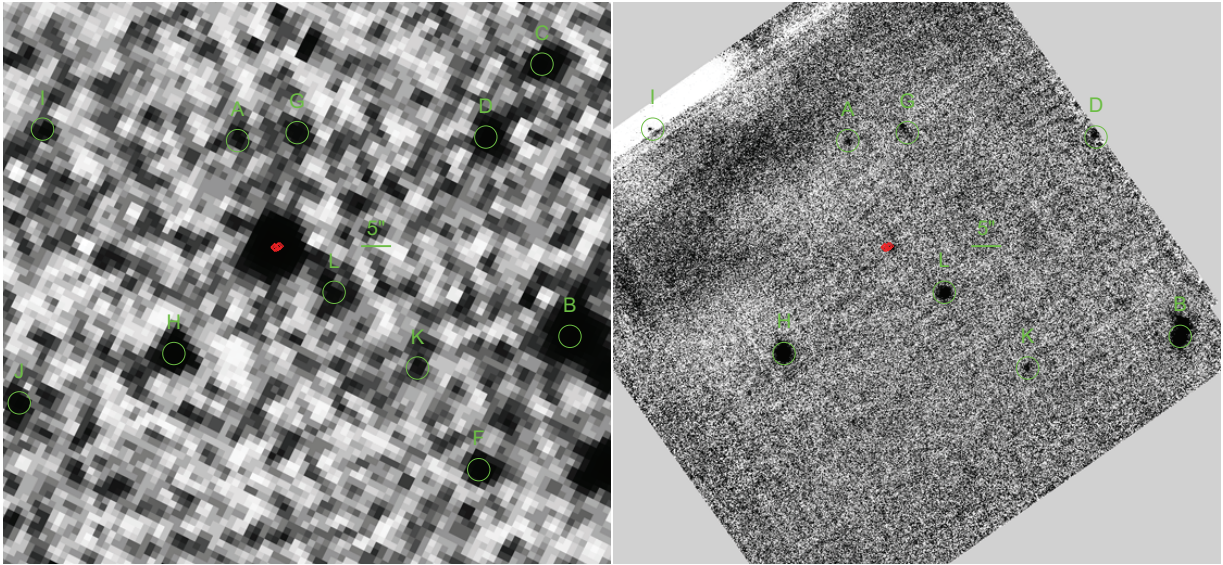}\caption{Spitzer (left) and Keck (right) images at 4.6 $\mu$m and 1.65 $\mu$m, respectively, of WISE 0146+4234 with the reference stars used for the co-registration of the fields circled in green. The positions of the brown dwarf are marked in red. A scale bar denotes 5\arcsec. \label{w0146image}}
 \end{figure*}

The Keck fields were referenced to the HST or HST/Spitzer reference frame using from 3 to 10 objects seen in common in the 40\arcsec\ field of view of NIRC2. The accuracy of this registration varied from 3-30 mas (Table~\ref{RefFrame}) with the number of reference objects and the quality of the night. Images showing HST, Spitzer and Keck fields are shown in Figures~\ref{w0146image}-Figure{w2220image} with the positions of some of the reference stars indicated in green. While the rotational orientation of the Spitzer and HST frames are well determined in their respective pipelines ($<$0.001$^o$) and thus has little effect on derived positions, the same cannot be said for the Keck images. We determined the rotation using the HST and/or Spitzer reference stars with an accuracy that varies between 0.005$^o$ up to 0.05$^o$, depending on the number of stars and the quality of the night. The effect of this rotational uncertainty is included in the assignment of the uncertainty in the position of the brown dwarf.

Finally, although the absolute coordinate system is not directly relevant to the determination of parallax and proper motions, we note that we have adopted the Spitzer frame in our quoted positions. The Spitzer positions are based on the 2MASS catalog as is the WISE coordinate system \citep{Cutri2011}. The estimated global accuracy of the 2MASS frame is estimated to be approximately 80 mas \citep{Skrutskie2006}.

%\section{Results \label{results}}

\subsection{Determination of Parallax and Proper Motion}

Table~\ref{AstrometryData} lists the positions for the WISE targets for each available epoch. The right ascension and declination data were fitted to a model incorporating proper motion and parallax \citep{Smart1977, Green1985}:

\begin{align*}
\alpha^\prime\equiv&\, \alpha_0+\mu_\alpha(t-T_0)/cos(\delta^\prime)\\
\delta^\prime\equiv&\, \delta_0+\mu_\delta(t-T_0)\tag{1}\\
\alpha(t)=&\, \alpha^\prime +\pi\Big( X(t) sin\,\alpha^\prime - Y(t) cos\,\alpha^\prime \Big)/cos\,\delta^\prime \\
\delta(t)=&\, \delta^\prime +\pi\Big( X(t) cos\,\alpha^\prime sin \,\delta^\prime+ Y(t) sin\,\alpha^\prime sin\,\delta^\prime -Z(t) cos\,\delta^\prime \Big)\tag{2} \\
\end{align*}

\noindent where $(\alpha_0, \delta_0)$ are the source position for equinox and epoch $T_0=$J2000.0,
$\mu_{\alpha,\delta}$ are proper motion in the two coordinates in \arcsec/yr, and $\pi$ is the annual parallax in arcsec. 
The coefficients $X(t),\ Y(t)$, and $Z(t)$ are the rectangular coordinates of the observatory as seen from the Sun in AU. 
Values of $X,Y,Z$ for the terrestrial or Earth-orbiting observatories are taken from the IDL ASTRO routine 
XYZ while $X,Y,Z$ values for the earth-trailing {\it Spitzer} observatory are obtained from the image headers provided by the SSC. Equations (1) and (2) are solved simultaneously using the {\it Mathematica} routine {\it NonLinearModelFit}\footnote{http://reference.wolfram.com/mathematica/ref/NonlinearModelFit.html} incorporating appropriate uncertainties for each data-point. 

The solutions are given in Table~\ref{Parallax} with precisions for the derived parallax values ranging from 5\% (WISE1541-2250) up to indeterminate values with uncertainties of 50\% (WISE 0836-1859). Figures~\ref{w0146motion} to \ref{w2220motion} show the fit to the total motion of the sources (proper motion plus parallax) as well as the fit to the motion with both proper motion and the effect of observatory location (terrestrial or Earth-trailing) removed. Our determinations are robust for 12 objects (uncertainties $<$ 15\%) with an average distance of 8.7 pc and a maximum distance for a well determined distance of 15 pc. Three objects (WISE0836-1859, WISE1542+2230, WISE2220-3628) have low precision parallaxes  due to either  a small number of measurements, in particular with Keck or HST and/or a sparse set of reference stars. For the first two objects it is likely that the true distance for  these T dwarfs is greater than 15 pc  and thus more challenging to determine. For the 13 objects showing uncertainties less than 20\% (and 12 objects with uncertainties $<$15\%) we are relatively immune to the Lutz-Kelker bias in our determination of absolute magnitudes or other derived quantities. The bias occurs when more objects at larger distances are scattered into a sample than when objects of smaller distances are scattered out of the sample \citep{Lutz1973}. 

These parallaxes are determined relative to small groups of objects (typically stars) and not tied directly to an absolute reference frame. Thus our parallaxes are relative measurements and may have biases at the $<$ 5 mas level \citep{Dupuy2012} which represents a limiting floor to the accuracy of our quoted distances. Mitigating against this problem are the large parallactic values ($\sim$100 mas) for sources located within 10-15 pc as well as the fact that each field typically contains one or more extragalactic sources which help to anchor the coordinate system in an absolute sense \citep{Mahmud2008}.

The distance estimates  determined herein are, on average, close to those  presented in \citet{Kirkpatrick2012}. A source-by-source comparison (Table~\ref{PlxComp}) gives the ratio of the Kirkpatrick (2012) values to the ones determined here. For the late T and Y dwarfs, \citet{Kirkpatrick2012} list only a few trig parallaxes \citep{Marsh2013} with the majority coming from photometric distances determined by comparing source brightness in the H and WISE W2 bands with color-magnitude diagrams for those few T and Y dwarf objects with measured parallaxes. For the 12 sources with Keck distance errors $<$20\%, the Kirkpatrick/Keck distance ratio is 0.9  with a dispersion 0.2 and mean uncertainty of 0.06. This close agreement indicates that the photometric parallaxes are, in general, adequate to predict a distance within 25\%. More importantly, the agreement in the average distances implies that the conclusions about the luminosity and mass functions for these ultra-low mass late T and Y dwarfs presented in \citet{Kirkpatrick2012} remain valid and  now rest on  more solid footing with  these more precise distances.

Finally, we note that \citet{Dupuy2013} have recently published parallaxes for 6 objects in common with our sample. Table~\ref{DupuyPlx} demonstrates good agreement (1-2 $\sigma$) between their parallax and proper motions  in all but one  case, WISE 1541-2250, which differs by 3 $\sigma$. Examination of the Spitzer data for this object shows significant contamination with a nearby star as the WISE object approaches the star. We simultaneously fitted   Gaussian profiles of the same width to the two sources for sightings when the sources were far enough apart to distinguish cleanly. For observations after MJD=56066, we were unable to make an accurate determination and did not use Spitzer data in our fitting. However, the WISE object and the star are cleanly delineated in the early Spitzer observations and most importantly in our high resolution Keck and HST data, leading us to trust our solution which puts the object at 5.7$\pm0.3$ pc instead of  Dupuy \& Kraus's more distant 13.5$\pm$5.6 pc. A few more observations, especially after the object clears the offending star, will put the distance to this object on a firm footing.

\section{Discussion\label{discussion}}

\subsection{Spectral Energy Distributions\label{SED}}

We have used published models which predict the spectral energy distributions (SED) of our sources to investigate their physical properties. We acknowledge at the outset that this discussion is fraught with danger given the known difficulties with modeling brown dwarfs with effective temperatures T$_{eff}<<1000$K and sometimes $<$400 K. Developing models at these low temperatures is very much an on-going task requiring new gas and dust opacities as well as incorporating clouds of water and metallic precipitates, and possibly non-equilibrium chemistry \citep{Baraffe2003, Morley2012, Marley2007}. In addition to the intrinsic model uncertainties, the models are degenerate between mass and age since the temperature and luminosity of a brown dwarf  decrease slowly with time. Thus a source with a particular SED, i.e. with some T$_{eff}$, could be either a young, low mass object or an older, more massive one. With these caveats in mind we examined two different sets of models, the dust-free  BT-Settl models \citep{Allard2003,Allard2010}  with opacities  updated relative to the older COND models \citep{Baraffe2003} and a series of models (hereafter denoted ``Morley" models) incorporating sulfide and chloride clouds as well as a cloud-free case \citep{Morley2012, Leggett2012, Saumon2008}.  The Morley  models are characterized by the amount of sedimentation of the precipitated material, according to a parameter $f_{sed}$, ranging from $2<f_{sed}<5$ and as well as a cloud-free case. A higher value of $f_{sed}$ corresponds to optically thinner clouds while a lower $f_{sed}$ corresponds to optically thicker clouds. Neither of these models  include   non-equilibrium chemistry or the influence of water clouds, although the effects of  water condensation are included in the model. 

The models  tabulate absolute magnitudes for a variety of  filters, including ground-based (MKO) $J$     and $H$, HST F125W and F140W, as well as Spitzer Channels 1 and 2 ([3.6] and [4.5] $\mu$m). We calculated a $\chi^2$ value based on absolute [4.5] $\mu$m flux density using the Spitzer Ch2 photometry and our distance estimate as well as up to 5 photometric colors: J-[4.5], H-[4.5], [F125W]-[4.5],[F140W]-[4.5] and [3.6]-[4.5].

\begin{align*}
\chi^2&=\frac{(Abs[4.5]_{obs}-Abs[4.5]_{model})^2}{(5/ln(10) \sigma_{D}/D)^2+\sigma([4.5])_{obs})^2}+\\
&\sum\limits_i\frac{((mag_i-[4.5])_{obs}-(mag_i-[4.5])_{model})^2}{\sigma(mag_i)^2+\sigma([4.5])_{obs}^2} )\tag{3}\\
\end{align*}

\noindent where $D$ is the distance to the source, $Abs[4.5]=[4.5]-5 \times \rm log(D/10 pc)$ is the absolute 4.5 $\mu$m magnitude, and $mag_i$ is the magnitude in the relevant band. The minimum $\chi^2$ values for each source were determined through the interpolated (mass, age) grid with (0.1 Gyr $<$ Age $<$ 10 Gyr, 5 $<$Mass$<$ 80 M$_{Jup}$) for the BT-Settl models, yielding the model parameters in Table~\ref{SettlMass}. For the coldest Y dwarfs, the data suggest T$_{eff}<$ 400K and in these cases we used a coarser grid of BT-Settl models, sampling ($300 K <T_{eff}<400 K$ and $3.0<log \, g<5.5$) for an assumed radius of 1 $R_{Jup}$ and where $log \, g$ is the log of the surface gravity. For the Morley models, we interpolated in a (T$_{eff}$,$log \, g$) grid for discrete values of $f_{sed}$. The solution spaces for each source, Log($\chi^2)$ as a function of model parameters,   are shown in Figure~\ref{AllSourceModelBTFit} and  Figure~\ref{AllSourceModelMorleyFit}. Tables~\ref{SettlMass} and ~\ref{MorleyMass} give the  fitted values  for each source   with their associated uncertainties derived from a Monte Carlo analysis in which the distances and photometric values were varied according to their nominal uncertainties. For the cold BT-Settl cases, the uncertainties reflect the coarseness of the grid, not the observational uncertainties. The tables include  values of  radius and {\it log g}  from the appropriate evolutionary tracks  as well as the $\chi^2$ of the fits.  Table~\ref{MorleyMass} also includes the differences in the derived values of T$_{eff}$, mass, and age between the Morley and BTSettl models.

The BT-Settl models (Figure~\ref{AllSourceModelBTFit})  show  a valley of preferred values in the (mass,age) plane with quite  good fits ($\chi^2<10$ with 2-3 degrees of freedom) for some of the sources with a median value of $\chi^2=$22 for 2-3 degrees of freedom. For the coolest sources, i.e. WISE 1828+2650 ($\ge$Y2),  WISE1541-2250 (Y0.5)  and WISE 2209+2711 (Y0:), the fits converge on $T_{eff}$= 350K and $log\,g$=4.5 with $\chi^2>400$. Figure~\ref{SampleSpecFits}a-o show the best fitting BT-Settl models.  Generally, the  BT-Settl solutions have a broad range  of masses from 12-28  M$_{Jup}$ with an average of 20 $\pm$ 6 M$_{Jup}$ and ages from 3.4-8.8 Gyr with an average of 7 $\pm$ 2 Gyr.  The Y dwarfs have lower masses and temperatures than the T dwarfs, 15  vs 25 M$_{Jup}$, and  390 K vs. 580 K. Figure~\ref{TeffCompare} shows the range in temperature for the late T and Y dwarfs derived from the two sets of models.

Overall, the Morley models fit the data less well  with a  median value of $\chi^2$ (with 2 or 3 degrees of freedom) of 47  compared with 22  for the BT-Settl models.  These models have uniformly high surface gravities, {\it log g}$\sim$ 5, at the high end of the input grid and thus yield  higher masses than BT-Settl cases, $\sim$30 M$_{Jup}$ in many cases. In fact, if the model grid is allowed to extend to {\it log g}=5.5, then the masses  approach 60 M$_{Jup}$ with  ages of 15 Gyr which do not seem reasonable. The difficult sources to fit with the BT-Settl models, e.g. WISE1828+2650 and WISE1541-2250, have  high $\chi^2$values with the Morley models as well. The derived effective temperatures in the two sets of models (Morley vs. BT-Settl) are  similar with the Morley models being  80 K warmer (Figure~\ref{TeffCompare}).

As noted in \citet{Beichman2013}, WISE1828+2650 resists simple modeling due to the large disparity between the short and long wavelength magnitudes with H-[4.5]=8.1 mag. While the 3-5 $\mu$m data alone yield a good fit to a  BTSettl model (T${eff}$=440 K, $log \, g=$4.5), such a model fails by  $\sim$3 magnitudes to fit the shorter wavelength data. Similarly, fitting only the short wavelength data yields a BT-Settl model  (T${eff}$=300, $log \, g=$4.5) which fails to reproduce the longer wavelength observations by comparable amounts. Adding extinction due to a very thick cloud layer with the absorption properties of ``interstellar grains"  suppresses the near-IR bands relative to the longer wavelengths and results a model (T${eff}$=474 K, $ log \, g=4.6,\, A_V=19$ mag)  with a significantly better $\chi^2$=202  (3 d.o.f.) than the model without  extinction, $\chi^2=3700$ (4 d.o.f.). Adding a cloud layer also improved the BT-Settl fit for  WISE 2209+2711 (T${eff}$=420 K, $ log \, g=4.8,\, A_V=15$ mag) with a $\chi^2$=123 (2 d.o.f.). Figure~\ref{SampleSpecFits} shows these extincted models as dotted lines for these two objects. Some previously  unmodeled aspect of atmospheric  physics or  evolutionary status that results in  a strongly absorbing cloud layer  may prove necessary to understanding these  objects.  

It is worth noting that  the poor model fits for the coldest sources are  not  improved by invoking a binary brown dwarf system. Fits to objects of common age but disparate masses did not show any improvement relative to the single object solutions. Nor is the juxtaposition of two unrelated sources a palatable solution since there is no evidence for one stationary and one moving object in the imaging data. similar to the putative Y dwarf companion to WD 0806-661 \citep{Luhman2012}, objects like WISE 1828+2650 and WISE1541-2250 must be underluminous at short wavelengths (or over-luminous in the long-wavelength bands) due to some as yet poorly understood aspect of these very cold atmospheres.

Figure~ \ref{MarleyCMD} compares the data with a number of models in two color-magnitude diagrams, J vs. J-H and [4.5] vs. [3.6]-[4.5]. Deviations in both color spaces are apparent with the BT-Settl models (orange, dot-dashed) and the Morley models bracketing most of the objects in Spitzer/WISE bands. The BT-Settl models tend to be $\sim$1 mag bluer at a given absolute magnitude than is observed, or 1-2 mag underluminous than observed at a given [3.6]-[4.5] color. Three varieties of Morley models are shown, one cloud-free, one with sulfide clouds ($f_{sed}$=5, Table ~\ref{MorleyMass}) and one incorporating water clouds (Morley et al., in preparation). The Morley models tend to be 1 mag redder than the observations at a given absolute magnitude, or 1 mag overluminous at a given color. Taken as a group, the Morley and BT-Settl models straddle the observations, but few of the models can be taken as providing a good fit, particularly for the less luminous, colder cases. There is a much wider divergence between the models and the observations in the JH color-magnitude diagram. These figures also include models incorporating water vapor clouds (Morley et al., in prep).  The BT-Settl models provide a good fit to the J-H colors and absolute magnitudes for the warmer objects, while the Morley objects do a better job on the colder objects at these wavelengths. WISE 1828+2650 stands out as extremely red in J-H and is poorly fitted in any of the models.

Finally, there are a number of conclusions to be drawn from this discussion. The BT and Morley models provide reasonable fits to the properties of the warmer T dwarfs, with the cloud-free BT-Settl models providing the best representation of the absolute magnitudes and colors. But the coldest objects are  difficult to fit and thus properties such as mass and age are quite uncertain. In some cases masses $\sim$ 10-15 M$_{Jup}$ are close to the range inferred for the objects (''planets'') found to be orbiting nearby stars, but precise determinations of masses and other properties  may simply be impossible using broad photometric bands. Even determining an effective temperature using a bolometric luminosity \citep{Dupuy2013} requires a bolometric correction that is model dependent and, as we have seen, quite uncertain. High resolution spectroscopy with JWST across the 1-10 $\mu$m band would yield unambiguous information on surface gravity and composition and would greatly improve our understanding of these objects. In addition, anchoring these models with a few sources with known ages and masses is absolutely critical. This can be accomplished by studying brown dwarfs in binary systems or investigating objects with higher mass companions of known ages, e.g. the potential Y dwarf companion to the white dwarf  WD 0806-661 \citep{Luhman2012}.

\subsection{Age of Sample \label{ages}}

We have no direct indication of the ages of our sample. The BT-Settl and Morley models are consistent with higher surface gravity, higher mass and thus older ages of a few Gyr or more. Because independent age estimates are important because of the Mass-Age degeneracy, we use the kinematic information to make a crude estimate of the ages of these stars. The tangential velocity of each object comes from its proper motion and distance: $v_{tan}=4.74 \mu/\Pi$ km s$^{-1}$ where $\mu$ is the total proper motion and $\Pi$ the parallax \citep{Smart1977}. For 12 objects with distance uncertainty less than 15\%, the average value of $v_{tan}$ is 34 km s$^{-1}$ with a dispersion of 24 km s$^{-1}$, which falls within the distribution of tangential velocities measured for nearby ($<$ 20 pc) L and T brown dwarfs \citep{Faherty2009}. There are, however, significant outliers in this distribution. WISE 0313+7807 has a remarkably small proper motion, 110 mas yr$^{-1}$, and thus a very small $v_{tan}=3\pm1$ km s$^{-1}$. At the other extreme, WISE 0410+1502 and WISE 2209+2711 have $v_{tan}=72\pm4$ km s$^{-1}$ and 59$\pm$4 km s$^{-1}$, respectively. WISE0335+4310 has the most extreme proper motion, $v_{tan}=78\pm10$ km s$^{-1}$.

These tangential velocities are consistent with M7-T9 objects studied by \citet{Faherty2009} suggesting that the extreme T and Y dwarfs studied here are drawn from the same kinematic population. For their 20 pc sample, \citet{Faherty2009} suggested ages of 2-4 Gyr for the objects with $v_{tan}<100$ km s$^{-1}$. An object with a high $v_{tan}$ like WISE 0335+4310 might be somewhat older, up to 8 Gyr. \citet{Faherty2009} suggested that a subset of their sample with low proper motions were younger than the average, perhaps $<$1 Gyr. Thus, the object with the lowest $v_{tan}$, WISE 0313+7807, might be  younger than the other sources. Yet its BT-Settl model age is 9 Gyr and a low SNR spectrum of WISE 0313+7807 \citep{Kirkpatrick2011} does not reveal any obvious peculiarities. The BT-Settl ages are all around 4-9 Gyr and thus consistent with the ages suggested by the kinematics. 

Without radial velocity (RV) information it is impossible to rule definitively on the association of any of these objects with nearby clusters. \citet{Beichman2013} described a search in RV space for $V_z=\pm$100 km s$^{-1}$ to look for potential associations with nearby, young clusters\footnote{Argus/IC2391, TW Hydrae, Tucana/Horologium, $\beta$ Pictoris, AB Doradus, hCha, Cha-near, Columba and the Hyades. See \citet{Beichman2013}} \citep{Zuckerman2004}. With one exception, none of the sample show a plausible kinematic membership with nearby clusters. If WISE 1804+3117 were to have $V_z\sim-20$ km s$^{-1}$, an association with Tucanae/Horologium would be possible, but since the age of this object from model fitting is $\sim$5 Gyr, an association with this 30 Myr old cluster would be problematical. 

\subsection{Brown Dwarfs or Free Floating Planets?}

The incidence of planetary-mass, field brown dwarfs is small. Within 10 pc the RECONS database \citep{Henry2006} shows 376 objects in 259 systems as of 2012. Of these objects, 248 are M stars, 16 are T8-T9.5 objects \citep{Kirkpatrick2012} and 11 are Y dwarfs. Thus extremely low mass objects represent just 7\% of the local population of objects. For the best-fitting BT-Settl models (Table~\ref{SettlMass}) there are only 5 Y dwarfs with masses $<$ 15 M$_{Jup}$. While these mass estimates are obviously speculative and model dependent, it is clear that objects with masses less than $\sim$ 15 M$_{Jup}$ form only a small percentage of the local population. The ratio of local ($<$10 pc) M dwarfs ($75<M<600$) M$_{Jup}$ to low mass brown dwarfs ($5<M<15$) M$_{Jup}$ in logarithmic mass units, $N(M_1\rightarrow M_2)/log(M_1/M_2)$, is large $\sim10:1$ with an obviously large uncertainty due to the uncertain mass estimates. \citet{Kirkpatrick2012} cite a similar number, 6:1, from their volume limited brown dwarf sample. Evidently the star formation processes responsible for populating the local solar neighborhood did not produce large numbers of $<15 $ M$_{Jup}$ objects. This same effect is seen in young clusters where the ratio of stars to brown dwarfs is more precisely estimated to be $\sim$ 6:1 \citep{Anderson2008} and references therein.

It is interesting to note that objects with $<15$ M$_{Jup}$ appear to be difficult to create in the protostellar environments as well. Radial velocity studies find that massive objects are rare in the inner reaches of planetary systems with objects $>$ 5 M$_{Jup}$ accounting for fewer than 79 out of 882 or 9\% presently cataloged planets within 10 AU of their host stars \citep{Cumming2008,Howard2012}. There are only 26 10 M$_{Jup}$ objects out of 882 or just 3\%. Here we have ignored the differences between M sin(i) and M which statistically reduces the number of low mass objects. Imaging surveys targeting the outer reaches of nearby A-F stars as well as lower mass M stars are beginning to either find objects of $\sim5-10$ M$_{Jup}$ or set limits on their occurrence. These coronagraphic studies are typically sensitive to 5-20 M$_{Jup}$ objects with ages $<$ 1 Gyr and located at orbital distances of tens to a few hundreds of AU. Apart from a dramatic examples like HR8799, Fomalhaut, and $\beta$ Pictoris, the success rate of these surveys has been limited, typically a few percent. Around A stars \citet{Vigan2012} find the occurrence rate of a ``planet" in the (3-14 M$_{Jup}$, 5-320 AU) range is $5.9-18$\% ($1\sigma$), nominally a factor of two higher than the incidence of a ``brown dwarf" in the (15-75 M$_{Jup}$, 5-320 AU) range. \citet{Nielsen2013} find the occurrence rate $<20$\% for ($>4$ M$_{Jup}$, 59 and 460 AU) at 95\% confidence, and $<$10\% ($>$ 10 M$_{Jup}$, 38 - 650 AU). They conclude by noting that ``fewer than 10\% of B and A stars can have an analog to the HR 8799 b (7 M$_{Jup}$, 68 AU) planet at 95\% confidence." Around M stars, \citet{Montet2013} find an occurrence rate of $6.5\pm3.0$\% for companions in the (1-13 M$_{Jup}$, 1-20 AU) range. 

Imaging studies are in their infancy with significant advances in sensitivity and angular resolution coming in the next few years with the Gemini Planet Imager\citep{Macintosh2012} and     P1640 \citep{Oppenheimer2012}. The improvements in contrast and sensitivity will increase the completeness of imaging surveys in terms of their mass limit. Improvements in Inner Working Angle will increase survey completeness for as yet unexplored orbital separations and may thus find many more ``super-Jupiters".

With these (uncertain) mass estimates in hand we can speculate as to the formation mechanism of these free floating planetary-mass objects. Observational evidence suggests at least two methods for brown dwarf formation: starlike formation from fragmentation of a molecular cloud \citep[e.g.]{bate03}, possibly aided by turbulence \citep{padoan04}, and protostellar disk fragmentation \citep{boss00, stamatellos07, stamatellos09}. \citet{huard06} and \citet{andre12} have both discovered proto-brown dwarf cores, indicating a starlike formation mechanism for at least some brown dwarfs. Young brown dwarfs have a similar disk fraction to young stars \citep{luhman07} and show the same scaling between mass and accretion rate as stars, $\dot{M} \propto M^2$ \citep{muzerolle03, muzerolle05, mohanty05}, again suggesting a common formation mechanism for stars and brown dwarfs. On the disk fragmentation side, \citet{Thies07} argue that there is a discontinuity in the IMF at the hydrogen-burning limit if unresolved binaries are taken into account, implying that brown dwarfs form differently from stars. Turbulent fragmentation has trouble explaining low-mass binaries: brown dwarf-brown dwarf pairs have not been observed in the numbers predicted \citep{reggiani11}, indicating that a different formation mechanism may be at work.

There are numerous reasons why both molecular cloud fragmentation and disk fragmentation produce fewer brown dwarfs than stars. While the opacity-limited minimum mass of fragments (either disk-born or cloud-born) is only 1--10 $M_{Jup}$ \citep{Low1976, Larson2005, Whitworth2006}, such fragments typically accrete mass and become stars \citep{bate03, kratter10} given typical masses for molecular cloud cores and onset times for protostellar collapse \citep{Myers2009}. \citet{vorobyov13} argues that the probability of fragment survival in gravitationally unstable disks is low, as inward migration and subsequent ejection of fragments is efficient. Vorobyov also shows that fragment survival requires that the instability must happen in the T-Tauri phase of disk evolution rather than the embedded phase, yet the necessary conditions for T-Tauri disk fragmentation may occur only rarely. The median disk/star mass ratio of Class II YSOs inferred from dust continuum observations is only 0.9\% \citep{andrews05, andrews07}. Even when gas is observed directly, as in the deuterated H$_2$ (HD) observations of TW Hydrae \citep{bergin13}, the masses inferred are almost always less than the $0.1 M_{disk} / M_*$ threshold required for disk fragmentation \citep{rafikov05}.

If molecular cloud formation and disk fragmentation are both unlikely, it makes sense to consider whether core accretion---the planet formation process in which a solid core eventually grows large enough to hydrodynamically accrete gas from a disk---might form low-mass field brown dwarfs. Numerical simulations by \citet{ford01} show that 30\% of the interactions between two giant planets near the stability boundary result in ejection, while microlensing measurements by \citet{Sumi11} reveal a population of possibly unbound $1 M_{Jup}$ planetary-mass objects in the galactic bulge. \citet{Mordasini2012} find that the planet mass produced by core accretion falls off dramatically for $M > 3 M_{Jup}$ in disks with $M < 0.06 M_{\odot}$, which would explain the dearth of high-mass planets and field brown dwarfs. Yet \citet{veras12} argue that planet-planet scattering alone cannot explain the large number of unbound planets discovered by Sumi et al., who estimate two free-floating planets per solar-type star. Veras \& Raymond instead suggest that other mechanisms for forming free floaters must be at work.

The likely formation mechanism for the free-floating objects presented here depends sensitively on their mass and velocity dispersion. In most cases the masses inferred from the BTSettl and Morley models are consistent with either disk fragmentation or starlike formation as most of the objects are above the rolloff in the planetary mass function predicted by \citet{Mordasini2012}. For the lowest mass Y dwarfs such as WISE 1828+2650 ($\sim$5-10 M$_{Jup}$), core accretion followed by ejection from a planetary system might be the more favored mechanism as such low mass objects are at or below the opacity-limited minimum mass. However, core accretion near the star (where formation of a massive planet is favorable) followed by planet-planet scattering produces objects with a high velocity dispersion. Our objects have tangential velocities consistent with stars in the Solar neighborhood and inconsistent with an origin in nearby young clusters, i.e. $<$ 100 pc and $<$ 100 Myr ($\S$~\ref{ages}), implying that core accretion and ejection from a close-packed planetary system is unlikely. Both starlike formation and disk fragmentation followed by ejection of partially contracted clumps \citep{basu12} both produce objects with low velocity dispersion, as observed for young brown dwarfs in Cha I\citep{joergens01}. Distinguishing between starlike formation and disk instability is difficult as both mechanisms are consistent with the observed IMF \citep{hennebelle09, stamatellos09}. In a review of brown dwarf observations to date, \citet{luhman07} conclude that starlike formation is the most likely origin for low-mass free floaters, while \citet{bate12} argues that starlike formation, disk fragmentation, and ejection of collapsing cores from molecular clouds probably operate together. The existence of mass solutions that are typical for starlike formation or disk fragmentation, combined with the low velocity dispersion, suggests that our objects are brown dwarfs rather than free-floating planets.

\section{Conclusions}

We have carried out a program of imaging a selection of the coldest brown dwarfs detected by the WISE satellite, including 6 late T and and 9 Y dwarfs to obtain multi-epoch astrometry over a 2-3 year baseline. From these data we have determined parallax and proper motions with better than 15\% accuracy for most of the sample with well determined distances ranging from 6 to 14 pc. By comparing absolute [4.5] magnitudes and a variety of colors from our Keck, HST and Spitzer photometry with models for low mass objects we can estimate masses and ages for this sample ranging between 3.4-8.8 Gyr and 12-30 M$_{Jup}$ for the best fitting BT-Settl models. The fits for the coldest objects, e.g. WISE 1828+2650, are quite poor so these values remain highly uncertain. On the modeling side there is an urgent need for Y dwarf models with a broad range of metallicity,  non-equilibrium chemistry, and effective temperatures as low as 300 K. Highly optically thick dust clouds ($A_V>10$ mag)  may be required to suppress the short wavelength emission and improve the agreement with the models. Observationally, it is critical to anchor these models with a few T or Y dwarf binaries for which dynamical masses can be obtained. In the future, long wavelength photometry out to $>$10 $\mu$m with JWST will provide model-independent bolometric luminosities and effective temperatures. Moderate resolution spectroscopy from 1-10 $\mu$m will provide diagnostic spectral lines which can give much more precise information on physical conditions, especially surface gravity, than can broad band photometry.

Our parallaxes are similar to those estimated by other authors and confirm that local population of coldest brown dwarfs is sparse. The relative lack of brown dwarfs with masses below $\sim$15 M$_{Jup}$ or exoplanets with masses above 10 M$_{Jup}$ suggest this is a difficult mass range for the formation of objects in either environment. The dispersion in tangential velocities for our objects suggest that the objects detected by WISE are, however, likely to represent the lowest mass end of the star formation process rather than a population of objects formed by core accretion in a protoplanetary disk that we subsequently ejected (at high velocity) from their parent system.

\acknowledgments

The research described in this publication was carried out in part at the Jet Propulsion Laboratory, California Institute of Technology, under a contract with the National Aeronautics and Space Administration. 
This publication makes use of data products from the Wide-field Infrared Survey Explorer, which is a 
joint project of the University of California, Los Angeles, and the Jet Propulsion Laboratory/California 
Institute of Technology, funded by the National Aeronautics and Space Administration.
This research has made use of the NASA/IPAC Infrared Science Archive (IRSA) and the NASA Exoplanet Archive
which are operated by the Jet Propulsion Laboratory, California Institute of Technology, under contract
with the National Aeronautics and Space Administration. This work is based in part on observations made with the {\it Spitzer} Space Telescope, which is operated by the Jet Propulsion Laboratory, California Institute of Technology, under a contract with NASA. This work is also based in part on observations made with the NASA/ESA {\it Hubble} Space Telescope, obtained at the Space Telescope Science Institute, which is operated by the Association of Universities for Research in Astronomy, Inc., under NASA contract NAS 5-26555. These observations are associated with program 12330. Support for program \#12330 was provided by NASA through a grant from the Space Telescope Science Institute. Some data presented herein were obtained at the W. M. Keck Observatory from telescope time allocated to the National Aeronautics and Space Administration through the agency's scientific partnership with the California Institute of Technology and the University of California. The Observatory was made possible by the generous financial support of the W. M. Keck Foundation. The authors wish to recognize and acknowledge the very significant cultural role and reverence that the summit of Mauna Kea has always had within the indigenous Hawaiian community. We are most fortunate to have the opportunity to conduct observations from this mountain. The 2MASS catalog and the RECONS database of nearby stars remain invaluable resources. We acknowledge the assistance of Mr. Tahina Ramiaramanantsoa with the early stages of the reduction of these data and Ms. Dimitra Touli with the model fitting. Finally, we would like to thank the anonymous referee for a careful reading of our paper which led to a number of valuable improvements to both its  content and presentation.

%\end{document}
\clearpage

\begin{deluxetable}{lccccccc}
\tabletypesize{\tiny}
\setlength{\tabcolsep}{0.02in}
\tablecaption{ Astrometric Targets \label{Objects}}
\tablehead{WISE Designation & Spectral Type&Sp. Ref & Detections (M/N)$^a$ &\# Keck Obs.&\# Hubble Obs. &\# Spitzer Obs.&Baseline (yr)}
\startdata
J014656.66+423410.0		(WISE0146+42)	& 	Y0	&1&13/39	&7&0&8&2.5\\
J031325.94+780744.2		(WISE0313+78)	& 	T8.5 &3&16/16	&4&0&5&3.6\\
J033515.01+431045.1		(WISE0335+43)	& 	T9	&4&9/12		&5&1&8&2.4\\
J041022.71+150248.4$^b$		(WISE0410+15)	&       Y0	&2&12/12	&2&1&11&2.3\\
J071322.55-291751.9		(WISE0713-29)	& 	Y0	&1&11/15	&5&0&5&1.3\\
J083641.10-185947.0		(WISE0836-18)	& 	T8p	&3&7/15		&4&0&3&2.1\\
J131106.20+012254.3		(WISE1311+01)	& 	T9:	&3&9/17		&5&0&4&2.2\\
J154151.65-225024.9		(WISE1541-22)	& 	Y0.5	&2&10/10	&4&2&4&2.1\\
J154214.00+223005.2		(WISE1542+22)	& 	T9.5	&4&22/45	&1&2&3&1.8\\
J173835.53+273259.0$^b$		(WISE1738+27)	&	Y0	&2&16/18	&3&1&10&2.7\\
J180435.37+311706.4		(WISE1804+31)	&	T9.5:	&3&15/19	&5&0&9&3.0\\
J182831.08+265037.7$^b$		(WISE1828+26)	&	$\ge$Y2	&1&12/18	&5&4&11&2.9\\
J205628.91+145953.2		(WISE2056+14)	&	Y0	&2&12/12	&6&1&11&2.9\\
J220905.73+271143.9		(WISE2209+27)	&	Y1	&5&13/15	&4&1&6&2.4\\
J222055.31-362817.4		(WISE2220-36)	&	Y0	&1&11/17	&2&1&6&1.8\\
\enddata
\tablecomments{$^a$Number of {\it actual} detections, $M$, relative to number of {\it possible} detections, $N$ in WISE W2 band.
$^1$\citet{Kirkpatrick2012}; $^2$\citet{Cushing2011};$^3$\citet{Kirkpatrick2011};$^4$\citet{Mace2013};$^5$\citet{Cushing2013}}
\end{deluxetable}

\clearpage

\begin{deluxetable}{lrcccccccc}
\tabletypesize{\tiny}
%\rotate
\setlength{\tabcolsep}{0.02in}
\centering
\tablecaption{ Observing Log and Astrometric Data \label{AstrometryData}}
\tablehead{
\colhead{WISE Designation}&\colhead{Observatory}&\colhead{Date (UT)}&\colhead{Filter}&\colhead{AOR}&\colhead{PI}&\colhead{MJD}&\colhead{RA (J2000)}&\colhead{DEC (J2000)}&\colhead{Uncertainty (mas)}}
\startdata
J014656.66+423410.0&WISE&2010-Jan-27&&&&55223.14&26.7361144&42.5694586&250\\ 
&Spitzer&2011-Apr-05&Ch1&41808128&Kirkpatrick&55656.09&26.7359654&42.5694282&60\\ 
&Spitzer&2011-Apr-05&Ch2&41808128&Kirkpatrick&55656.09&26.7359584&42.5693949&60\\ 
&Keck&2011-Dec-19&H&&Beichman&55914.28&26.7358852&42.5694054&50\\ 
&Spitzer&2012-Mar-07&H&&Beichman&55993.04&26.7358141&42.5694177&60\\ 
&Spitzer&2012-Mar-07&Ch2&44544000&Kirkpatrick&55993.04&26.7358141&42.5694177&60\\ 
&Spitzer&2012-Oct-15&Ch2&44588544&Kirkpatrick&56215.07&26.7358141&42.5694177&50\\ 
&Keck&2013-Jan-25&H&&Beichman&56317.22&26.7356976&42.5694103&30\\ 
&Spitzer&2013-Mar-13&H&&Beichman&56364.25&26.7356437&42.5693903&60\\ 
&Spitzer&2013-Mar-13&Ch2&46549760&Kirkpatrick&56364.25&26.7356437&42.5693903&60\\ 
&Spitzer&2013-Mar-21&Ch2&46549504&Kirkpatrick&56372.31&26.7356303&42.5694013&60\\ 
&Spitzer&2013-Apr-06&Ch2&46549248&Kirkpatrick&56388.81&26.7356659&42.5693978&60\\ 
&Spitzer&2013-Apr-11&Ch2&46548992&Kirkpatrick&56393.13&26.7356847&42.5693894&60\\ 
&Keck&2013-Sep-20&H&&Beichman&56555.42&26.7356234&42.5694212&20\\ 
&Keck&2013-Nov-19&H&&Beichman&56615.29&26.7355696&42.569411&30\\ \hline
J031358.93+780748.9&WISE&2010-Dec-21&&&&55256.99&48.3581137&78.1289762&250\\ 
&WISE&2010-Dec-21&&&&55448.08&48.35846&78.1289878&250\\ 
&Spitzer&2010-Dec-21&Ch1&41443840&Kirkpatrick&55551.35&48.3586706&78.1290368&60\\ 
&Spitzer&2010-Dec-21&Ch2&41443840&Kirkpatrick&55551.35&48.3586761&78.1290121&60\\ 
&Spitzer&2011-Apr-23&Ch2&41735936&Kirkpatrick&55674.71&48.3584757&78.128978&60\\ 
&Keck&2011-Oct-16&H&&Beichman&55850.57&48.3588985&78.1290222&20\\ 
&Spitzer&2011-Dec-02&Ch2&44803072&Kirkpatrick&55897.22&48.3588009&78.1290122&60\\ 
&Spitzer&2012-Apr-24&Ch2&44798464&Kirkpatrick&56041.15&48.3585303&78.1290161&60\\ 
&Keck&2012-Oct-07&H&&Beichman&56207.51&48.3589982&78.1290526&20\\ 
&Keck&2013-Jan-25&H&&Beichman&56317.25&48.3587497&78.1290253&30\\ 
&Keck&2013-Sep-20&H&&Beichman&56555.53&48.3591015&78.129052&30\\ \hline
J033515.01+431045.1&WISE1&2010-Feb-15&&&&55242.16&53.8125634&43.1791225&310\\ 
&WISE2&2010-Aug-27&&&&55435.86&53.8127677&43.1791506&150\\ 
&Spitzer&2011-Apr-19&Ch1&41838848&Kirkpatrick&55670.15&53.8129519&43.1789742&60\\ 
&Spitzer&2011-Apr-19&Ch2&41838848&Kirkpatrick&55670.15&53.8129111&43.1789762&60\\ 
&Spitzer&2011-Nov-17&Ch2&44573696&Kirkpatrick&55882.78&53.8131682&43.1788608&60\\ 
&Keck&2012-Oct-07&H&&Beichman&56207.59&53.8134114&43.17867&20\\ 
&Spitzer&2012-Nov-22&Ch2&46436096&Kirkpatrick&56253.19&53.8134568&43.1786443&60\\ 
&Keck&2012-Nov-29&H&&Beichman&56260.38&53.8134285&43.1786345&20\\ 
&Keck&2013-Jan-25&H&&Beichman&56317.32&53.8134738&43.1785834&20\\ 
&HST&2013-Mar-29&F125W&&Cushing&56380.74&53.8135311&43.1785361&20\\ 
&Spitzer&2013-Apr-07&Ch2&46595328&Kirkpatrick&56389.02&53.8135567&43.178527&60\\ 
&Spitzer&2013-Apr-17&Ch2&46595072&Kirkpatrick&56399.8&53.8135846&43.1785371&60\\ 
&Spitzer&2013-Apr-22&Ch2&46594816&Kirkpatrick&56404.5&53.8135702&43.1785451&60\\ 
&Spitzer&2013-May-05&Ch2&46594560&Kirkpatrick&56417.21&53.8135371&43.1785424&60\\ 
&Keck&2013-Sep-20&H&&Beichman&56555.56&53.813732&43.1784515&20\\ 
&Keck&2013-Nov-19&H&&Beichman&56615.32&53.813756&43.1784148&20\\ \hline
J041022.71+150248.4&WISE&2010-Feb-16&&&&55243.6&62.5946547&15.046819&250\\ 
&WISE&2010-Aug-26&&&&55434.09&62.594941&15.0464875&250\\ 
&Spitzer&2010-Oct-21&Ch1&40828160&Kirkpatrick&55490.06&62.5949777&15.0464452&55\\ 
&Spitzer&2010-Oct-21&Ch2&40828160&Kirkpatrick&55490.06&62.5949953&15.0464292&55\\ 
&Spitzer&2011-Apr-14&Ch2&41442304&Kirkpatrick&55665.88&62.5950177&15.0460896&55\\ 
&HST&2012-Sep-01&F140W&&Cushing&56171.83&62.5954954&15.0452734&20\\ 
&Spitzer&2011-Nov-19&Ch2&44567808&Kirkpatrick&55884.56&62.5952786&15.0457531&55\\ 
&Spitzer&2011-Nov-24&Ch1&44508160&Dupuy&55889.76&62.5952814&15.0457285&55\\ 
&Spitzer&2012-Mar-29&Ch1&44508416&Dupuy&56015.06&62.5952928&15.0455135&55\\ 
&Spitzer&2012-Mar-30&Ch2&44564480&Kirkpatrick&56016.76&62.5952956&15.0455307&55\\ 
&Spitzer&2012-Apr-29&Ch1&44508672&Dupuy&56046.9&62.5953018&15.0454548&55\\ 
&Spitzer&2012-Oct-30&Ch1&44508672&Dupuy&56230.96&62.5955446&15.0451638&55\\ 
&Spitzer0&2012-Nov-19&Ch2&46443008&Kirkpatrick&56250.9&62.595579&15.0451303&55\\ 
&Spitzer1&2012-Nov-30&Ch2&46442752&Kirkpatrick&56261.93&62.5955494&15.0451248&55\\ 
&Keck&2013-Jan-25&H&&Beichman&56317.28&62.5955394&15.0450125&40\\ 
&Keck&2013-Feb-20&H&&Beichman&56343.24&62.5955423&15.0449683&20\\ \hline
J071322.55-291751.9&WISE&2010-Apr-09&&&&55296.64&108.3439684&-29.2977331&160\\ 
&WISE&2010-Oct-18&&&&55488.21&108.3441041&-29.2978282&200\\ 
&Keck&2011-Oct-16&H&&Beichman&55850.64&108.3442071&-29.2979174&30\\ 
&Spitzer&2012-Jan-02&Ch1&44568064&Kirkpatrick&55928.89&108.344187&-29.2979651&80\\ 
&Spitzer&2012-Jan-02&Ch2&44568064&Kirkpatrick&55928.89&108.3442477&-29.2979653&55\\ 
&Keck&2012-Mar-31&H&&Beichman&56017.24&108.3441896&-29.2979665&30\\ 
&Keck&2012-Oct-07&H&&Beichman&56207.63&108.3443149&-29.2980289&30\\ 
&Spitzer&2012-Dec-25&Ch2&46439936&Kirkpatrick&56286.71&108.3443274&-29.2980777&55\\ 
&Spitzer&2013-Jan-17&Ch2&46439680&Kirkpatrick&56309.98&108.3443808&-29.2980687&55\\ 
&Keck&2013-Jan-25&H&&Beichman&56317.35&108.3443186&-29.2980846&20\\ 
&Spitzer&2013-Feb-06&Ch2&46439424&Kirkpatrick&56329.14&108.3443683&-29.2981035&55\\ 
&Keck&2013-Feb-20&H&&Beichman&56343.28&108.3443092&-29.2980885&30\\ \hline
J083641.10-185947.0&WISE&2010-May-02&&&&55319.67&129.1712834&-18.9963895&1000\\ 
&WISE&2010-Nov-10&&&&55510.55&129.1714539&-18.996376&1220\\ 
&Spitzer&2011-Jan-01&Ch2&40833536&Kirkpatrick&55563&129.1715552&-18.9962973&50\\ 
&Spitzer&2011-May-31&Ch2&41701888&Kirkpatrick&55712.03&129.1715494&-18.9963169&50\\ 
&Spitzer&2012-Jan-17&Ch2&44556032&Kirkpatrick&55943.76&129.1715477&-18.9963443&50\\ 
&Keck&2012-Nov-29&H&&Beichman&56260.57&129.1715439&-18.9963665&30\\ 
&Keck&2013-Jan-25&H&&Beichman&56317.41&129.17153&-18.9963856&20\\ 
&Keck&2013-Feb-20&H&&Beichman&56343.33&129.1715283&-18.9963813&20\\ 
&Keck&2013-Nov-19&H&&Beichman&56615.59&129.1715276&-18.996414&20\\ \hline
J131106.20+012254.3&WISE&2010-Jan-09&&&&55206.33&197.7760137&1.3817997&350\\ 
&WISE&2010-Jul-02&&&&55380.12&197.7759224&1.3817217&340\\ 
&Spitzer&2011-Mar-29&Ch1&40826368&Kirkpatrick&55649.37&197.7761222&1.3814907&60\\ 
&Spitzer&2011-Mar-29&Ch2&40826368&Kirkpatrick&55649.37&197.7760981&1.3815201&60\\ 
&Spitzer&2012-Mar-29&Ch2&44575232&Kirkpatrick&56015.51&197.7762172&1.3812636&60\\ 
&Keck&2012-Mar-31&H&&Beichman&56017.4&197.7761705&1.3812709&30\\ 
&Keck&2012-Jul-09&H&&Beichman&56117.26&197.7761937&1.3812213&30\\ 
&Spitzer&41143&Ch2&44571904&Kirkpatrick&56161.13&197.7761852&1.3811918&50\\ 
&Keck&2013-Jan-25&H&&Beichman&56317.5&197.7762561&1.3810702&30\\ 
&Keck&2013-Feb-20&H&&Beichman&56343.45&197.776254&1.3810576&20\\ 
&Keck&2013-May-27&H&&Beichman&56439.25&197.7762548&1.3810158&20\\ \hline
J154151.65-225024.9&WISE&2010-Feb-16&&&&55244.84&235.4651965&-22.840523&500\\ 
&Spitzer&2011-Apr-13&Ch1&41788672&Kirkpatrick&55664.91&235.4648435&-22.8404433&66\\ 
&Spitzer&2011-Apr-13&Ch2&41788672&Kirkpatrick&55664.91&235.4648328&-22.8404443&60\\ 
&WISE&2010-Aug-15&&&&55424&235.4650457&-22.8400781&500\\ 
&Spitzer&2012-Apr-22&Ch1&44512512&Dupuy&56039.24&235.4645941&-22.8404542&80\\ 
&Spitzer&2012-Apr-28&Ch2&44550144&Kirkpatrick&56045.83&235.4646137&-22.840462&60\\ 
&Keck&2012-Mar-31&H&&Beichman&56017.51&235.464582&-22.840456&20\\ 
&Spitzer&2012-May-19&Ch1&44512768&Dupuy&56066.22&235.464575&-22.8404522&85\\ 
&Keck&2012-Jul-09&H&&Beichman&56117.28&235.464431&-22.840446&20\\ 
&Keck&2013-Jan-25&H&&Beichman&56317.63&235.4643821&-22.840476&20\\ 
&HST&2013-Feb-12&F125W&&Cushing&56335.74&235.4643635&-22.8404761&20\\ 
&HST&2013-May-09&F105W&&Cushing&56421.55&235.4642595&-22.8404771&20\\ 
&Keck&2013-May-27&H&&Beichman&56439.33&235.4642458&-22.8404802&20\\ \hline
J154214.00+223005.2&WISE&2010-Feb-04&&&&55232.37&235.558604&22.5015172&400\\ 
&WISE&2010-Aug-03&&&&55412.02&235.5583999&22.5015432&400\\ 
&Spitzer&2011-Apr-18&Ch1&41058816&Kirkpatrick&55669.41&235.5579949&22.5013517&60\\ 
&Spitzer&2011-Apr-18&Ch2&41058816&Kirkpatrick&55669.41&235.5580421&22.5013728&60\\ 
&Spitzer&2012-Apr-15&Ch2&44559616&Kirkpatrick&56032.02&235.5577765&22.5012418&60\\ 
&HST&2012-Mar-04&F140W&&Kirkpatrick&55990.91&235.5577799&22.5012605&15\\ 
&Spitzer&2012-Sep-21&Ch2&44557568&Kirkpatrick&56191.18&235.5575565&22.5012083&60\\ 
&HST&2013-Feb-13&F125W&&Cushing&56336.82&235.557504&22.5011602&15\\ 
&Keck&2013-Feb-20&H&&Beichman&56343.52&235.5575356&22.5011738&60\\ \hline
J173835.53+273259.0&WISE&2010-Mar-13&&&&55269.03&264.6480543&27.5496933&250\\ 
&Spitzer&2010-Sep-18&Ch1&40828416&Kirkpatrick&55457.58&264.6480843&27.549658&50\\ 
&Spitzer&2010-Sep-18&Ch2&40828416&Kirkpatrick&55457.58&264.6480788&27.5496439&50\\ 
&WISE&2010-Sep-09&&&&55448.65&264.6481684&27.5496833&250\\ 
&HST&2011-May-12&F140W&&Kirkpatick&55693.81&264.6481914&27.5495878&15\\ 
&Spitzer&2011-May-20&Ch2&41515264&Kirkpatrick&55701.63&264.6482049&27.549556&50\\ 
&Spitzer&2011-Nov-26&Ch2&41515264&Kirkpatrick&55891.28&264.648178&27.5495077&50\\ 
&Keck&2012-Mar-31&H&&Beichman&56017.55&264.6482856&27.5495029&25\\ 
&Spitzer&2012-May-08&Ch1&44513536&Dupuy&56055.9&264.6482997&27.549458&50\\ 
&Spitzer&2012-May-12&Ch2&44558336&Kirkpatrick&56059.9&264.6483229&27.5494919&50\\ 
&Keck&2012-Jul-09&H&&Beichman&56117.26&264.6482643&27.5495025&30\\ 
&Spitzer&2012-Jul-10&Ch1&44513792&Dupuy&56118.85&264.6483209&27.5494664&50\\ 
&Spitzer&2012-Sep-27&Ch1&44513024&Dupuy&56197.4&264.6482847&27.5494635&50\\ 
&Spitzer&2012-Nov-19&Ch2&46437888&Kirkpatrick&56250.73&264.6482591&27.5494299&50\\ 
&Spitzer&2012-Nov-27&Ch1&44513280&Dupuy&56258.77&264.6482799&27.5494348&50\\ 
&Keck&2013-May-27&H&&Beichman&56439.43&264.6483932&27.5494055&30\\ \hline
J180435.37+311706.4&WISE&2010-Mar-21&&&&55277.1&271.1472306&31.2851638&340\\ 
&WISE&2010-Nov-09&&&&55509.91&271.1471832&31.2852385&280\\ 
&Spitzer&2010-Sep-26&Ch1&40836352&Kirkpatrick&55465.2&271.1472408&31.2851226&50\\ 
&Spitzer&2010-Sep-26&Ch2&40836352&Kirkpatrick&55465.2&271.1472431&31.2851484&50\\ 
&Spitzer&2011-May-25&Ch2&41565696&Kirkpatrick&55706.84&271.1472367&31.2851427&50\\ 
&Spitzer&2011-Nov-29&Ch2&44571136&Kirkpatrick&55894.05&271.14717&31.2851347&50\\ 
&Keck&2012-Jul-09&H&&Beichman&56117.37&271.1470991&31.2851768&30\\ 
&Spitzer&2011-Dec-01&Ch1&44515328&Dupuy&55896.97&271.1471423&31.2851566&50\\ 
&Spitzer&2012-May-16&Ch1&44515584&Dupuy&56063.73&271.147159&31.2851443&50\\ 
&Spitzer&2012-May-16&Ch2&44515584&Dupuy&56063.75&271.1471621&31.2851558&50\\ 
&Spitzer&2012-Jul-25&Ch1&44515840&Dupuy&56133.39&271.1471254&31.2851738&50\\ 
&Spitzer&2012-Oct-03&Ch1&44515072&Dupuy&56203.43&271.1470927&31.2851711&50\\ 
&Keck&2013-Apr-22&H&&Beichman&56404.53&271.1470676&31.2851603&20\\ 
&Keck&2013-May-27&H&&Beichman&56439.4&271.147049&31.2851677&20\\ \hline
J182831.08+265037.7&WISE1&2010-Mar-30&&&&55285.66&277.1295162&26.8438&170\\ 
&WISE&2010-Sep-28&&&&55467.55&277.1295247&26.8439192&210\\ 
&Keck&2010-Jul-01&H&&Beichman&55378.44&277.1296241&26.8438953&100\\ 
&Spitzer&2010-Jul-10&Ch1&39526656&Mainzer&55387.29&277.1296029&26.8438554&60\\ 
&Spitzer&2010-Jul-10&Ch2&39526656&Mainzer&55387.34&277.1296042&26.8438808&60\\ 
&Spitzer&2010-Dec-04&Ch2&41027328&Kirkpatrick&55534.27&277.1296675&26.8438286&60\\ 
&HST&2011-May-09&F140W&&Kirkpatrick&55690.89&277.1298806&26.8439048&30\\ 
&Keck&2011-Oct-16&H&&Beichman&55850.21&277.1299543&26.8439071&10\\ 
&Spitzer&2011-Nov-29&Ch2&44586752&Kirkpatrick&55894.04&277.1300176&26.8438958&60\\ 
&Spitzer&2011-Dec-02&Ch1&44516352&Dupuy&55897.48&277.1300065&26.8439088&60\\ 
&Spitzer&2012-May-25&Ch1&44516608&Dupuy&56072.2&277.1302159&26.8439439&60\\ 
&Spitzer&2012-May-25&Ch2&44516608&Dupuy&56072.25&277.1301923&26.8439382&60\\ 
&Keck&2012-Jul-09&H&&Beichman&56117.32&277.1302146&26.8439617&10\\ 
&Spitzer&2012-Jul-23&Ch1&44516864&Dupuy&56131.04&277.1302484&26.8439671&60\\ 
&Keck&2012-Oct-07&H&&Beichman&56207.22&277.1302611&26.8439344&50\\ 
&Spitzer&2012-Oct-18&Ch2&44516096&Dupuy&56218.2&277.1302737&26.84398&60\\ 
&Spitzer0&2012-Nov-18&Ch2&46439168&Kirkpatrick&56249.43&277.1302789&26.8439821&60\\ 
&Spitzer1&2012-Dec-08&Ch2&46438912&Kirkpatrick&56269.92&277.1303385&26.8439822&60\\ 
&HST&2013-Apr-22&F105W&&Cushing&56404.88&277.1305007&26.8439937&10\\ 
&HST&2013-May-06&F125W&&Cushing&56418.83&277.1305121&26.844001&10\\ 
&HST&2013-May-08&F105W&&Cushing&56420.76&277.13051&26.8440029&10\\ 
&Keck&2013-May-27&H&&Beichman&56439.36&277.1305206&26.8440076&10\\ \hline
J205628.91+145953.2&WISE-1&2010-May-13&&&&55329.29&314.1204976&14.9981178&290\\ 
&Keck&2010-Jul-01&H&&Beichman&55378.6&314.1204617&14.9981905&30\\ 
&WISE&2010-Nov-08&&&&55514.2&314.1204976&14.9981178&290\\ 
&Spitzer&2010-Dec-10&Ch1&40836608&Kirkpatrick&55540.03&314.1205267&14.9982425&60\\ 
&Spitzer&2010-Dec-10&Ch2&40836608&Kirkpatrick&55540.03&314.1205241&14.998241&60\\ 
&Spitzer&2011-Jul-06&Ch2&41831424&Kirkpatrick&55748.1&314.1207526&14.9983505&60\\ 
&HST&2011-Sep-05&F140W&&Kirkpatrick&55808.36&314.1207034&14.9983548&20\\ 
&Keck&2011-Oct-16&H&&Beichman&55850.35&314.1207055&14.9983614&20\\ 
&Keck&2011-Dec-19&H&&Beichman&55914.2&314.1207544&14.9983705&40\\ 
&Spitzer&2012-Jan-06&Ch2&44573184&Kirkpatrick&55932.56&314.1207682&14.998396&60\\ 
&Spitzer&2012-Jan-22&Ch1&44517376&Dupuy&55948.98&314.1207601&14.9983875&60\\ 
&Keck&2012-Jul-09&H&&Beichman&56117.46&314.1209349&14.9984907&20\\ 
&Spitzer&2012-Jul-10&Ch1&44517632&Dupuy&56118.83&314.1209268&14.9984825&60\\ 
&Keck&2012-Oct-07&H&&Beichman&56207.28&314.120941&14.9985341&50\\ 
&Spitzer&2012-Jul-18&Ch2&44569600&Kirkpatrick&56126.76&314.1209601&14.9984791&60\\ 
&Spitzer&2012-Aug-21&Ch1&44517888&Dupuy&56160.05&314.1209851&14.9984985&60\\ 
&Spitzer&2012-Dec-22&Ch2&46464000&Kirkpatrick&56283.4&314.1209624&14.9985174&60\\ 
&Spitzer&2013-Jan-04&Ch2&46463488&Kirkpatrick&56296.19&314.1210188&14.9985212&60\\ 
&Spitzer&2013-Jan-22&Ch2&46462720&Kirkpatrick&56314.75&314.1210053&14.9985509&60\\ 
&Keck&2013-May-27&H&&Beichman&56439.46&314.1211613&14.998618&20\\ \hline
J220905.73+271143.9&WISE&2010-Jun-06&&&&55354.86&332.2739012&27.1955919&250\\ 
&Spitzer&2010-Dec-31&Ch2&40821248&Kirkpatrick&55561.94&332.2740681&27.1953371&60\\ 
&Keck&2011-Jul-20&H&&Beichman&55762.5&332.2743368&27.1951698&30\\ 
&Spitzer&2011-Jul-27&Ch2&41698816&Kirkpatrick&55769.86&332.2743509&27.1951224&60\\ 
&Spitzer&2012-Jan-14&Ch2&44548352&Kirkpatrick&55940.6&332.2744675&27.1949265&60\\ 
&Keck&2012-Jul-09&H&&Beichman&56117.52&332.2747063&27.1948078&30\\ 
&Keck&2012-Oct-07&H&&Beichman&56207.32&332.2747329&27.1946743&30\\ 
&HST&2012-Sep-15&F140W&&Cushing&56185.58&332.2747399&27.1947115&20\\ 
&Spitzer&2013-Jan-10&Ch2&46543616&Kirkpatrick&56302.15&332.2748377&27.1945693&60\\ 
&Spitzer&2013-Jan-31&Ch2&46543360&Kirkpatrick&56323.38&332.2748291&27.1945492&60\\ 
&Spitzer&2013-Feb-14&Ch2&46543104&Kirkpatrick&56337.87&332.2748893&27.1945119&60\\ 
&Keck&2013-May-27&H&&Beichman&56439.53&332.2750607&27.1944435&20\\ \hline
J222055.31-362817.4&WISE&2010-May-14&&&&55330.96&335.2304846&-36.4713796&332\\ 
&WISE&2010-Nov-09&&&&55509.91&335.23058743&-36.4715195&281\\ 
&Spitzer&2012-Jan-23&Ch1&44552448&Kirkpatrick&55949.11&335.23056511&-36.4715078&60\\ 
&Spitzer&2012-Jan-23&Ch2&44552448&Kirkpatrick&55949.11&335.23056635&-36.4715455&60\\ 
&Spitzer&2012-Jul-15&Ch2&44574464&Kirkpatrick&56123.9&335.23068028&-36.4715003&60\\ 
&HST&2012-Nov-23&F125W&&Cushing&56254.33&335.23066052&-36.4715666&20\\ 
&Spitzer&2012-Dec-24&Ch2&46460928&Kirkpatrick&56285.09&335.23068603&-36.471558&60\\ 
&Spitzer&2013-Jan-06&Ch2&46460160&Kirkpatrick&56298.03&335.23065602&-36.4715587&60\\ 
&Spitzer&2013-Jan-26&Ch2&46459392&Kirkpatrick&56318.93&335.2307343&-36.4715553&60\\ 
&Keck&2013-Sep-21&H&&Beichman&56556.32&335.23076402&-36.4715891&10\\ 
&Keck&2013-Nov-19&H&&Beichman&56615.2&335.2307524&-36.4715821&10\\ \hline
\enddata
\end{deluxetable}

\clearpage

\begin{deluxetable}{lccccccccc}
\tabletypesize{\tiny}
%\rotate
\setlength{\tabcolsep}{0.02in}
\centering
\tablecaption{ Photometric Data (Magnitudes) \label{MagData}}
\tablehead{
\colhead{WISE Designation}&\colhead{F105W}&\colhead{J}&\colhead{F125W}&\colhead{F140W}&\colhead{H$^a$}&\colhead{WISE [3.35]}&\colhead{Spitzer [3.6]}&\colhead{Spitzer [4.5]}&\colhead{WISE [4.6]}}
\startdata
J014656.66+423410.0&&19.40$\pm$0.25$^b$&&&20.91$\pm$0.21&$>$18.99&17.42$\pm$0.05&15.05$\pm$0.03&15.08$\pm$0.068\\
J031325.94+780744.2&&17.67$\pm$0.07$^b$&&&17.67$\pm$0.07&15.87$\pm$0.058&15.31$\pm$0.05&13.23$\pm$0.03&13.18$\pm$0.03\\
J033515.01+431045.1&&20.07$\pm$0.30$^d$&20.23$\pm$0.05&&19.76$\pm$0.13&$>$18.15&16.58$\pm$0.05&14.39$\pm$0.03&14.60$\pm$0.08\\
J041022.71+150248.4&&19.44$\pm$0.03$^e$&&19.74$\pm$0.03&20.02$\pm$0.05$^e$&$>$18.25&16.62$\pm$0.05&14.10$\pm$0.03&14.18$\pm$0.055\\
J071322.55-291751.9&&19.64$\pm$0.15$^b$&&&19.85$\pm$0.05&$>$18.35&16.67$\pm$0.05&14.22$\pm$0.03&14.48$\pm$0.06\\
J083641.10-185947.0&&18.99$\pm$0.22$^d$&&&19.49$\pm$0.24&$>$18.41&16.85$\pm$0.05&15.06$\pm$0.03&15.18$\pm$0.098\\
J131106.20+012254.3&&18.75$\pm$0.07$^c$&&&19.09$\pm$0.07&$>$18.27&16.81$\pm$0.05&14.64$\pm$0.03&14.76$\pm$0.086\\
J154151.65-225024.9&21.41$\pm$0.01&21.12$\pm$0.06$^e$&21.69$\pm$0.05&&21.54$\pm$0.11&16.74$\pm$0.16&16.70$\pm$0.05&14.21$\pm$0.03&14.26$\pm$0.06\\
J154214.00+223005.2&&20.25$\pm$0.13$^d$&20.73$\pm$0.03&20.46$\pm$0.03&20.34$\pm$0.06&$>$18.88&17.27$\pm$0.05&15.02$\pm$0.03&15.02$\pm$0.06\\
J173835.53+273259.0&&20.05$\pm$0.09$^e$&&19.89$\pm$0.05&20.45$\pm$0.09$^e$&$>$18.40&16.94$\pm$0.05&14.49$\pm$0.03&14.55$\pm$0.06\\
J180435.37+311706.4&&18.67$\pm$0.04$^f$&&&19.21$\pm$0.11$^b$&$>$18.64&16.55$\pm$0.05&14.59$\pm$0.03&14.74$\pm$0.06\\
J182831.08+265037.7&23.96$\pm$0.10&23.57$\pm$0.35$^g$&23.83$\pm$0.05&23.36$\pm$0.05&22.45$\pm$0.08$^g$&$>$18.47&16.88$\pm$0.05&14.30$\pm$0.03&14.39$\pm$0.06\\
J205628.91+145953.2&&19.43$\pm$0.04$^e$&&19.57$\pm$0.04&19.96$\pm$0.04$^e$&$>$18.25&16.07$\pm$0.05&13.92$\pm$0.03&13.98$\pm$0.05\\
J220905.73+271143.9&&22.58$\pm$0.14$^h$&&23.17$\pm$0.03&22.98$\pm$0.31$^h$&$>$18.47&N/A&14.71$\pm$0.03&14.79$\pm$0.07\\
J222055.31-362817.4&&20.38$\pm$0.17$^b$&21.21$\pm$0.05&&20.81$\pm$0.30$^b$&$>$18.65&17.17$\pm$0.05&14.75$\pm$0.03&14.66$\pm$0.06\\
\enddata
\tablecomments{$^a$Unless otherwise noted, H-band photometry is from NIRC2 from observations reported here. Photometry is on the MKO-NIR system; $^b$\citet{Kirkpatrick2012}; $^c$\citet{Kirkpatrick2011}; $^d$\citet{Mace2013}; $^e$\citet{Leggett2013}; 
$^f$unpublished Palomar WIRC data; $^g$\citet{Beichman2013}; $^h$\citet{Cushing2013}.}
\end{deluxetable}

\clearpage

\begin{deluxetable}{lcc}
%\tabletypesize{\tiny}
%\rotate
%\setlength{\tabcolsep}{0.02in}
\centering
\tablecaption{ Spitzer Photometric Variability (Channel 2) \label{VarData}}
\tablehead{
\colhead{WISE Designation}&\colhead{\# Observations}&\colhead{$\sigma_{pop}$\,(mag)}}
\startdata
J014656.66+423410.0&7&0.013\\
J031325.94+780744.2&4&0.030\\
J033515.01+431045.1&7&0.012\\
J041022.71+150248.4&9&0.030\\
J071322.55-291751.9&4&0.007\\
J083641.10-185947.0&3&0.007\\
J131106.20+012254.3&3&0.010\\
J154151.65-225024.9&2&$<$0.1$^a$\\
J154214.00+223005.2&3&0.017\\
J173835.53+273259.0&5&0.024\\
J180435.37+311706.4&4&0.006\\
J182831.08+265037.7&6&0.013\\
J205628.91+145953.2&7&0.015\\
J220905.73+271143.9&6&0.019\\
J222055.31-362817.4&5&0.020\\
\enddata
\tablecomments{$^a$Confused with nearby star}
\end{deluxetable}

\begin{deluxetable}{lccc|cc|ccc}
\tabletypesize{\tiny}
\setlength{\tabcolsep}{0.02in}
\centering
\tablecaption{ Astrometric Reference Frames\label{RefFrame}}
\tablehead{
\colhead{WISE Designation}&\colhead{N$^1$}&\colhead{$\sigma$(Ref,mas)$^2$}&\colhead{$\sigma$(Limit,mas)$^3$}&\colhead{N$^4$}&\colhead{$\sigma$(Ref,mas)$^5$}&\colhead{N$^6$}&\colhead{$\sigma$(Ref,mas)$^7$}&\colhead{$\sigma$(Theta,deg)$^8$}}
\startdata
&&\hfil {\bf Spitzer}\hfil &&\hfil {\bf HST-Spitzer}\hfil &&\hfil {\bf Spitzer/HST-Keck}\hfil &&\\
J014656.66+423410.0&98&1&59&N/A&N/A&3&5-30&0.04-0.16\\
J031325.94+780744.2&103&2.5&56&N/A&N/A&4&6-28&0.044-0.066\\
J033515.01+431045.1&141&0.8&60&11&19&10&14-19&0.011-0.017\\
J041022.71+150248.4&63&0.3&55&9&24&5&10-30&0.02-0.08\\
J071322.55-291751.9&73&1.5&53&N/A&N/A&6&16-20&0.015-0.022\\
J083641.10-185947.0&46&2.3&47&N/A&N/A&5&27-29&0.06-0.12\\
J131106.20+012254.3&27&1.2&42&N/A&N/A&5&14-28&0.03-0.06\\
J154151.65-225024.9&107&0.5&47&10&5&7-10&8-17&0.007-0.022\\
J154214.00+223005.2&81&0.8&63&10&14&3&30&0.11\\
J173835.53+273259.0&28&0.5&54&N/A&N/A&6&10-20&0.011-0.018\\
J180435.37+311706.4&102&0.5&55&N/A&N/A&8&14-17&0.011-0.015\\
J182831.08+265037.7&27&1.5&48&16&5-15&9-10&4-15&0.003-0.018\\
J205628.91+145953.2&134&3&75&12&10&6-7&5-18&0.007-0.038\\
J220905.73+271143.9&105&0.5&62&10&9&5-8&5-22&0.021-0.045\\
J222055.31-362817.4&37&7&63&4&26&5-6&5-10&0.02-0.04\\
\enddata
\tablecomments{$^1$Number of sources in common between multiple Spitzer epochs. $^2$Standard deviation of the mean of the central position of the combined Spitzer frames. $^3$Limiting accuracy for any one source on single epoch, one axis. $^4$Number of sources in common between Spitzer and HST frame, if available. $^5$Standard deviation of the mean of the central positions between the Spitzer and HST frames. $^6$Number of sources in common between Keck and Spitzer/HST frames. $^7$Range in the standard deviation of the mean of the central positions between the Keck and Spitzer/HST frames. $^8$Range in the precision of the determination of rotation angle between Keck and Spitzer/HST frames.}
\end{deluxetable}

\begin{deluxetable}{lccccccccc}
\tabletypesize{\tiny}
%\rotate
\setlength{\tabcolsep}{0.015in}
\centering
\tablecaption{ Parallax and Proper Motion Solutions \label{Parallax}}
\tablehead{
\colhead{WISE Designation}&\colhead{RA (J2000.0)}&\colhead{DEC (J2000.0)}&\colhead{$\mu_\alpha(\arcsec yr^{-1})^a$}&\colhead{$\mu_\delta (\arcsec yr^{-1})$}&\colhead{$\pi(\arcsec)$}&\colhead{Dist (pc)}&\colhead{V$_{tan}$}&\colhead{$\chi^2$ $^b$}&\colhead{$\chi^2$ $^c$}}
\startdata
J014656.66+423410.0&1h46m57.0940s$\pm$0.0112s&42$^o$34$^\prime$10.214$^{\prime\prime}$$\pm$0.$^{\prime\prime}$215&-0.441$\pm$0.013&-0.026$\pm$0.016&0.094$\pm$0.014&10.6$\pm$1.5&22$\pm$3&23.0(27)&61.6(28)\\
J031358.93+780748.9&3h13m25.8000s$\pm$0.0097s&78$^o$7$^\prime$43.524$^{\prime\prime}$$\pm$0.$^{\prime\prime}$705&0.080$\pm$0.012&0.072$\pm$0.057&0.153$\pm$0.015&6.5$\pm$0.6&3$\pm$1&21.4(17)&104.7(18)\\
J033515.01+431045.1&3h35m14.2520s$\pm$0.0096s&43$^o$10$^\prime$53.405$^{\prime\prime}$$\pm$0.$^{\prime\prime}$196&0.826$\pm$0.011&-0.803$\pm$0.015&0.070$\pm$0.009&14.3$\pm$1.7&78$\pm$10&21.5(27)&71.4(28)\\
J041022.71+150248.4&4h10m22.0630s$\pm$0.0110s&15$^o$3$^\prime$11.053$^{\prime\prime}$$\pm$0.$^{\prime\prime}$158&0.966$\pm$0.013&-2.218$\pm$0.013&0.160$\pm$0.009&6.2$\pm$0.4&72$\pm$4&23.7(33)&232.7(34)\\
J071322.55-291751.9&7h13m22.2510s$\pm$0.0166s&-29$^o$17$^\prime$47.558$^{\prime\prime}$$\pm$0.$^{\prime\prime}$277&0.388$\pm$0.020&-0.419$\pm$0.022&0.106$\pm$0.013&9.4$\pm$1.2&26$\pm$3&17.4(19)&75.4(20)\\
J083641.10-185947.0&8h36m41.2030s$\pm$0.0061s&-18$^o$59$^\prime$45.080$^{\prime\prime}$$\pm$0.$^{\prime\prime}$086&-0.038$\pm$0.007&-0.144$\pm$0.006&0.020$\pm$0.008&48.9$\pm$20.0&35$\pm$14&3.9(13)&5.6(14)\\
J131106.20+012254.3&13h11m6.0538s$\pm$0.0135s&1$^o$23$^\prime$2.850$^{\prime\prime}$$\pm$0.$^{\prime\prime}$2&0.280$\pm$0.016&-0.838$\pm$0.016&0.062$\pm$0.012&16.1$\pm$3.0&68$\pm$13&13.0(17)&34.8(18)\\
J154151.65-225024.9&15h41m52.2500s$\pm$0.0100s&-22$^o$50$^\prime$24.540$^{\prime\prime}$$\pm$0.$^{\prime\prime}$162&-0.857$\pm$0.012&-0.087$\pm$0.013&0.176$\pm$0.009&5.7$\pm$0.3&23$\pm$1&16.8(19)&354.0(20)\\
J154214.00+223005.2&15h42m14.7040s$\pm$0.0200s&22$^o$30$^\prime$9.098$^{\prime\prime}$$\pm$0.$^{\prime\prime}$33&-0.960$\pm$0.024&-0.374$\pm$0.026&0.096$\pm$0.041&10.4$\pm$4.5&51$\pm$22&23.6(13)&32.9(14)\\
J173835.53+273259.0&17h38m35.2890s$\pm$0.0076s&27$^o$33$^\prime$2.091$^{\prime\prime}$$\pm$0.$^{\prime\prime}$128&0.317$\pm$0.009&-0.321$\pm$0.011&0.128$\pm$0.010&7.8$\pm$0.6&17$\pm$1&19.4(27)&122.9(28)\\
J180435.37+311706.4&18h4m35.5700s$\pm$0.0082s&31$^o$17$^\prime$6.105$^{\prime\prime}$$\pm$0.$^{\prime\prime}$143&-0.269$\pm$0.010&0.035$\pm$0.011&0.080$\pm$0.010&12.6$\pm$1.6&16$\pm$2&23.8(27)&69.7(28)\\
J182831.08+265037.7&18h28m30.2950s$\pm$0.0059s&26$^o$50$^\prime$36.030$^{\prime\prime}$$\pm$0.$^{\prime\prime}$075&1.024$\pm$0.007&0.174$\pm$0.006&0.106$\pm$0.007&9.4$\pm$0.6&46$\pm$3&34.5(39)&234.6(40)\\
J205628.91+145953.2&20h56m28.3190s$\pm$0.0072s&14$^o$59$^\prime$47.804$^{\prime\prime}$$\pm$0.$^{\prime\prime}$097&0.812$\pm$0.009&0.534$\pm$0.008&0.140$\pm$0.009&7.1$\pm$0.5&33$\pm$2&27.1(35)&201.7(36)\\
J220905.73+271143.9&22h9m4.7813s$\pm$0.0111s&27$^o$11$^\prime$58.336$^{\prime\prime}$$\pm$0.$^{\prime\prime}$185&1.217$\pm$0.013&-1.372$\pm$0.015&0.147$\pm$0.011&6.8$\pm$0.5&59$\pm$4&15.1(19)&148.1(20)\\
J222055.31-362817.4&22h20m55.0650s$\pm$0.0122s&-36$^o$28$^\prime$16.312$^{\prime\prime}$$\pm$0.$^{\prime\prime}$227&0.283$\pm$0.013&-0.097$\pm$0.017&0.136$\pm$0.017&7.4$\pm$0.9&10$\pm$1&16.8(17)&69.0(18)\\

\enddata
\tablecomments{$^a$Proper motion in right ascension is given in units of $\arcsec yr^{-1}$ and includes the correction for $cos(\delta)$. $^b$ $\chi^2$ value with degrees of freedom in parentheses. Fit includes parallax.$^c$ $\chi^2$ value with degrees of freedom in parentheses. Fit does not include parallax.}
\end{deluxetable}

\begin{deluxetable}{lccc}
%\tabletypesize{\tiny}
%\rotate
%\setlength{\tabcolsep}{0.02in}
\centering
\tablecaption{ Kirkpatrick et al. Parallax Comparison \label{PlxComp}}
\tablehead{
\colhead{WISE Designation}&\colhead{Kirkpatrick Distance (pc)$^1$ }&\colhead{Keck Distance (pc)}&\colhead{Kirkpatrick/Keck Ratio}}
\startdata
J014656.66+423410.0&6.3&10.6$\pm$1.5&0.6$\pm$0.1\\
J031325.96+780744.2&8.6&6.5$\pm$0.6&1.3$\pm$0.1\\
J033515.01+431045.1&14.0&14.3$\pm$1.7&1.0$\pm$0.1\\
J041022.71+150248.5$^2$&6.1&6.2$\pm$0.4&1.0$\pm$0.1\\
J071322.55-291751.9&7.1&9.4$\pm$1.2&0.8$\pm$0.1\\
J083641.12-185947.2&22.2&48.9$\pm$20.0&0.5$\pm$0.2\\
J131106.24+012252.4&13.6&16.1$\pm$3.0&0.8$\pm$0.2\\
J154151.66-225025.2&4.2&5.7$\pm$0.3&0.7$\pm$0.0\\
J154214.00+223005.2&12.6&10.4$\pm$4.5&1.2$\pm$0.5\\
J173835.53+273258.9$^2$ &9.0&7.8$\pm$0.6&1.2$\pm$0.1\\
J180435.40+311706.1&9.2&12.6$\pm$1.6&0.7$\pm$0.1\\
J182831.08+265037.8$^2$ &8.2&9.4$\pm$0.6&0.9$\pm$0.1\\
J205628.90+145953.3&5.2&7.1$\pm$0.5&0.7$\pm$0.0\\
J222055.32-362817.5&8.1&7.4$\pm$0.9&1.1$\pm$0.1\\ \hline
Average$^3$&&&0.9$\pm$0.2\\
\enddata
\tablecomments{$^1$``Adopted'' distance in Kirkpatrick et al. (2012); $^2$ Kirkpatrick et al (2012) distance was based on trig parallax \citep{Marsh2013}.$^3$Average value of distance ratio for 12 sources with fractional uncertainties $<$20\%.}
\end{deluxetable}

\begin{deluxetable}{lccccc}
\tabletypesize{\tiny}
%\rotate
%\setlength{\tabcolsep}{0.02in}
\centering
\tablecaption{ Dupuy \& Kraus Parallax Comparison \label{DupuyPlx}}
\tablehead{
\colhead{WISE Designation}&\colhead{$\Delta\ Parallax/\sigma_{tot}^1$ }&\colhead{$\Delta\ \mu RA/\sigma_{tot}^1$}&\colhead{$\Delta\ \mu Dec/\sigma_{tot}^a$}&\colhead{Dupuy Distance (pc)}&\colhead{This paper, (pc)}}
\startdata
J041022.71+150248.5&1.6&-0.2&0.0&7.6$\pm$0.9&6.2$\pm$0.4\\
J154151.65-225025.2$^2$&3.2&-0.7&-0.1&13.5$\pm$5.7&5.7$\pm$0.3\\
J173835.52+273258.9&1.3&-0.4&0.1&9.8$\pm$1.7&7.8$\pm$0.6\\
J180435.40+311706.1&1.3&1.0&0.0&16.7$\pm$3.1&12.6$\pm$1.6\\
J182831.08+265037.8&2.3&-0.3&0.0&14.3$\pm$2.9&9.4$\pm$0.6\\
J205628.90+145953.3&-0.1&-1.1&0.0&6.9$\pm$1.1&7.1$\pm$0.5\\
\enddata
\tablecomments{$^a$Difference between values in this paper and \citet{Dupuy2013} relative to the combined uncertainties.$^2$Obvious confusion in Spitzer data with neighboring star affects Spitzer-only parallax determination.}
\end{deluxetable}

\begin{deluxetable}{lc|ccccccc}
%\tabletypesize{\tiny}
%\rotate
%\setlength{\tabcolsep}{0.02in}
\centering
\tablecaption{BT-Settl Model Parameters$^a$ \label{SettlMass}}
\tablehead{
\colhead{WISE Designation}&\colhead{Spectral}&\colhead{Age }&\colhead{Mass}&\colhead{T$_{eff}$ }&\colhead{Radius }&\colhead{Log g }&\colhead{$\chi^2$}&\colhead{d.o.f}\\
& Type& \colhead{(Gyr)}& \colhead{(M$_{Jup}$)}&\colhead{(K)}&\colhead{(R$_{Jup}$) } &\colhead{(cm s$^{-2}$)}}
\startdata
J014656.66+423410.0&Y0&3.4$\pm$2.5&14.4$\pm$5.5&451$\pm$23&0.97&4.61&9.1&2\\
J031358.93+780748.9&T8.5&8.8$\pm$0.4&26.2$\pm$1.7&651$\pm$46&0.88&4.95&17.4&2\\
J033515.01+431045.1&T9&8.0$\pm$0.4&21.8$\pm$1.1&465$\pm$23&0.90&4.84&54.2&3\\
J041022.71+150248.4&Y0&8.0$\pm$0.4&18.2$\pm$0.9&409$\pm$20&0.92&4.75&25.4&3\\
J071322.55-291751.9&Y0&7.5$\pm$1.1&19.5$\pm$1.8&422$\pm$21&0.92&4.78&18.1&2\\
J083641.10-185947.0&T8p&4.2$\pm$3.1&26.2$\pm$9.1&662$\pm$52&0.90&4.93&2.9&2\\
J131106.20+012254.3&T9&7.6$\pm$2.6&27.0$\pm$3.5&641$\pm$53&0.88&4.96&7.2&2\\
J154151.65-225024.9$^b$&Y0.5&5.0$\pm$2.0&12.0$\pm$3.0&350$\pm$25&1.0&4.50&410&3\\
J154214.00+223005.2&T9.5&8.5$\pm$0.4&19.4$\pm$1.0&477$\pm$24&0.91&4.78&11.2&4\\
J173835.53+273259.0&Y0&8.2$\pm$0.4&18.6$\pm$0.9&409$\pm$20&0.92&4.76&47.1&3\\
J180435.37+311706.4&T9.5:&5.2$\pm$1.1&27.9$\pm$2.1&583$\pm$29&0.89&4.97&2.4&2\\
J182831.08+265037.7$^b$&$\ge$Y2&5.0$\pm$2.0&12.0$\pm$3.0&350$\pm$25&1.0&4.50&3,700&4\\
J205628.91+145953.2&Y0&8.0$\pm$2.0&17.0$\pm$0.9&407$\pm$20&0.93&4.71&112.5&3\\
J220905.73+271143.9$^b$&Y0:&5.0$\pm$2.0&12.0$\pm$0.6&350$\pm$25&1.0&4.10&1,000&2\\
J222055.31-362817.4&Y0&7.6$\pm$0.4&14.1$\pm$0.8&404$\pm$20&0.95&4.61&5.5&3\\ \hline
Average&&6.6&19.4&473&0.93&4.72&387.5\\
Dispersion&&1.9&5.7&114&0.04&0.23&993.6\\
Median&&7.6&19.0&437&0.92&4.76&21.8\\
\enddata
\tablecomments{$^a$Fits of photometry to BT-Settl model \citep{Allard2003,Allard2010}. Uncertainties in the model parameters are the larger of the dispersion in Monte Carlo calculations or 10\%. $^b$As discussed in the text, these models fits were derived using a coarse low temperature grid ($\leq$400 K) with uncertainties based on grid spacing. These model values should be regarded as quite uncertain.}
\end{deluxetable}

\begin{deluxetable}{lc|cccccccc|ccc}
\tabletypesize{\tiny}
%\rotate
%\setlength{\tabcolsep}{0.02in}
\centering
\tablecaption{Morley$^a$ Model Parameters \label{MorleyMass}}
\tablehead{
\colhead{WISE Designation}&\colhead{Spectral}&\colhead{Age$^b$ }&\colhead{Mass}&\colhead{T$_{eff}$}&\colhead{Rad}&\colhead{Log g}&\colhead{Sed}&\colhead{$\chi^2$}&\colhead{d.o.f}&\colhead{$\Delta T^c$}&\colhead{Mass}&\colhead{$\Delta$Age}\\
& \colhead{Type}& \colhead{(Gyr)}& \colhead{(M$_{Jup}$)}& \colhead{(K)}& \colhead{(R$_{Jup}$)}&\colhead{cm s$^{-2}$}& & &&\colhead{(K)}&\colhead{Ratio$^d$}&\colhead{Gyr$^e$}}
\startdata
J014656.66+423410.0&Y0&6&31.9$\pm$0.1&570$\pm$13&0.89&5.00$\pm$0.05&5&46.5&2&119&2.0&2.7\\
J031358.93+780748.9&T8.5&4&32.4$\pm$1.1&662$\pm$7&0.90&5.00$\pm$0.05&3&28.7&2&11&2.0&-5.0\\
J033515.01+431045.1&T9&3&25.2$\pm$3.9&605$\pm$10&0.95&4.85$\pm$0.22&2&24.6&3&140&3.0&-5.2\\
J041022.71+150248.4&Y0&6&25.3$\pm$1.8&491$\pm$5&0.92&4.87$\pm$0.10&5&137.6&3&82&3.0&-2.2\\
J071322.55-291751.9&Y0&8&31.5$\pm$0.1&513$\pm$7&0.88&5.00$\pm$0.00&4&28.0&2&90&2.0&0.9\\
J083641.10-185947.0&T8p&3&33.1$\pm$1.0&765$\pm$18&0.91&4.99$\pm$0.04&2&15.8&2&103&2.0&-1.7\\
J131106.20+012254.3&T9&3&31.0$\pm$3.4&672$\pm$12&0.91&4.97$\pm$0.16&5&9.2&2&31&2.0&-4.3\\
J154151.65-225024.9$^b$&Y0.5&14&30.8$\pm$0.0&441$\pm$4&0.87&5.00$\pm$0.05&2&193.2&3&91&3.0&8.9\\
J154214.00+223005.2&T9.5&6&31.8$\pm$0.1&563$\pm$5&0.89&5.00$\pm$0.00&4&36.6&4&86&4.0&-2.2\\
J173835.53+273259.0&Y0&8&31.3$\pm$0.6&514$\pm$6&0.89&5.00$\pm$0.03&5&120.2&3&105&3.0&0.1\\
J180435.37+311706.4&T9.5:&3&32.3$\pm$1.4&706$\pm$7&0.91&4.99$\pm$0.06&3&47.3&2&122&2.0&-2.1\\
J182831.08+265037.7$^b$&$\ge$Y2&15&22.0$\pm$1.0&400$\pm$40&0.74&5.00$\pm$0.05&2&1,468.9&4&50&4.0&10.0\\
J205628.91+145953.2&Y0&10&31.2$\pm$0.1&488$\pm$4&0.88&5.00$\pm$0.01&5&216.0&3&81&3.0&1.8\\
J220905.73+271143.9$^b$&Y0:&15&22.0$\pm$1.0&400$\pm$40&0.74&5.00$\pm$0.05&2&387.9&2&50&2.0&10.0\\
J222055.31-362817.4&Y0&8&31.3$\pm$1.4&525$\pm$6&0.89&4.99$\pm$0.06&2&57.4&3&127&3.0&0.3\\ \hline
Average&&7.4&29.4&556&0.88&4.98&&197&&83&1.6&0.8\\
Dispersion&&4.5&4.0&114&0.06&4.00&&381&&88&1.6&5.3\\
Median&&6.2&31.3&538&0.89&5.00&&47&&36&0.4&-0.8\\
\enddata
\tablecomments{$^a$Fits of photometry to Morley et al. models as described in \citet{Morley2012, Leggett2012,
 Saumon2008}.$^b$ Ages interpolated from Figure 4 in \citet{Saumon2008} for cloudy models with {\it sed}=2; $^c$ In the sense T$_{Morley}$-T$_{BTSettl}$;$^d$In the sense M$_{Morley}$/M$_{BTSettl}$;$^e$ In the sense Age$_{Morley}$-Age$_{BTSettl}$ }
\end{deluxetable}

\clearpage

\clearpage

%********************************
%HST/Spitzer/Keck Images
%*******************************

%********************************
%HST/Spitzer/Keck Images
%*******************************
%******WISE0146

%******WISE0313
 \begin{figure*}
 \includegraphics[scale=0.85]{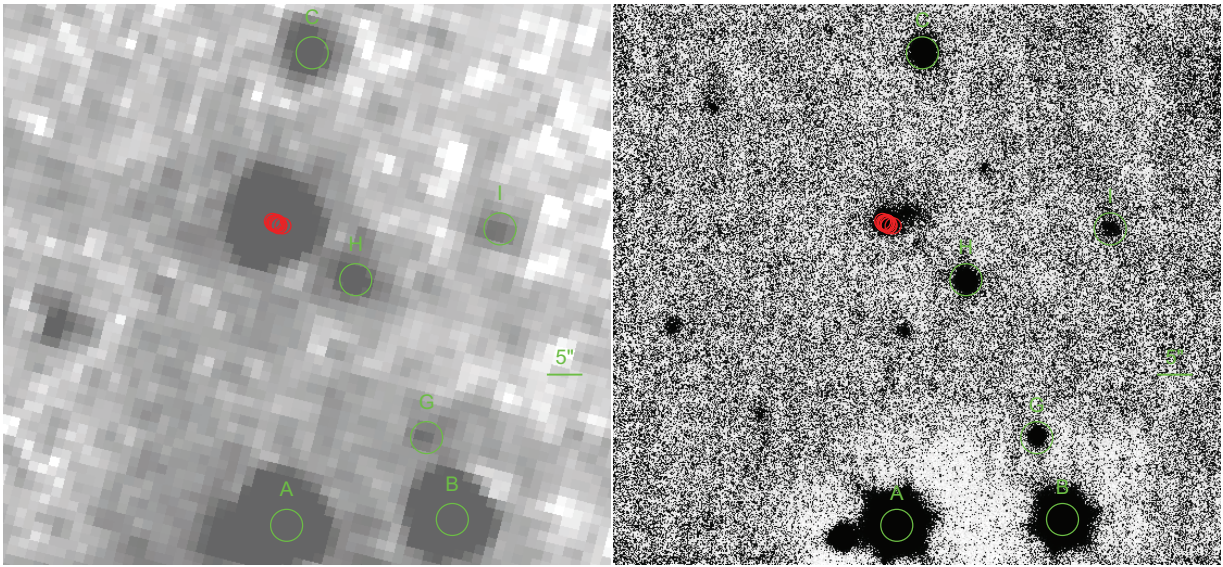}\caption{Spitzer (left) and Keck (right) images at 4.6 $\mu$m and 1.65 $\mu$m, respectively, of WISE 0313+7807 with the reference stars used for the co-registration of the fields circled in green. The positions of the brown dwarf are marked in red. A scale bar denotes 5\arcsec. A faint galaxy near the source does not affect the astrometry or the mid-IR photometry. \label{w0313image}}
 \end{figure*}

\clearpage
%****WISE0335+4310
 \begin{figure*}
 \includegraphics[scale=0.85]{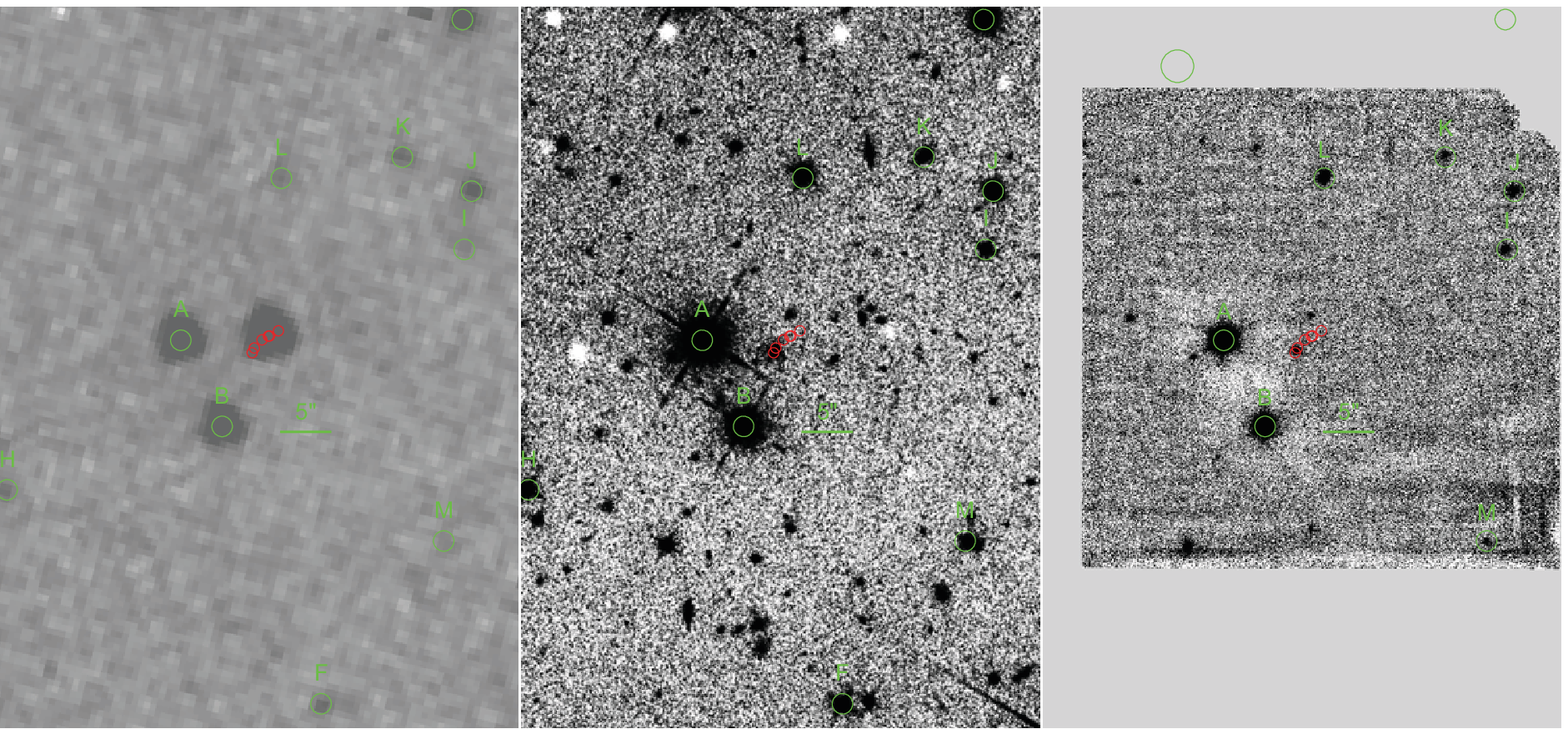}\caption{Spitzer Space Telescope (left), Hubble Space Telescope (HST,center) and Keck (right) images at 4.6 $\mu$m, F125W and H, respectively, of WISE 0335+4310 with the reference stars used for the co-registration of the fields circled in green. The positions of the brown dwarf are marked in red.	 A scale bar denotes 5\arcsec. \label{w0335image}}
 \end{figure*}

%****WISE0410+1502
 \begin{figure*}
 \includegraphics[scale=0.85]{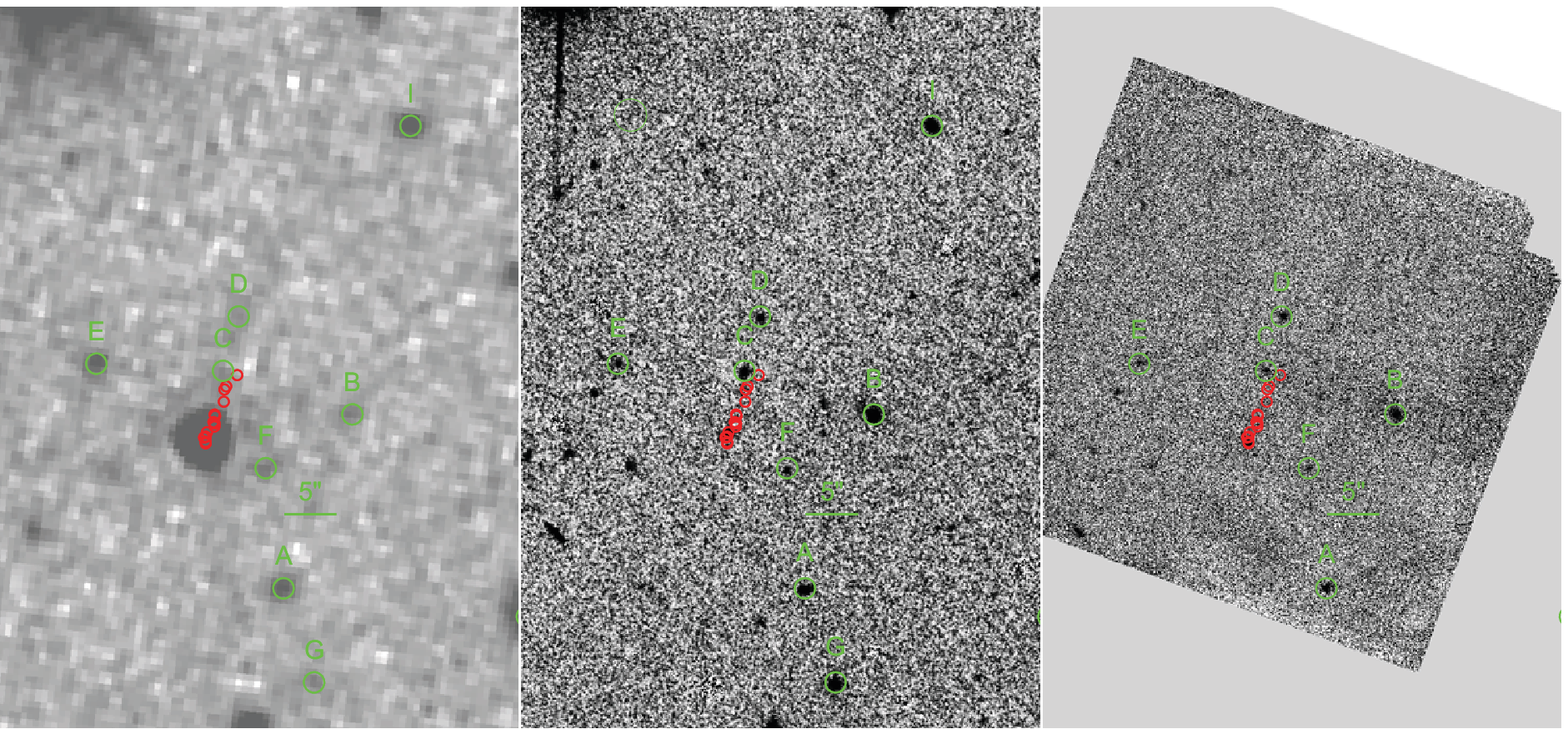}\caption{Spitzer Space Telescope (left), Hubble Space Telescope (HST,center) and Keck (right) images at 4.6 $\mu$m, F125W and H, respectively, of WISE 0410+1502 with the reference stars used for the co-registration of the fields circled in green. The positions of the brown dwarf are marked in red. A scale bar denotes 5\arcsec. \label{w0410image}}
 \end{figure*}

\clearpage

%***WISE0713-2917

 \begin{figure*}
 \includegraphics[scale=0.85]{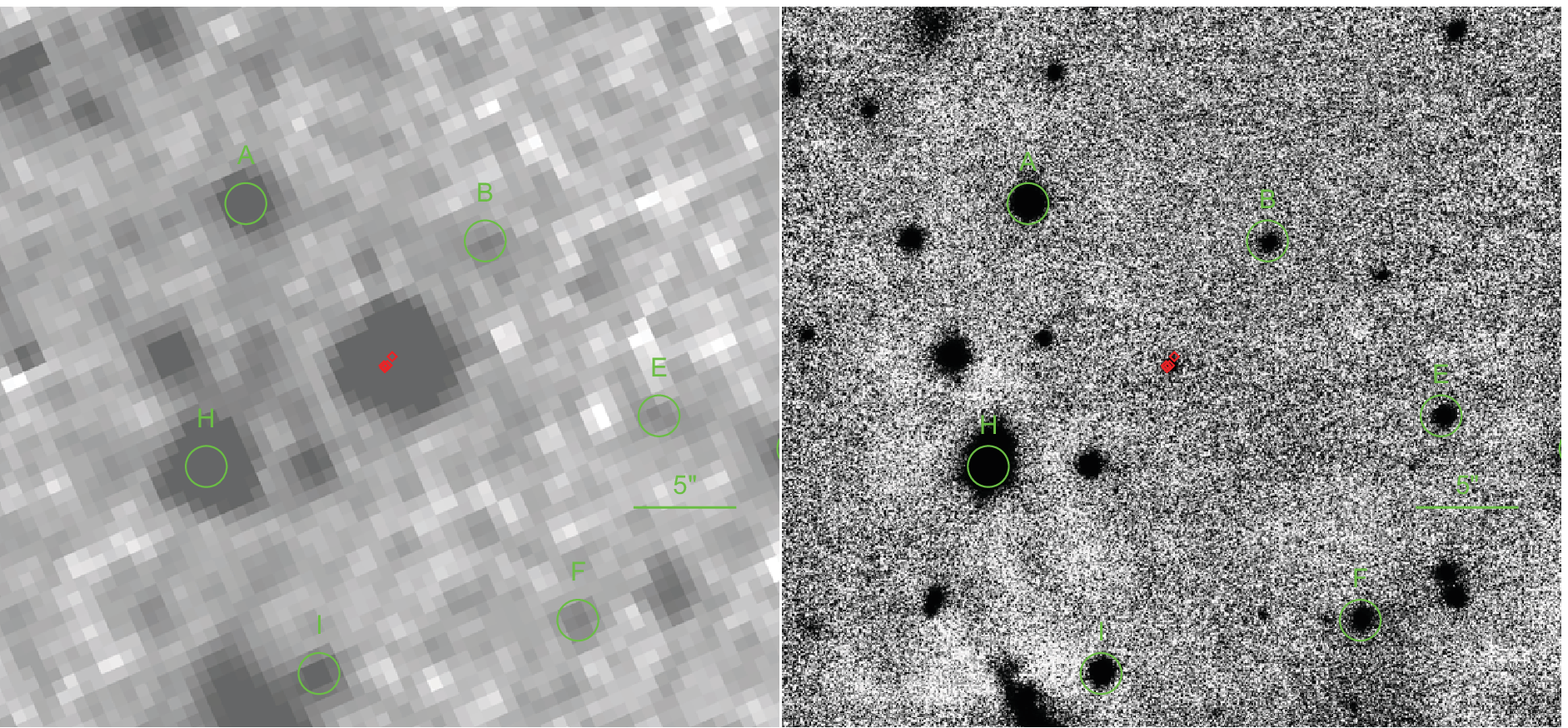}\caption{Spitzer (left) and Keck (right) images at 4.6 $\mu$m and 1.65 $\mu$m, respectively, of WISE 0713-2917 with the reference stars used for the co-registration of the fields circled in green. The positions of the brown dwarf are marked in red. North is up and East is to the left. A scale bar denotes 5\arcsec. \label{w0713image}}
 \end{figure*}

%***********WISE0836-1859

 \begin{figure*}
 \includegraphics[scale=0.85]{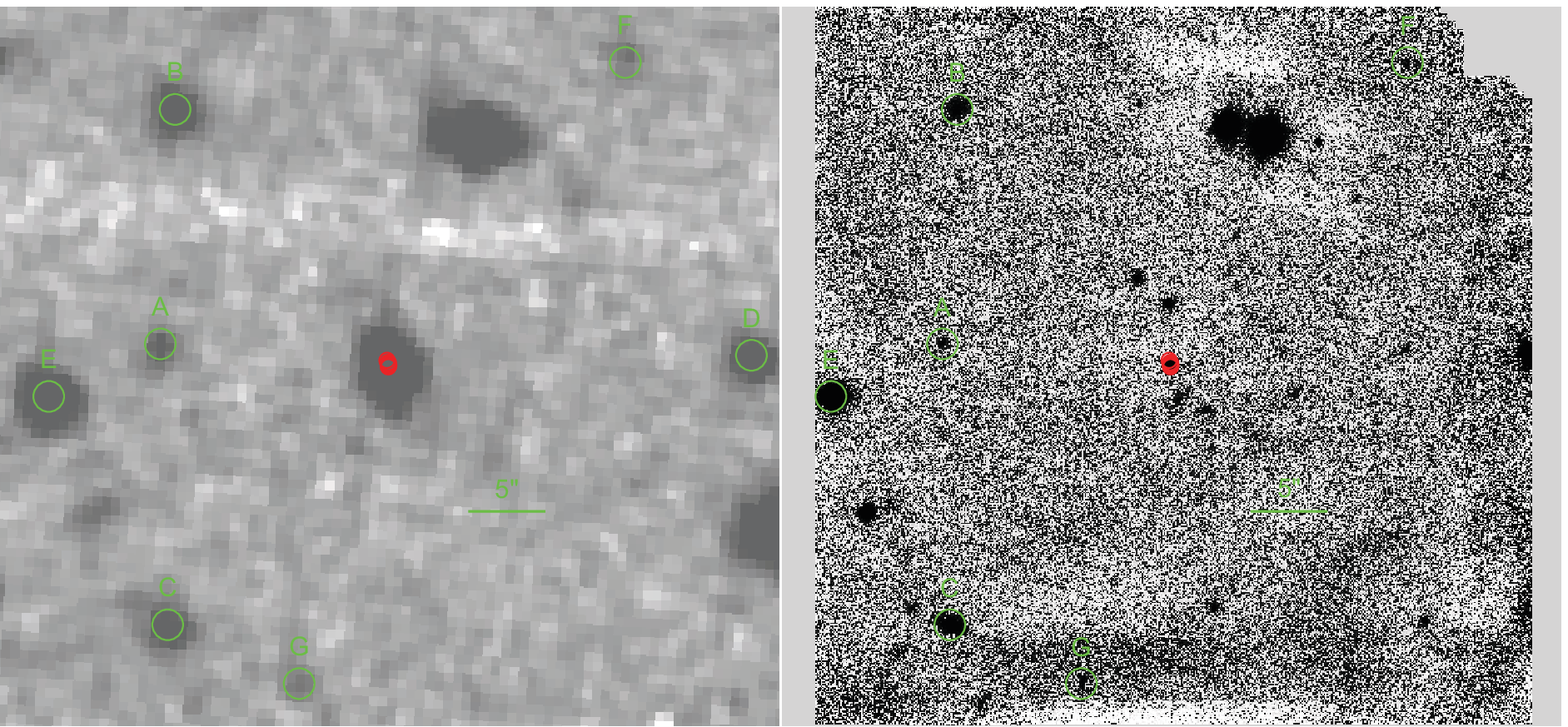}\caption{Spitzer (left) and Keck (right) images at 4.6 $\mu$m and 1.65 $\mu$m, respectively, of WISE 0836-1859 with the reference stars used for the co-registration of the fields circled in green. The positions of the brown dwarf are marked in red. North is up and East is to the left. A scale bar denotes 5\arcsec. \label{w0836image}}
 \end{figure*}
\clearpage

%***WISE1311+0122
 \begin{figure*}
 \includegraphics[scale=0.85]{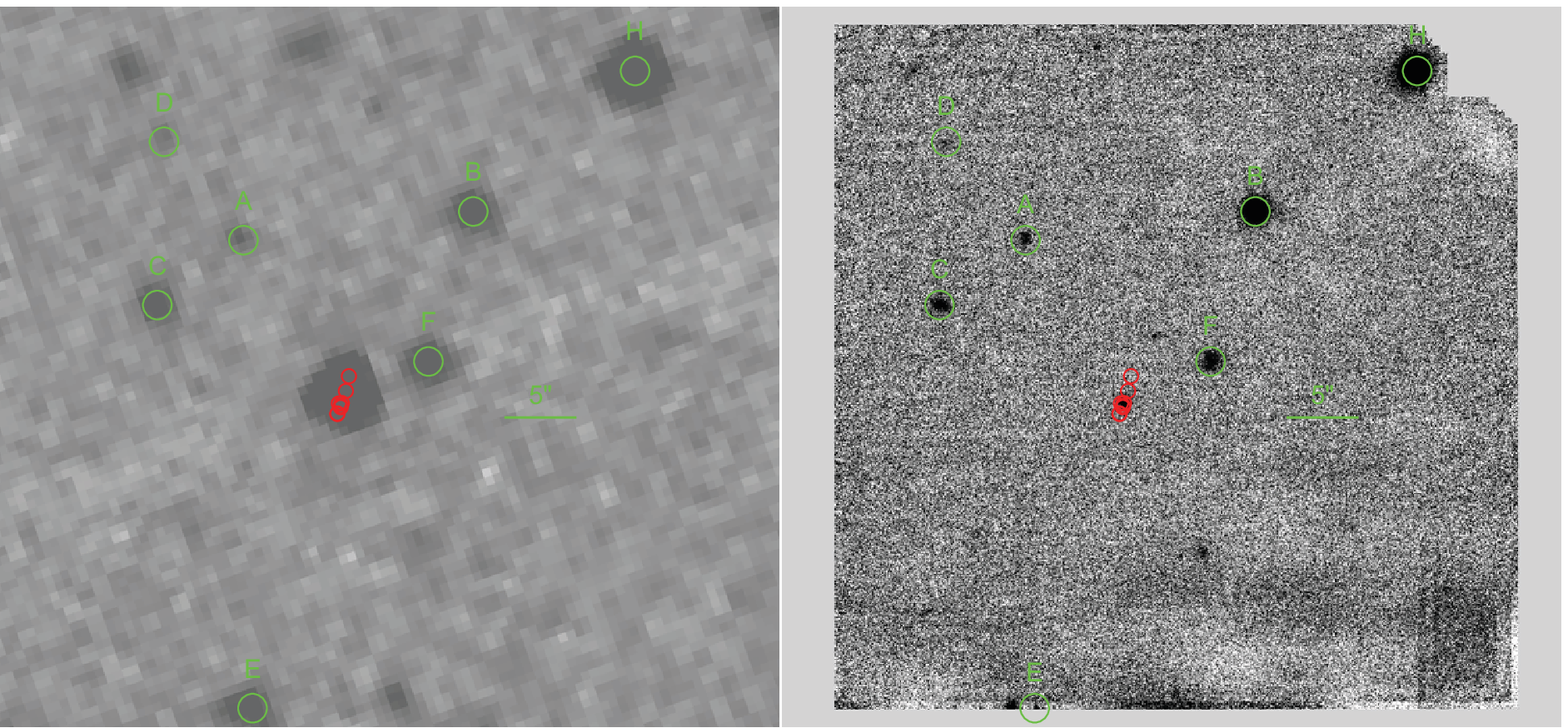}\caption{Spitzer (left) and Keck (right) images at 4.6 $\mu$m and 1.65 $\mu$m, respectively, of WISE 1311+0122 with the reference stars used for the co-registration of the fields circled in green. The positions of the brown dwarf are marked in red. North is up and East is to the left. A scale bar denotes 5\arcsec. \label{w1311image}}
 \end{figure*}

%****WISE1541-2250

 \begin{figure*}
 \includegraphics[scale=0.85]{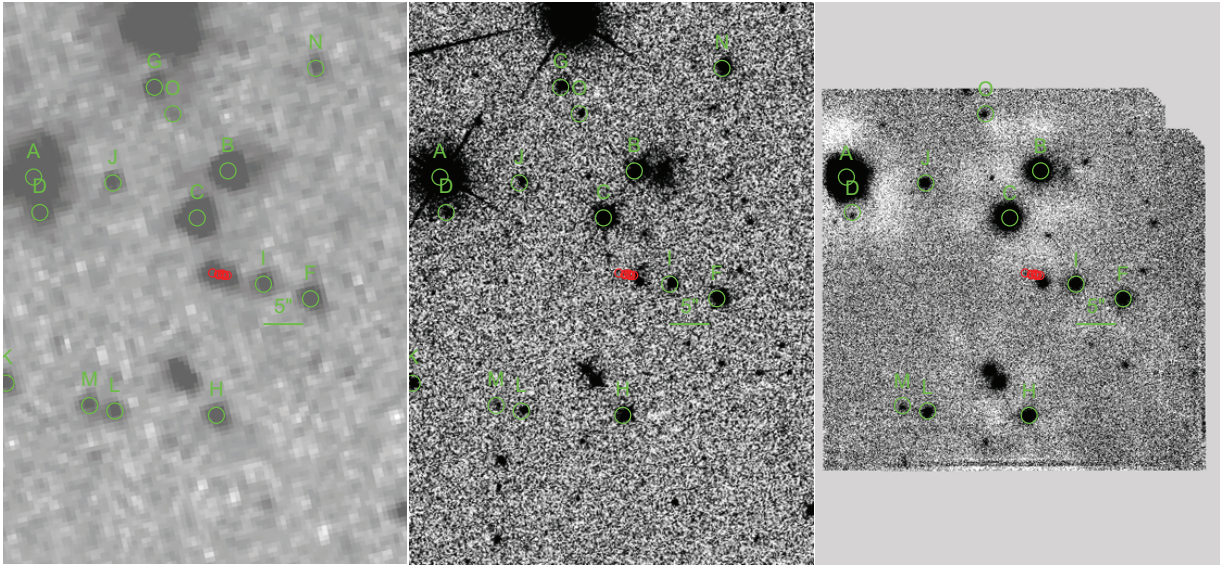}\caption{Spitzer Space Telescope (left), Hubble Space Telescope (HST,center) and Keck (right) images at 4.6 $\mu$m, F125W and H, respectively, of WISE1541-2250 with the reference stars used for the co-registration of the fields circled in green. The positions of the brown dwarf are marked in red. North is up and East is to the left. A scale bar denotes 5\arcsec. Confusion with the star close to the brown dwarf is a problem for later Spitzer epochs and accordingly were not used in the astrometric solution. \label{w1541image}}
 \end{figure*}

\clearpage

%****WISE1542+2230
\clearpage
 \begin{figure*}
 \includegraphics[scale=0.85]{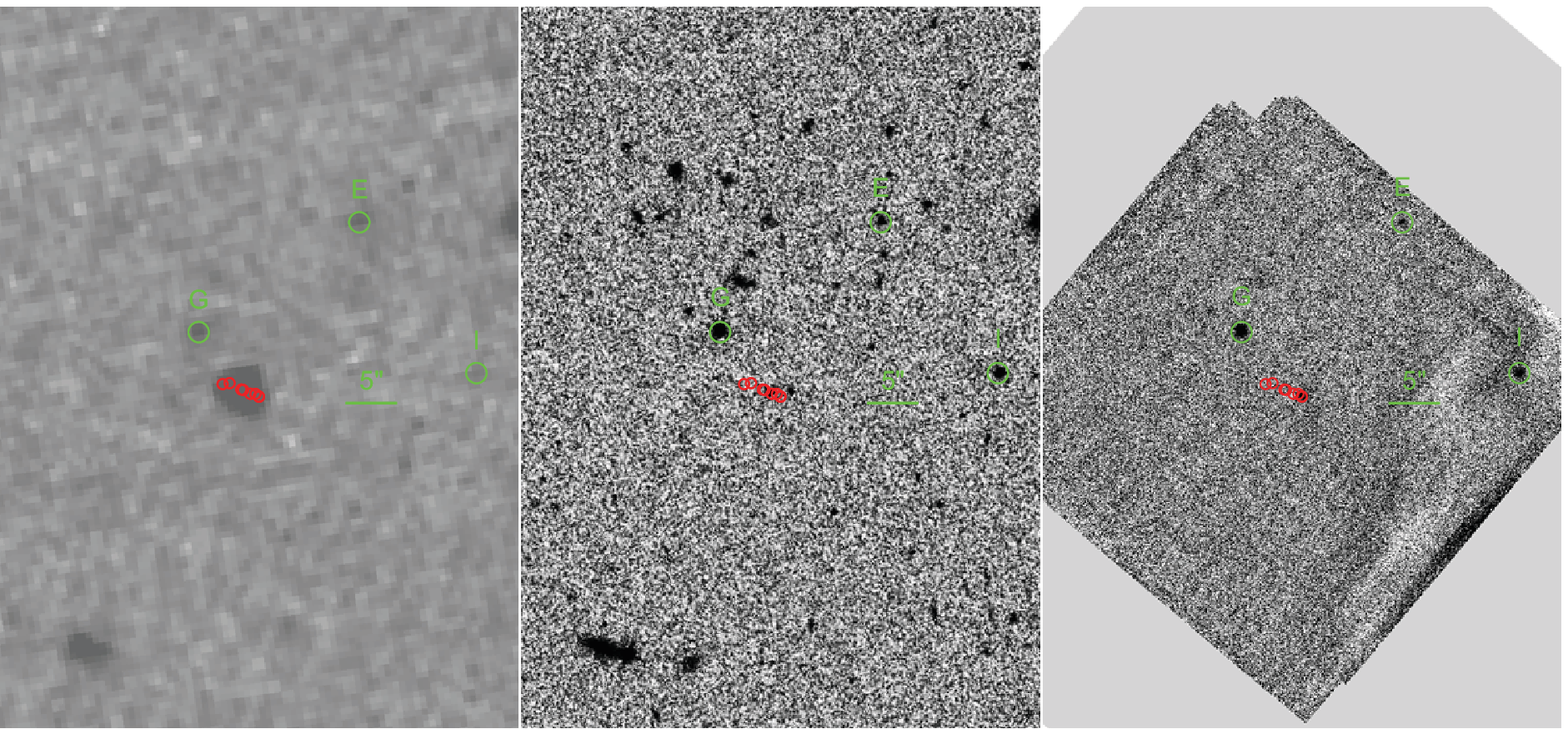}\caption{Spitzer Space Telescope (left), Hubble Space Telescope (HST,center) and Keck (right) images at 4.6 $\mu$m, F125W and H, respectively, of WISE 1542+2230 with the reference stars used for the co-registration of the fields circled in green. The positions of the brown dwarf are marked in red. North is up and East is to the left. A scale bar denotes 5\arcsec. \label{w1542image}}
\end{figure*}

%*****WISE 1738+2732
\begin{figure*}
 \includegraphics[scale=0.85]{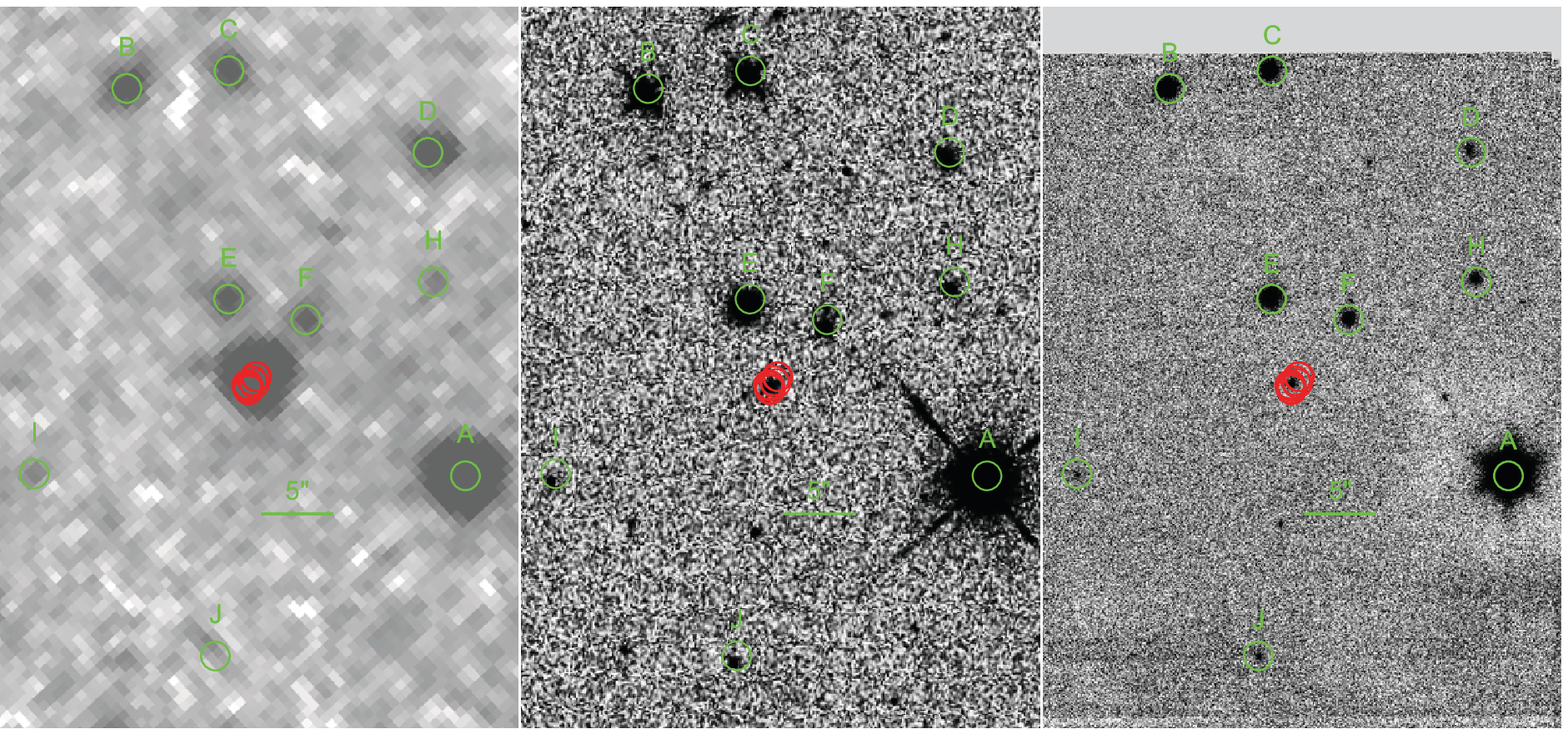}\caption{Spitzer Space Telescope (left), Hubble Space Telescope (HST,center) and Keck (right) images at 4.6 $\mu$m, F125W and H, respectively, of WISE 1738+2732 with the reference stars used for the co-registration of the fields circled in green. The positions of the brown dwarf are marked in red. North is up and East is to the left. A scale bar denotes 5\arcsec. \label{w1738image}}
 \end{figure*}
\clearpage

%******WISE1804+3117

\clearpage
 \begin{figure*}
 \includegraphics[scale=0.85]{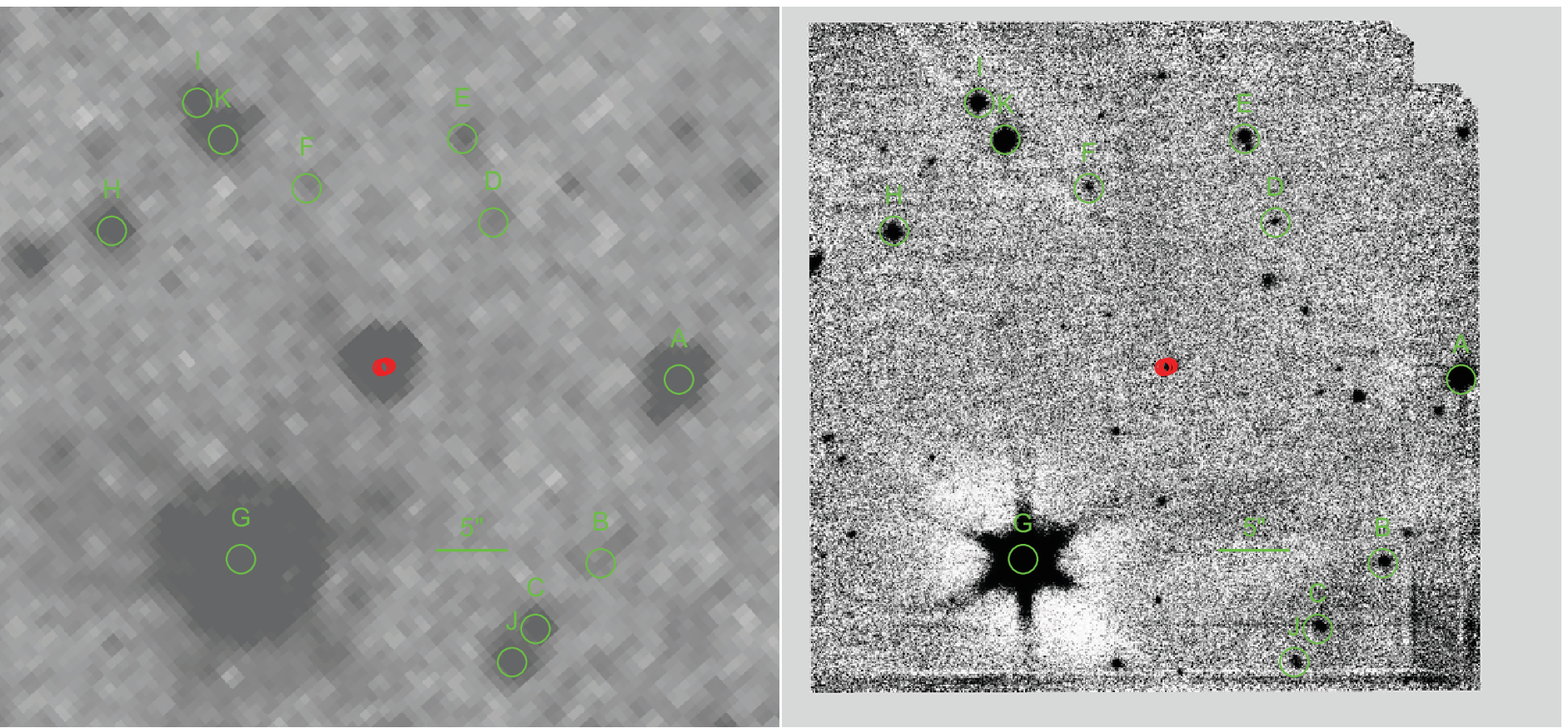}\caption{Spitzer (left) and Keck (right) images at 4.6 $\mu$m and 1.65 $\mu$m, respectively, of WISE1804+3117 with the reference stars used for the co-registration of the fields circled in green. The positions of the brown dwarf are marked in red. North is up and East is to the left. A scale bar denotes 5\arcsec. \label{w1804image}}
 \end{figure*}

%******WISE1828
 \begin{figure*}
 \includegraphics[scale=0.85]{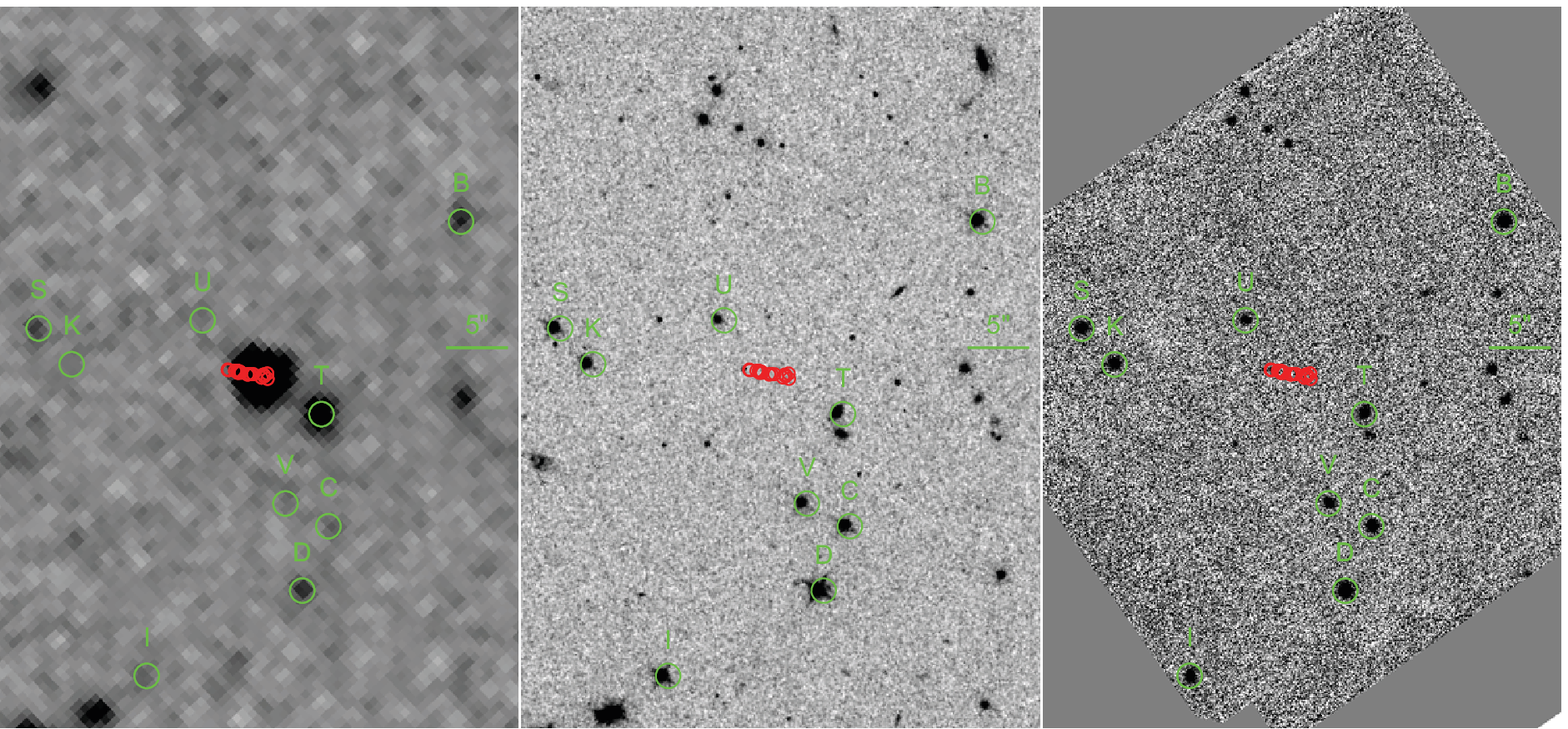}\caption{Spitzer Space Telescope (left), Hubble Space Telescope (HST,center) and Keck (right) images at 4.6 $\mu$m, F125W and H, respectively, of WISE 1828+2650 with the reference stars used for the co-registration of the fields circled in green. The positions of the brown dwarf are marked in red. North is up and East is to the left. A scale bar denotes 5\arcsec. \label{w1828image}}
 \end{figure*}
\clearpage

%**********WISE2056
 \begin{figure*}
 \includegraphics[scale=0.85]{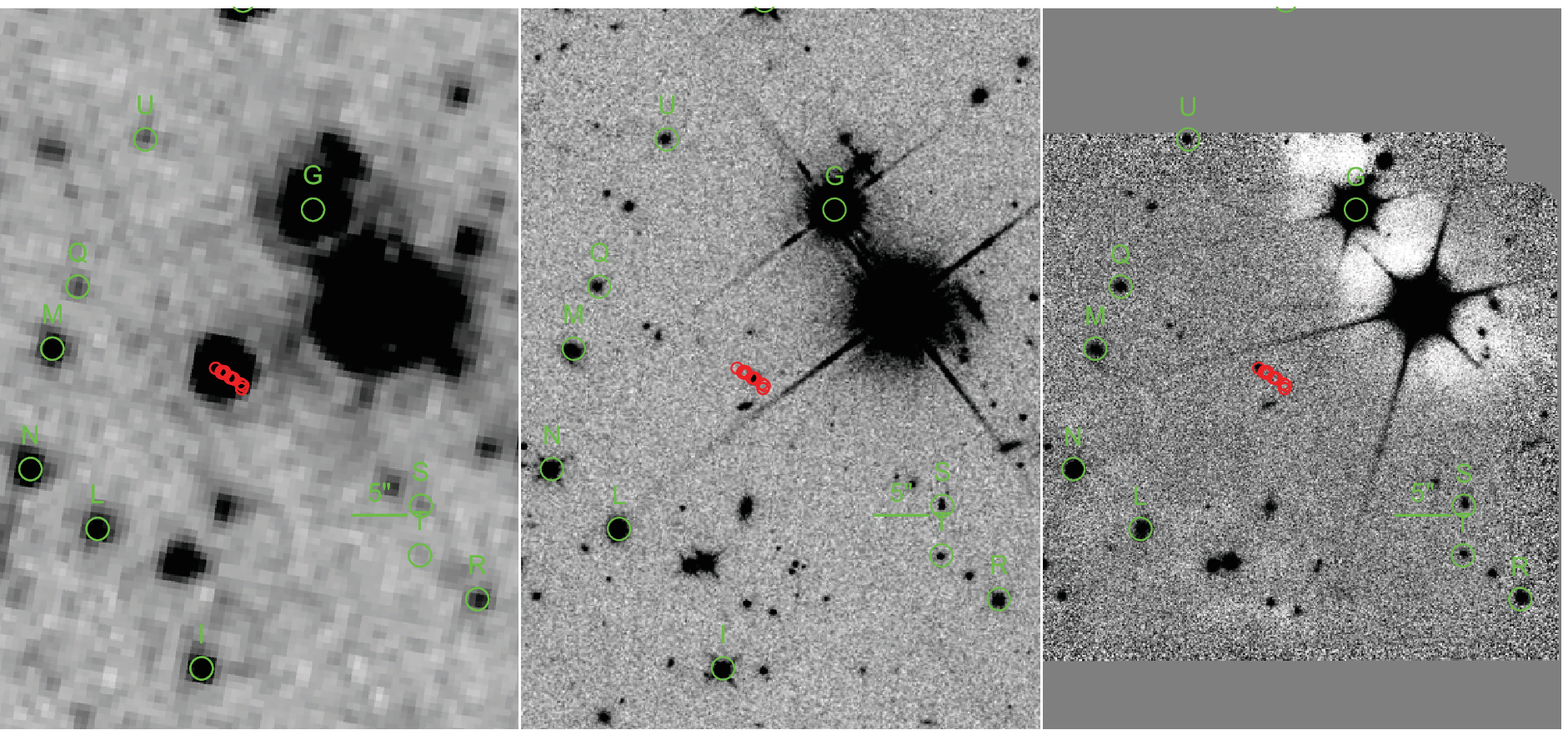}\caption{Spitzer Space Telescope (left), Hubble Space Telescope (HST,center) and Keck (right) images at 4.6 $\mu$m, F125W and H, respectively, of WISE2056+1459 with the reference stars used for the co-registration of the fields circled in green. The positions of the brown dwarf are marked in red. North is up and East is to the left. A scale bar denotes 5\arcsec. \label{w2056image}}
 \end{figure*}

 \begin{figure*}
 \includegraphics[scale=0.85]{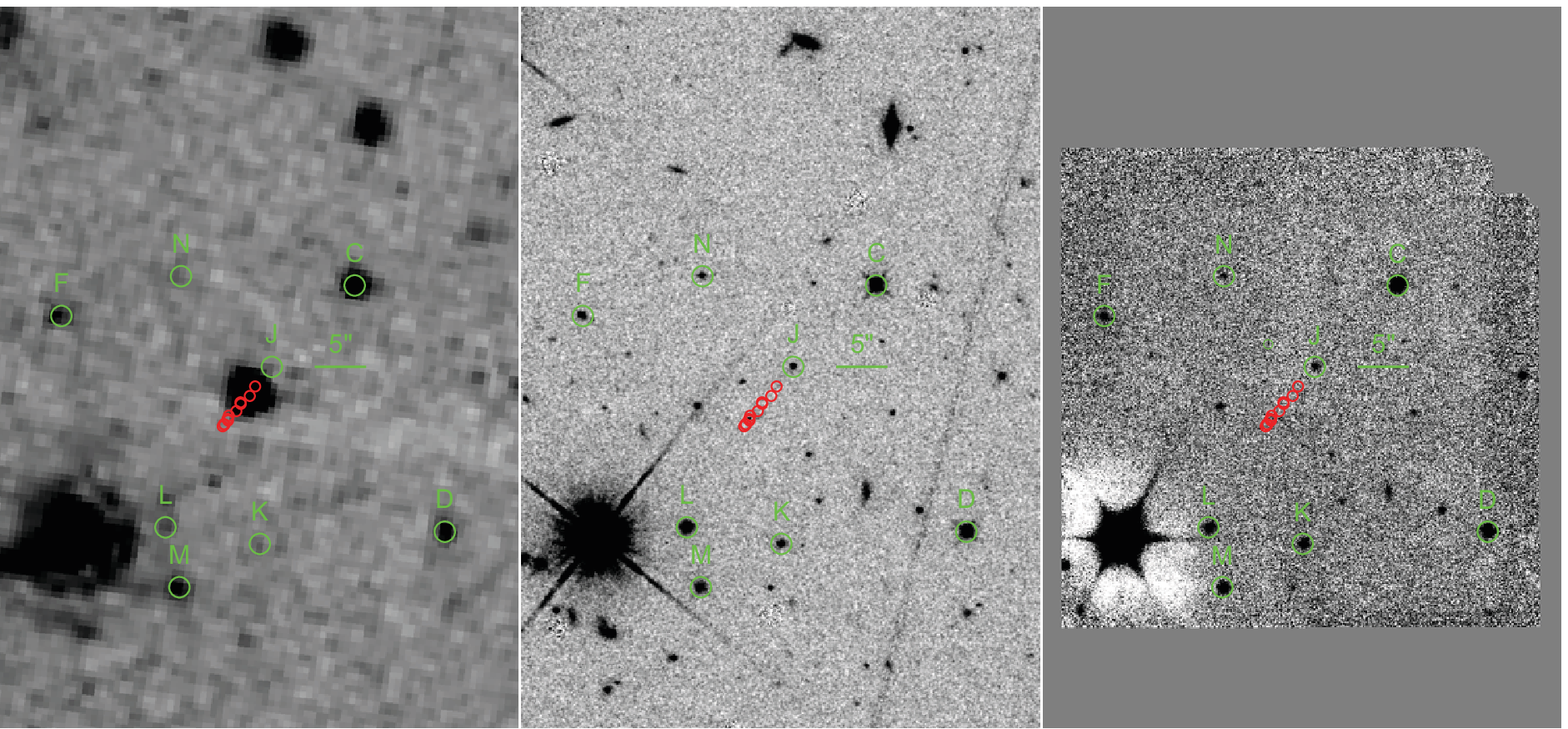}
\caption{Spitzer Space Telescope (left), Hubble Space Telescope (HST,center) and Keck (right) images at 4.6 $\mu$m, F125W and H, respectively, of WISE 2209+2711 with the reference stars used for the co-registration of the fields circled in green. The position of the brown dwarf is marked in red. North is up and East is to the left. A scale bar denotes 5\arcsec.
\label{WISE2209image}}
\end{figure*}
\clearpage

\clearpage
%****************WISE2220
 \begin{figure*}
 \includegraphics[scale=0.85]{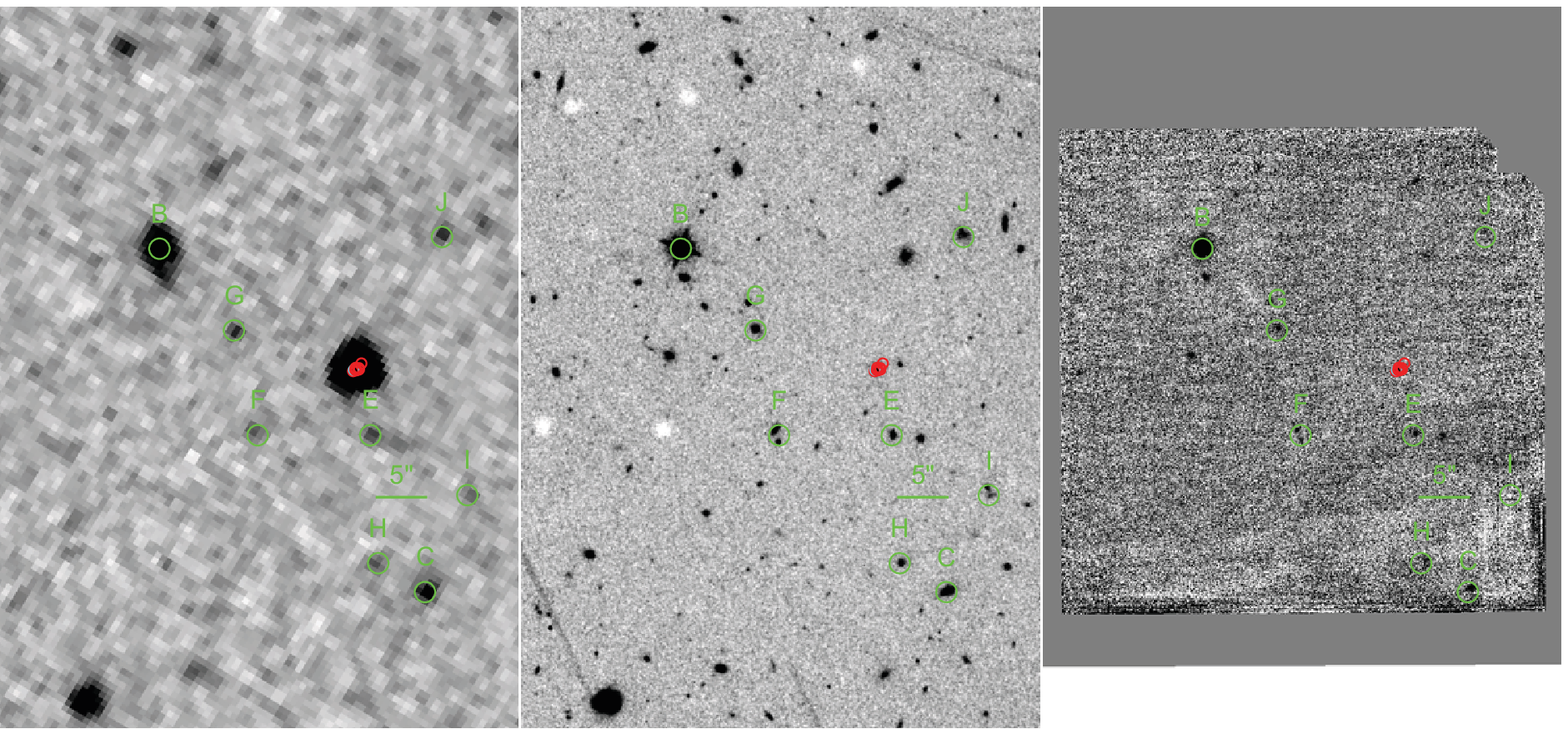}\caption{Spitzer Space Telescope (left), Hubble Space Telescope (HST,center) and Keck (right) images at 4.6 $\mu$m, F125W and H, respectively, of WISE 2220-3628 with the reference stars used for the co-registration of the fields circled in green. The position of the brown dwarf is marked in red. North is up and East is to the left. A scale bar denotes 5\arcsec. \label{w2220image}}
 \end{figure*}

\clearpage

%********************************
%Parallactic Motions
%*******************************

\begin{figure}
\begin{center}$
\begin{array}{c}
\includegraphics[width=6in]{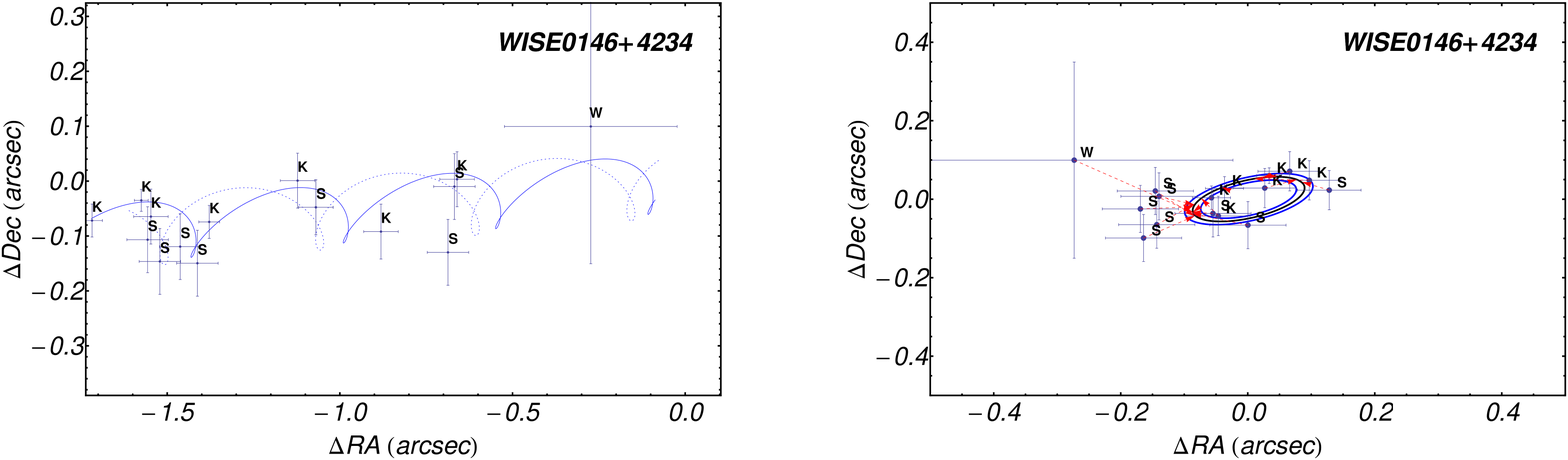}\\
\includegraphics[width=6in]{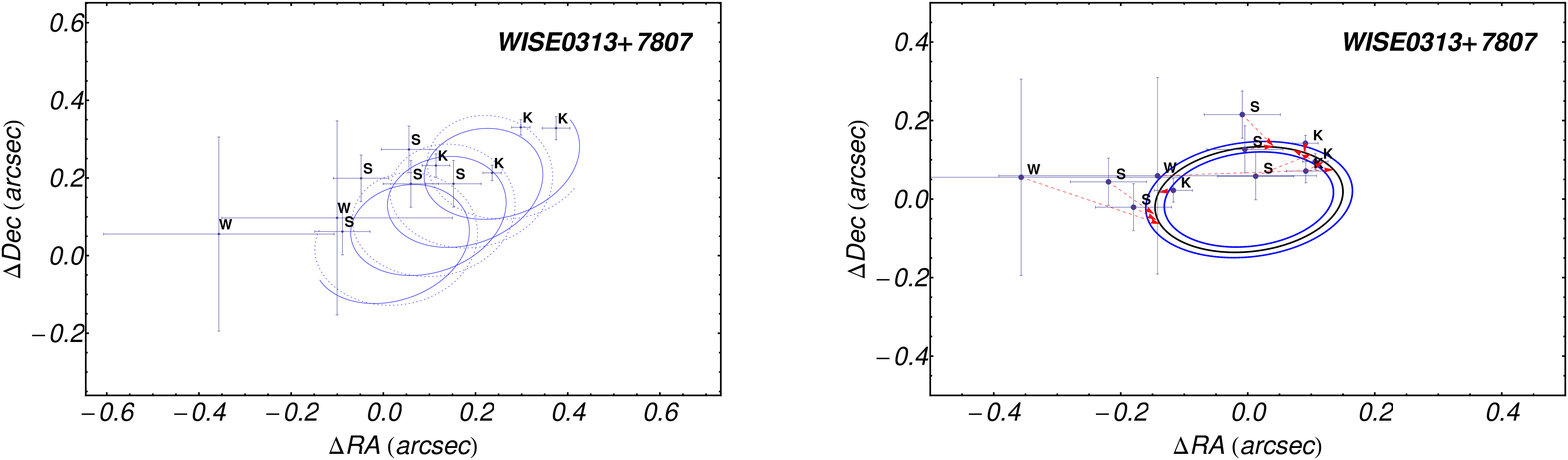}\\
\end{array}$
\end{center}
\caption{The parallactic solutions for two brown dwarfs, WISE0146+4234 (top) and WISE0313+7807 (bottom). In both figures, the left hand panel shows the total motion including both proper motion and parallax as seen from earth-centered observatories (solid line; WISE (W), Keck (K), or Hubble (H)) and the earth-trailing Spitzer (S) telescope (dotted line). The right-hand panel shows the derived parallactic ellipse with observations from the various facilities denoted with appropriate letter (K---Keck, S---Spitzer, W---WISE, H---Hubble. Arrows connect the data points to the points on ellipse appropriate to the observing epochs. Ellipses corresponding to $\pi\pm1\sigma$ are also shown. Motion in right ascension is given in units of $\arcsec yr^{-1}$ and includes the correction for $cos(\delta). $\label{w0146motion}}
\end{figure}

\clearpage

\begin{figure}
\begin{center}$
\begin{array}{c}
\includegraphics[width=6in]{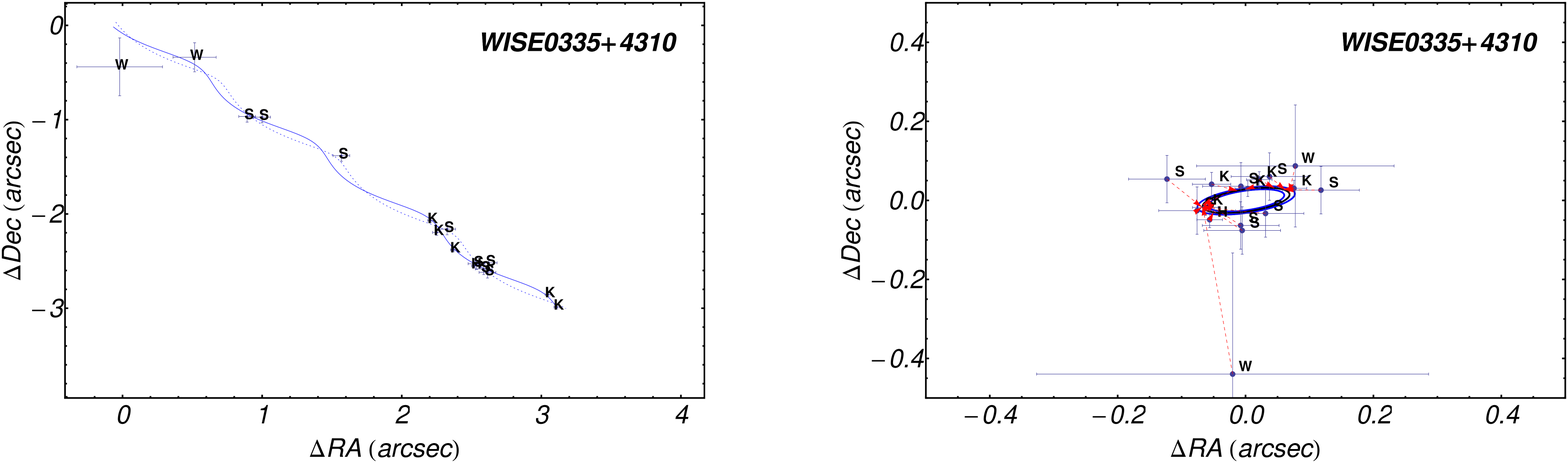}\\
\includegraphics[width=6in]{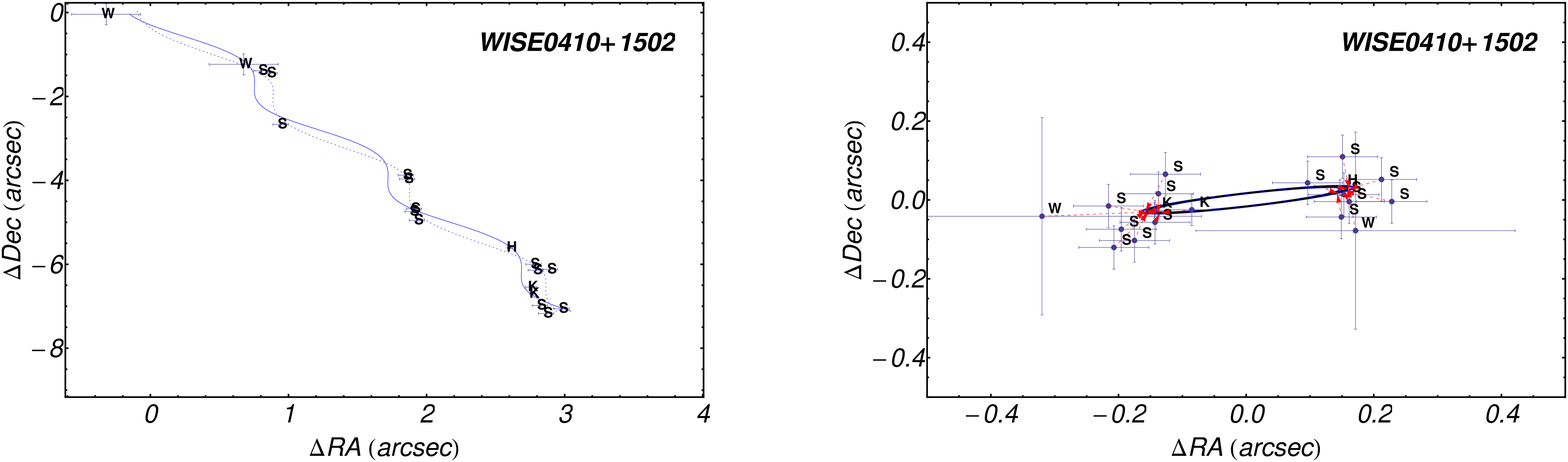}\\
\end{array}$
\end{center}
\caption{As described in Figure~\ref{w0146motion}, the figure shows the parallactic solutions for two brown dwarfs, WISE0335+4310 (top) and WISE0410+1502 (bottom). \label{w0335motion}}
\end{figure}
\clearpage

\begin{figure}
\begin{center}$
\begin{array}{c}
\includegraphics[width=6in]{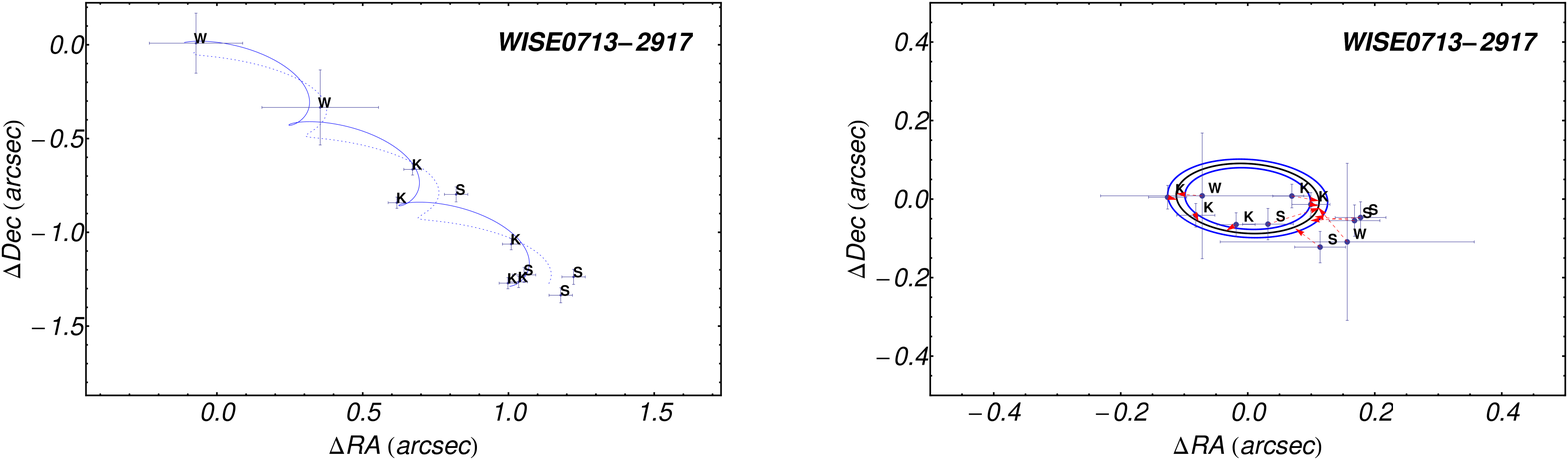}\\
\includegraphics[width=6in]{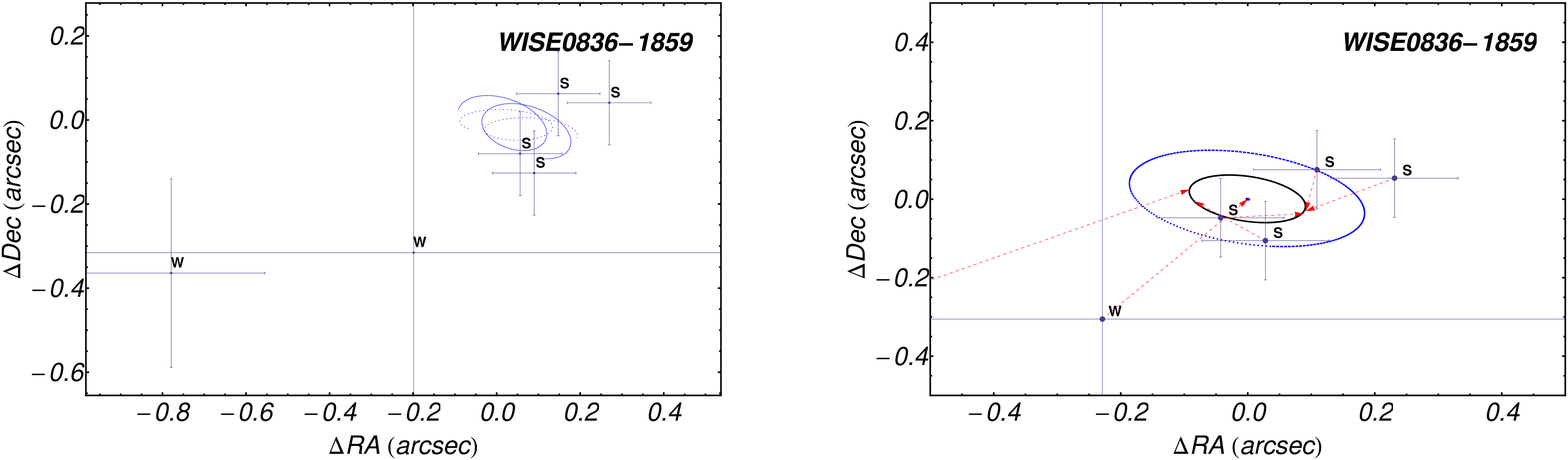}\\
 \end{array}$
 \end{center}
\caption{As described in Figure~\ref{w0146motion}, the figure shows the parallactic solutions for two brown dwarfs, WISE0713-2917 (top) and WISE0836-1859 (bottom). The solution for WISE0836-1859 has relatively few data points and is poorly constrained\label{w0713motion}}
\end{figure}

\clearpage

 \begin{figure}
 \begin{center}$
 \begin{array}{c}
 \includegraphics[width=6in]{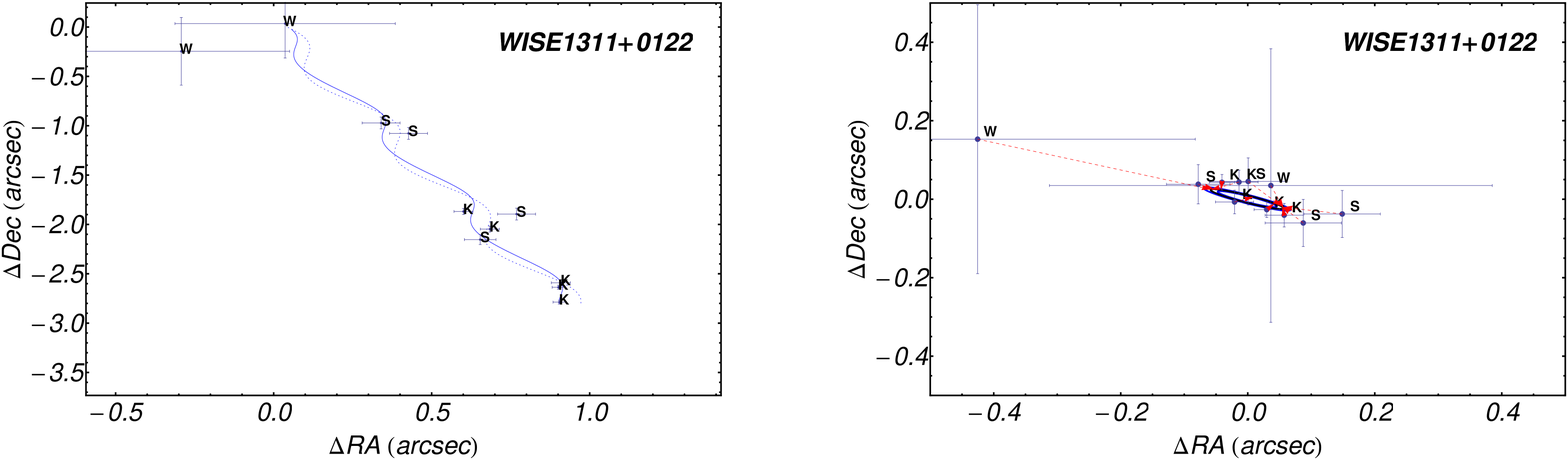}\\
 \includegraphics[width=6in]{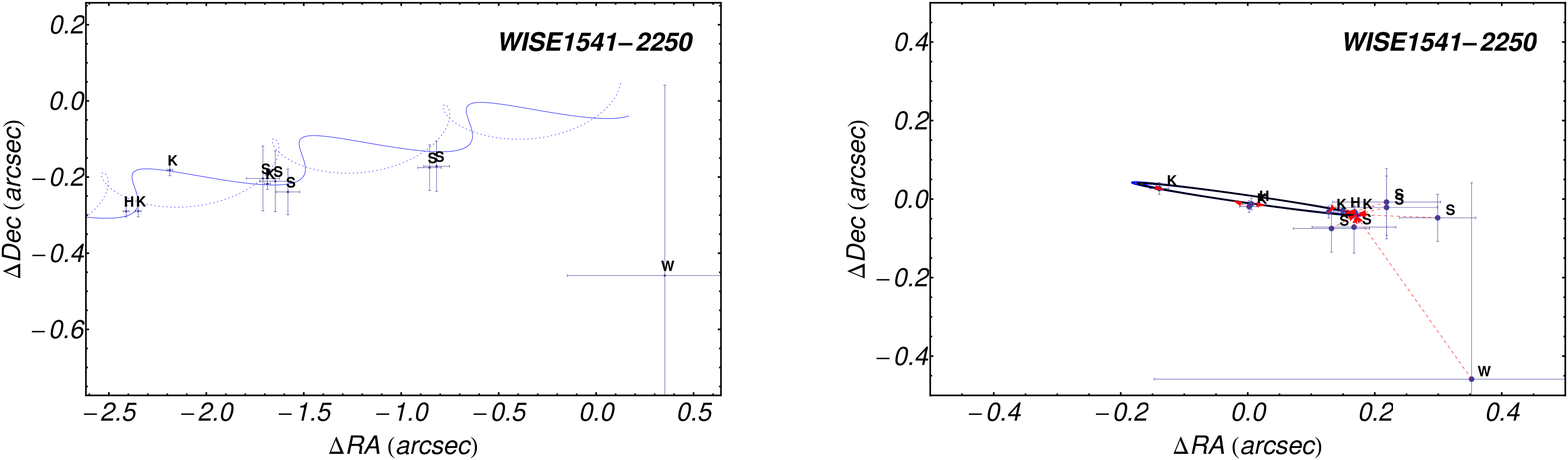}\\
 \end{array}$
 \end{center}
 \caption{As described in Figure~\ref{w0146motion}, the figure shows the parallactic solutions for two brown dwarfs, WISE1311+0122 (top) and WISE1541-2250 (bottom). \label{w1311motion}}
\end{figure}
 
 \clearpage
\begin{figure}
\begin{center}$
\begin{array}{c}
 \includegraphics[width=6in]{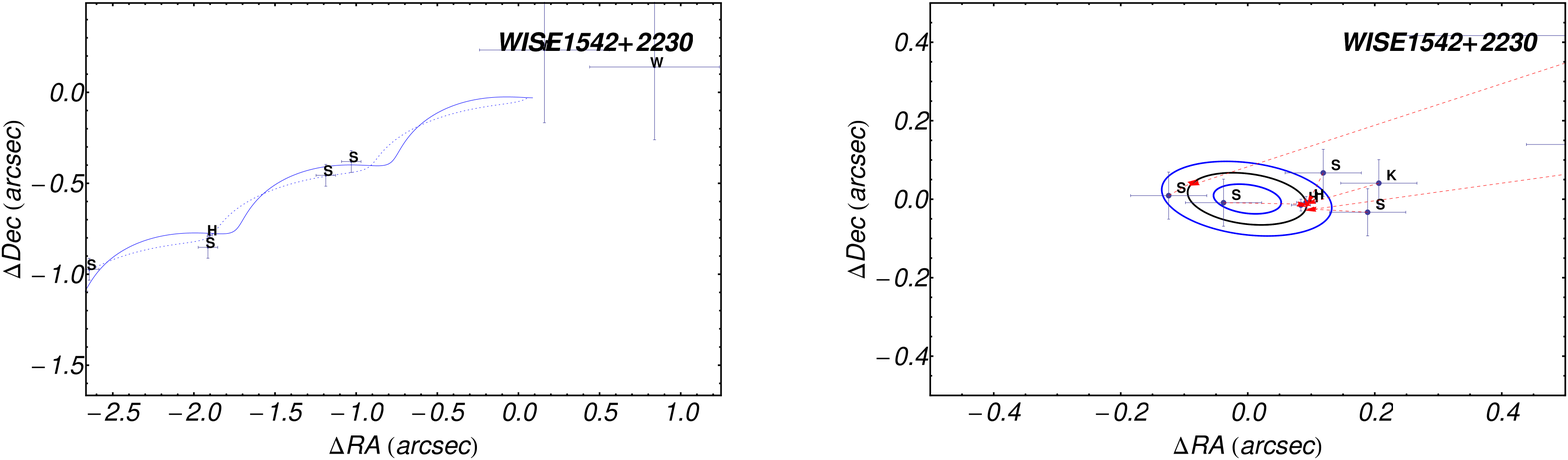}\\
\includegraphics[width=6in]{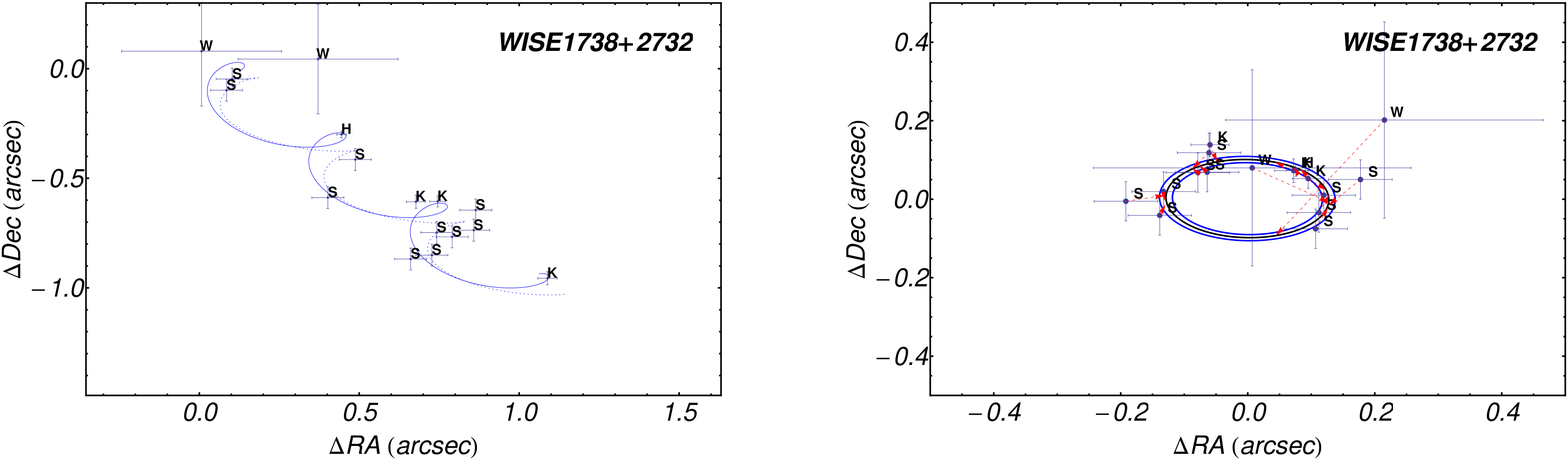}\\
\end{array}$
\end{center}
\caption{As described in Figure~\ref{w0146motion}, the figure shows the parallactic solutions for two brown dwarfs, WISE1542+2230 (top) and WISE1738+2732 (bottom). \label{w1542motion}}
\end{figure}
\clearpage

\begin{figure}
\begin{center}$
\begin{array}{c}
\includegraphics[width=6in]{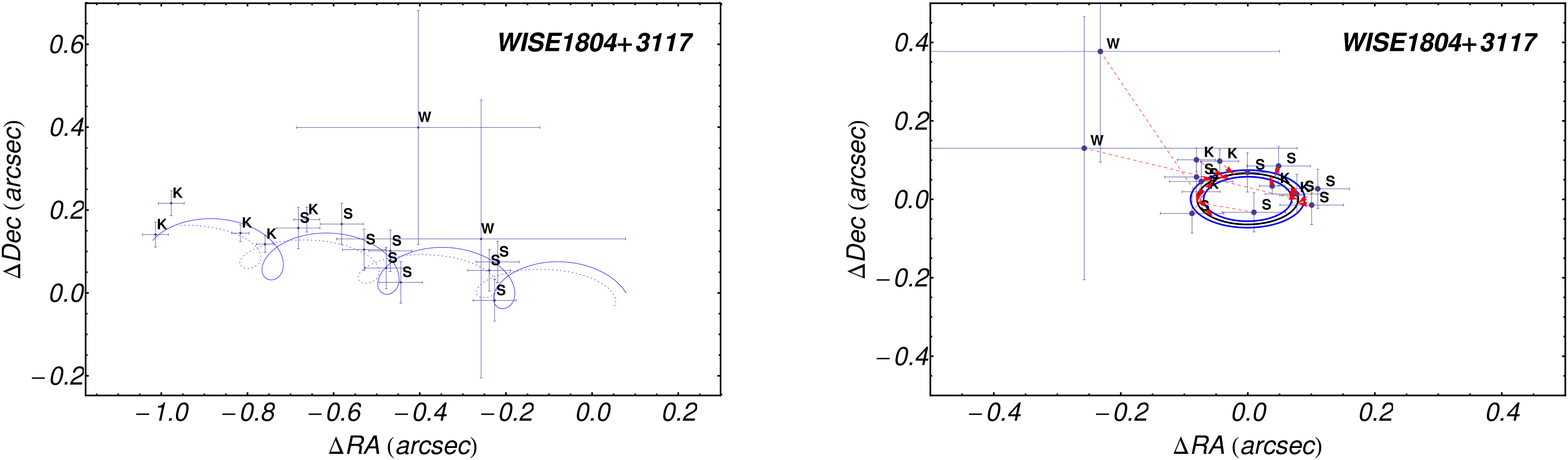}\\
\includegraphics[width=6in]{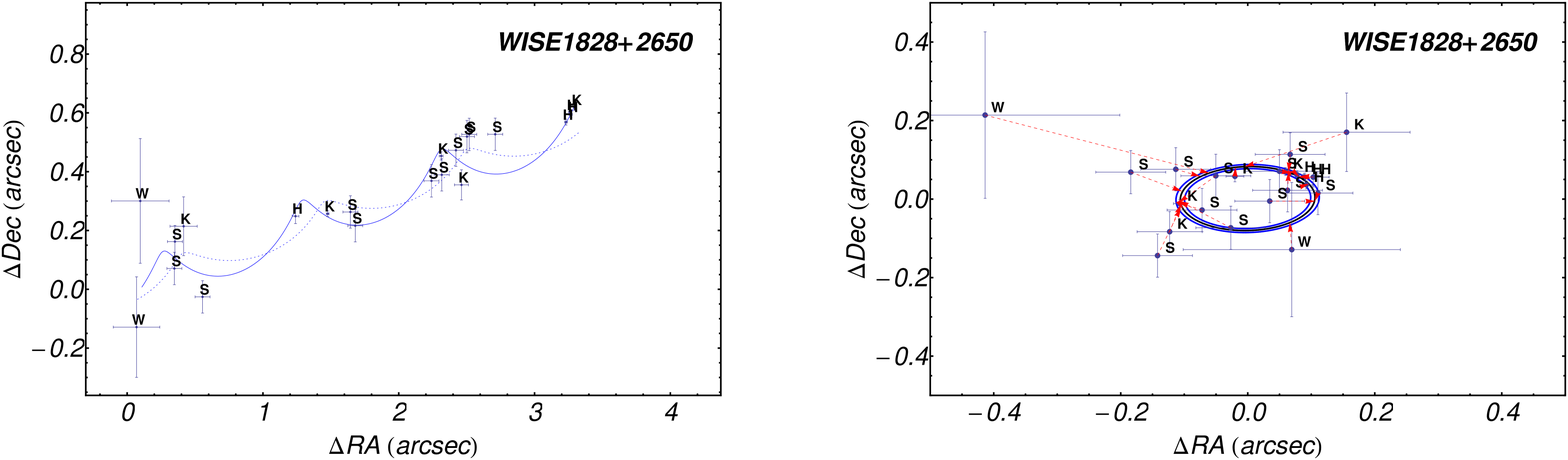}\\
\end{array}$
\end{center}
\caption{As described in Figure~\ref{w0146motion}, the figure shows the parallactic solutions for two brown dwarfs, WISE1804+3117 (top) and WISE1828+2650 (bottom). \label{w1804motion}}
\end{figure}

\clearpage

\begin{figure}
\begin{center}$
\begin{array}{c}
\includegraphics[width=6in]{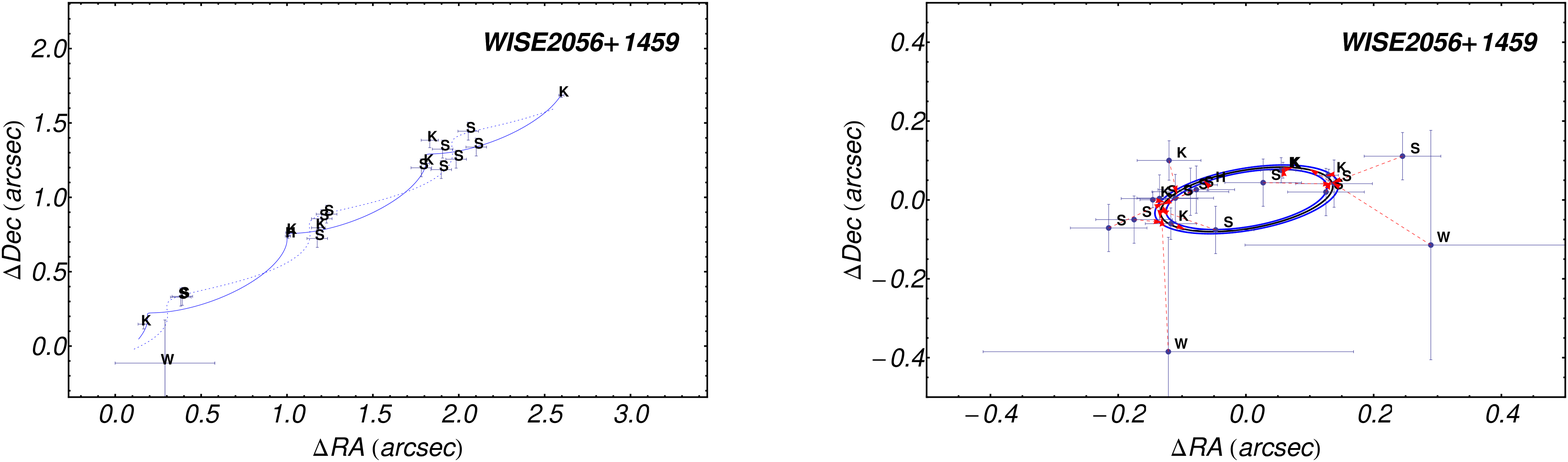}\\
\includegraphics[width=6in]{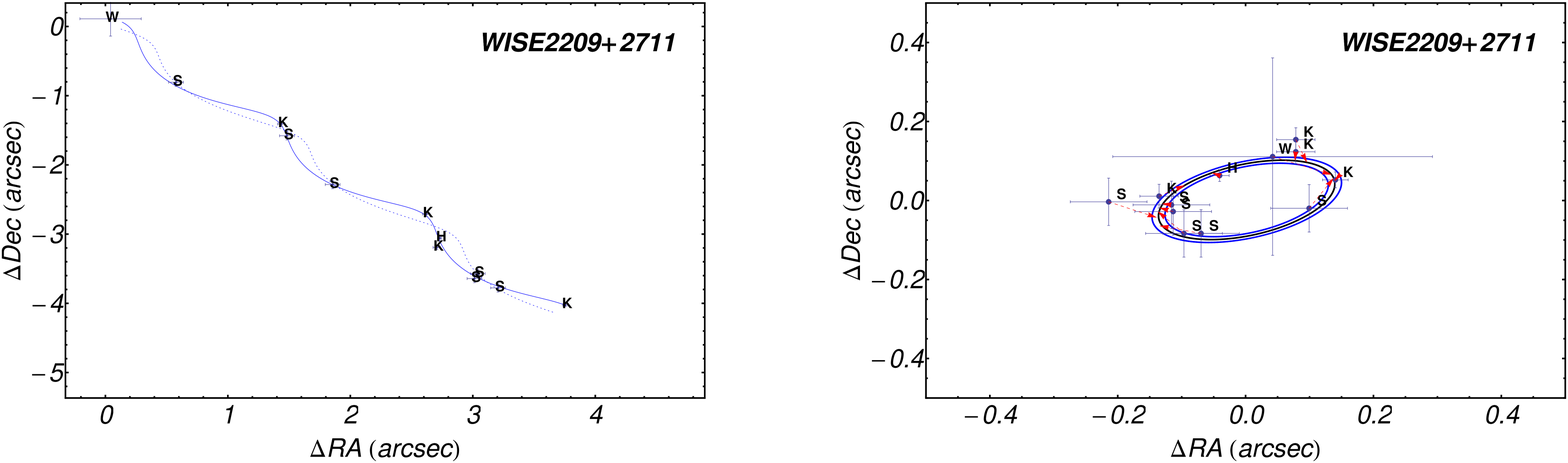}\\
\end{array}$
\end{center}
\caption{As described in Figure~\ref{w0146motion}, the figure shows the parallactic solutions for two brown dwarfs, WISE2056+1459 (top) and WISE2209+2711 (bottom). \label{w2056motion}}
\end{figure}

\clearpage

\begin{figure}
\begin{center}$
\begin{array}{c}
\includegraphics[width=6in]{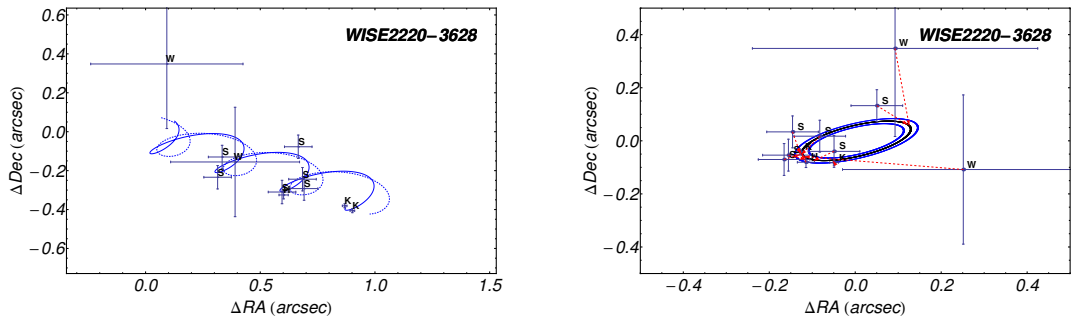}\\
\end{array}$
\end{center}
\caption{As described in Figure~\ref{w0146motion}, the figure shows the parallactic solutions for the brown dwarf WISE2220-3628. \label{w2220motion}}
\end{figure}

\clearpage

\begin{figure}
\includegraphics[width=6in]{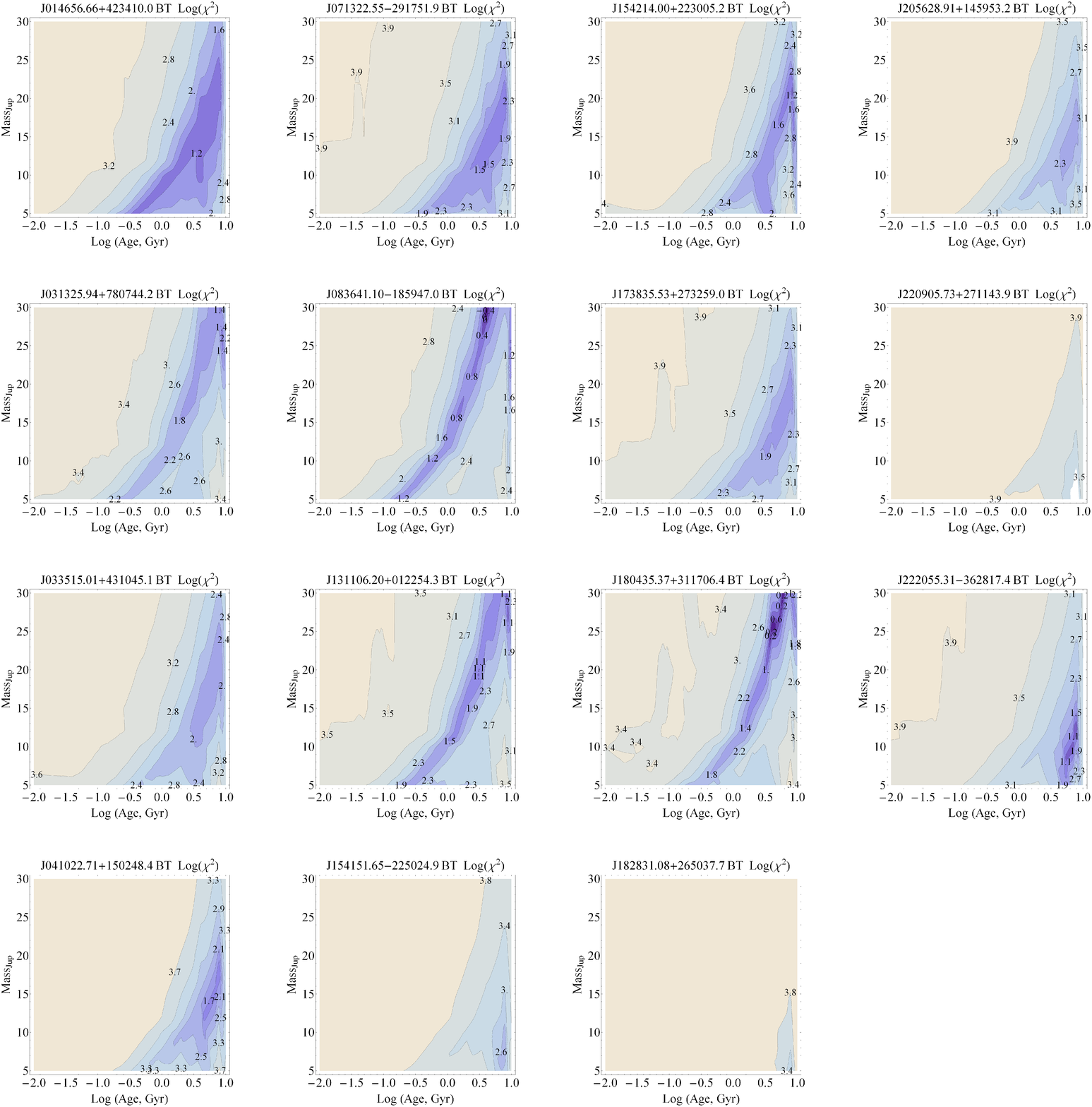}
\caption{A sequence of fits of BTSettl models \citep{Allard2003,Allard2010} to the absolute 4.5 $\mu$m brightness and to other $mag_i$-[4.5] colors for four of the late T and Y dwarfs in our sample. The plots show contours of the logarithm of the $\chi^2$ parameter defined in Equation (3). The high values of $\chi^2$ indicate that the BTSettl models are relatively poor fits to the spectral energy distributions of the very cold sources. \label{AllSourceModelBTFit}}
\end{figure}

\clearpage

\begin{figure}
\includegraphics[width=6in]{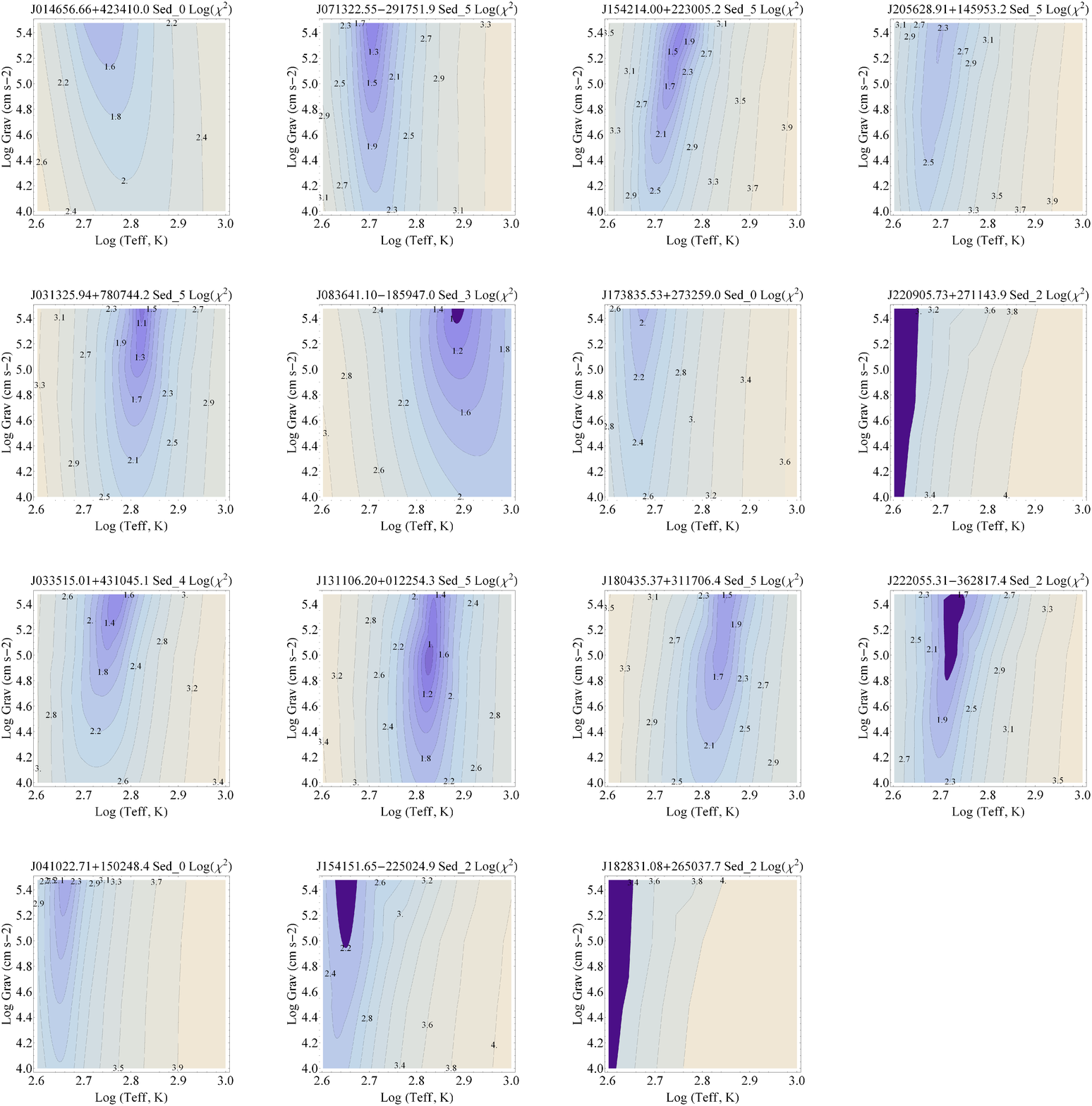}
\caption{A sequence of fits of models \citep{Morley2012}to the absolute 4.5 $\mu$m brightness and to other $mag_i$-[4.5] colors for four of the late T and Y dwarfs in our sample. The plots show contours of the logarithm of the $\chi^2$ parameter defined in Equation (3). In each case the model shown represents a slice through the 3-D parameter space for the value of the sedimentation parameter, $f_{sed}$, that best fits the data. The $f_{sed}$ value is given at the top of each plot. \label{AllSourceModelMorleyFit}}
\end{figure}
\clearpage

%
%\begin{figure*}
%\begin{center}$
%\begin{array}{c}
%\plottwo{fig27a.eps}{fig27b.eps}\\
%\plottwo{fig27c.eps}{fig27d.eps}\\
%\end{array}$
%\end{center}
%\caption{The result of fitting the photometric colors and absolute absolute 4.5 $\mu$m brightness to the BT-Settl models is shown for four representative sources. The top two plots for WISE 0146+4234 and WISE 0410+1502 show relatively good fits ($\chi^2$ of 9 and 25, respectively). The lower two plots show more problematical cases WISE 1541-2250 and WISE 1828+2650 ($\chi^2$ of 209 and 1214, respectively). For these latter two objects the primary area of disagreement is that the near-IR bands are greatly depressed relative to the model predictions. \label{SampleSpecFits}}
%\end{figure*}
%\clearpage

\begin{figure*}
\begin{center}$
\begin{array}{c}
\plottwo{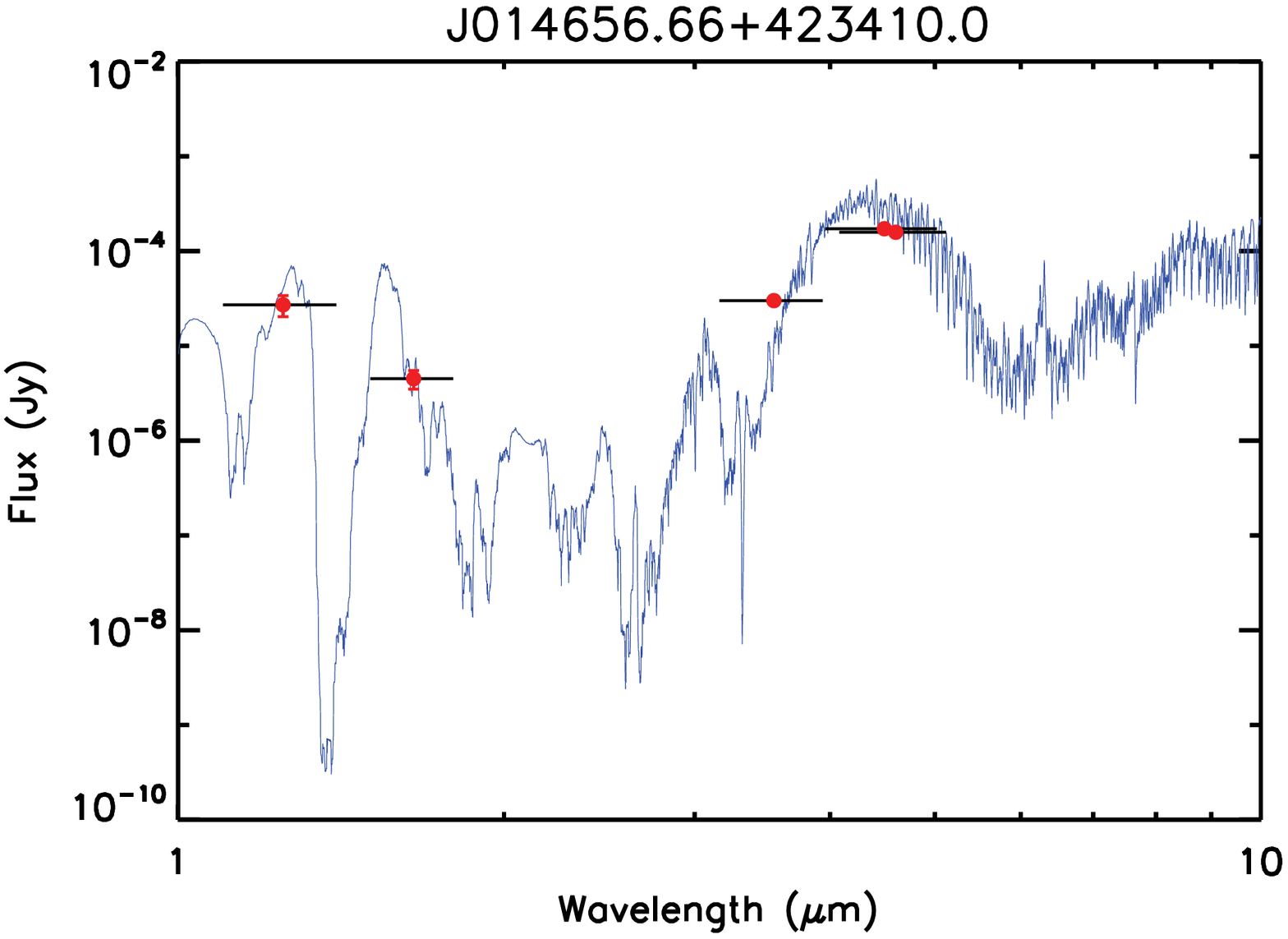}{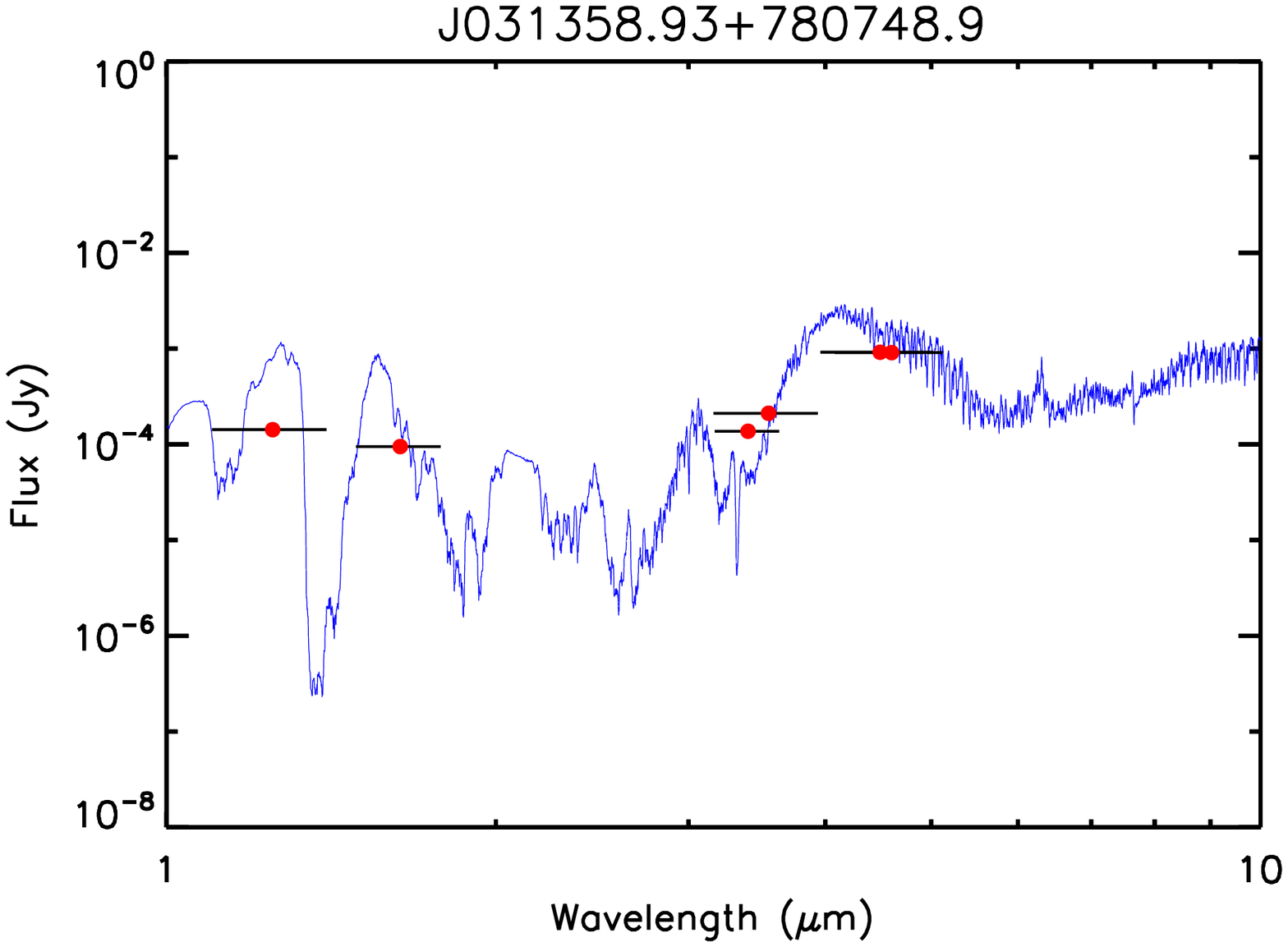}\\ %\plottwo{WISE0146.eps}{WISE0313.eps}\\
\plottwo{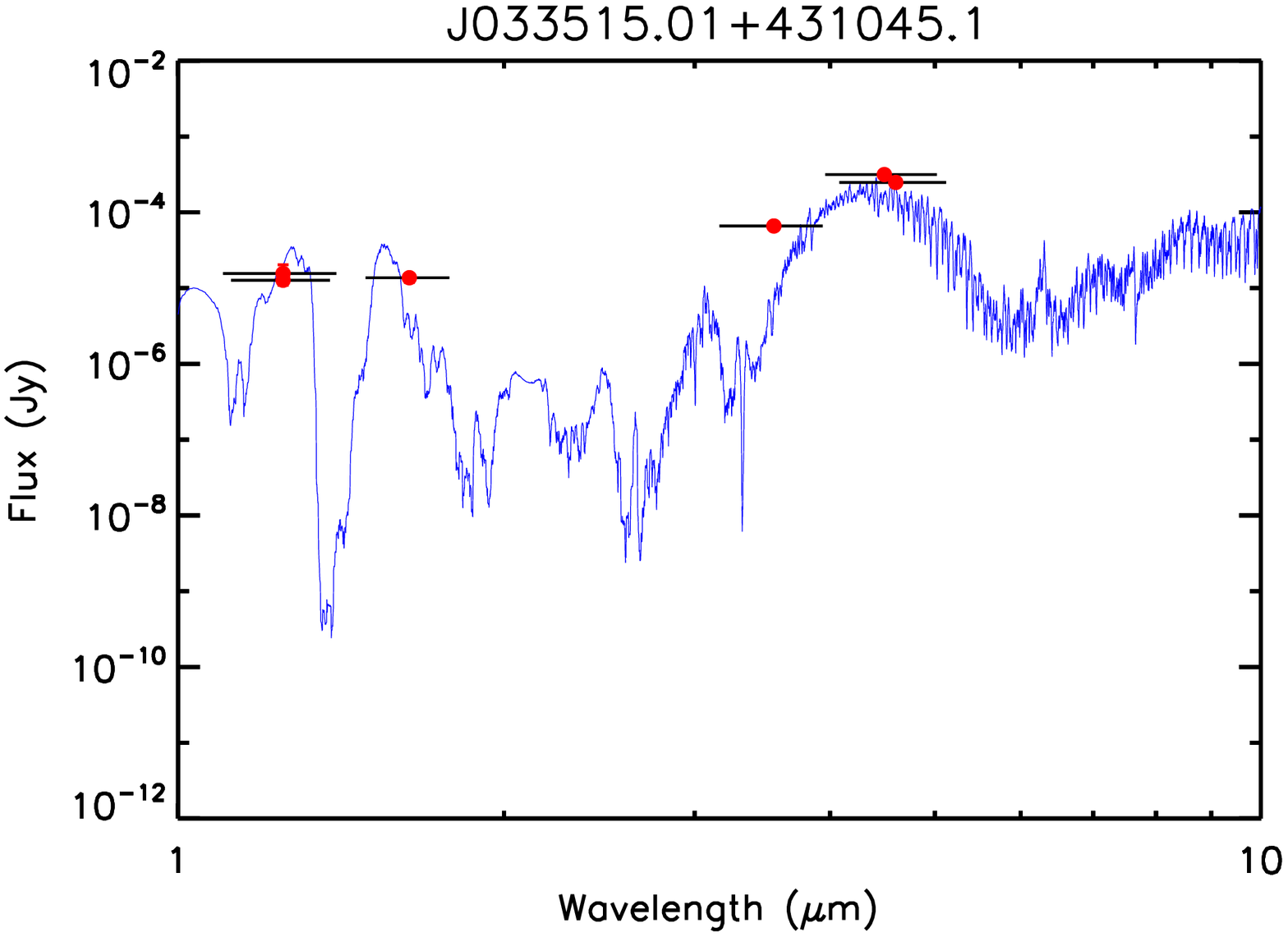}{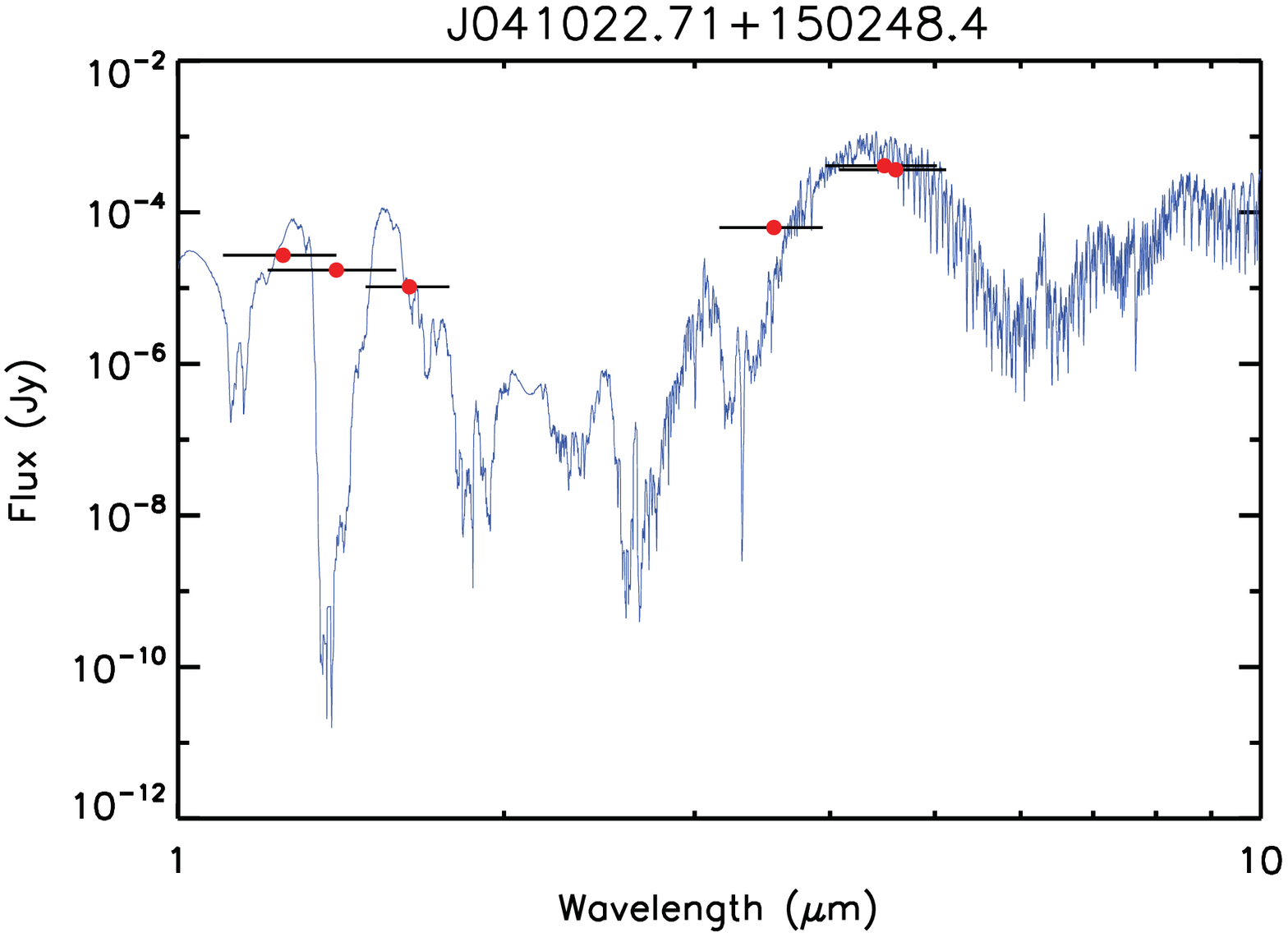}\\ %\plottwo{WISE0335.eps}{WISE0410.eps}\\
\plottwo{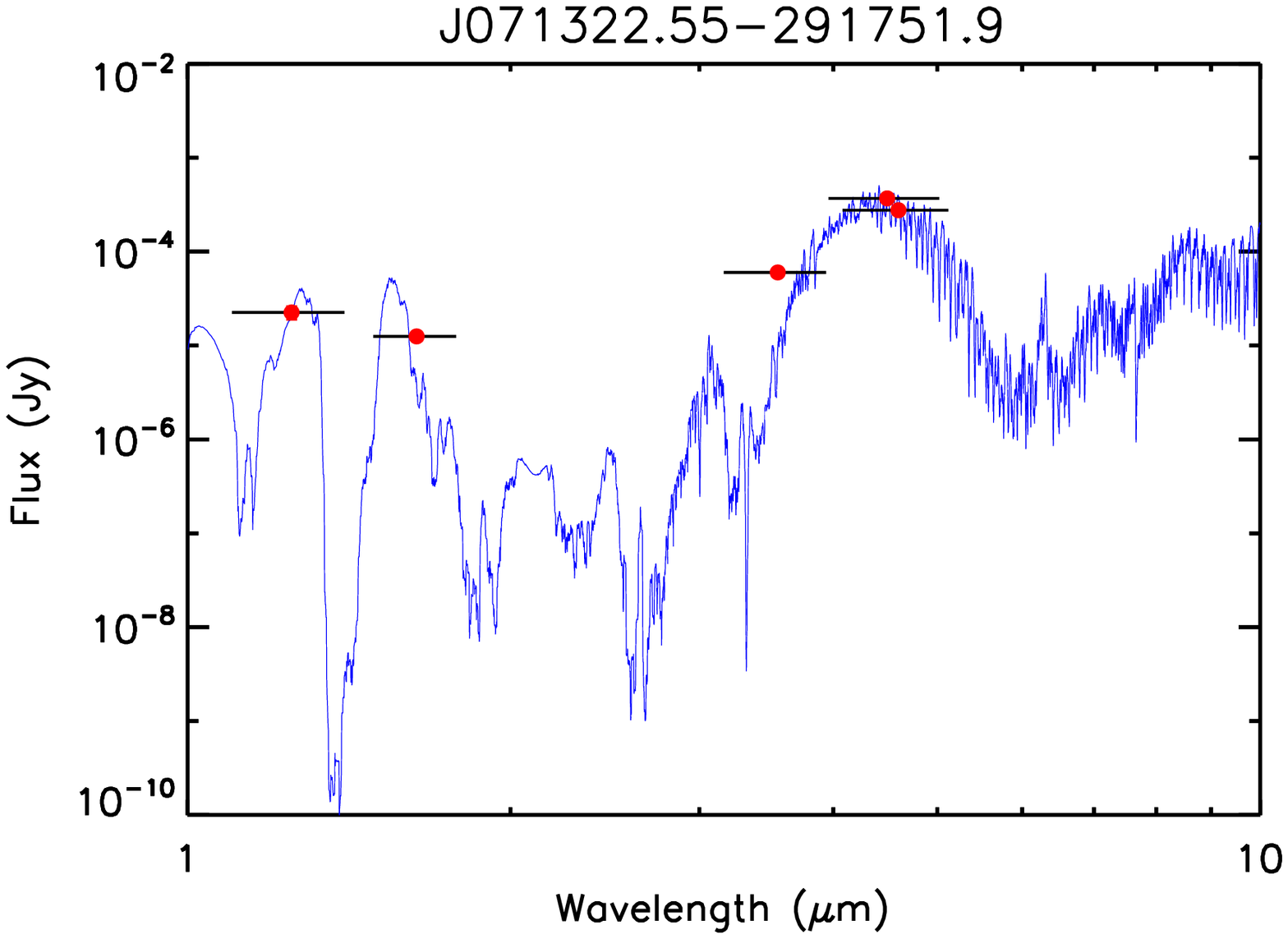}{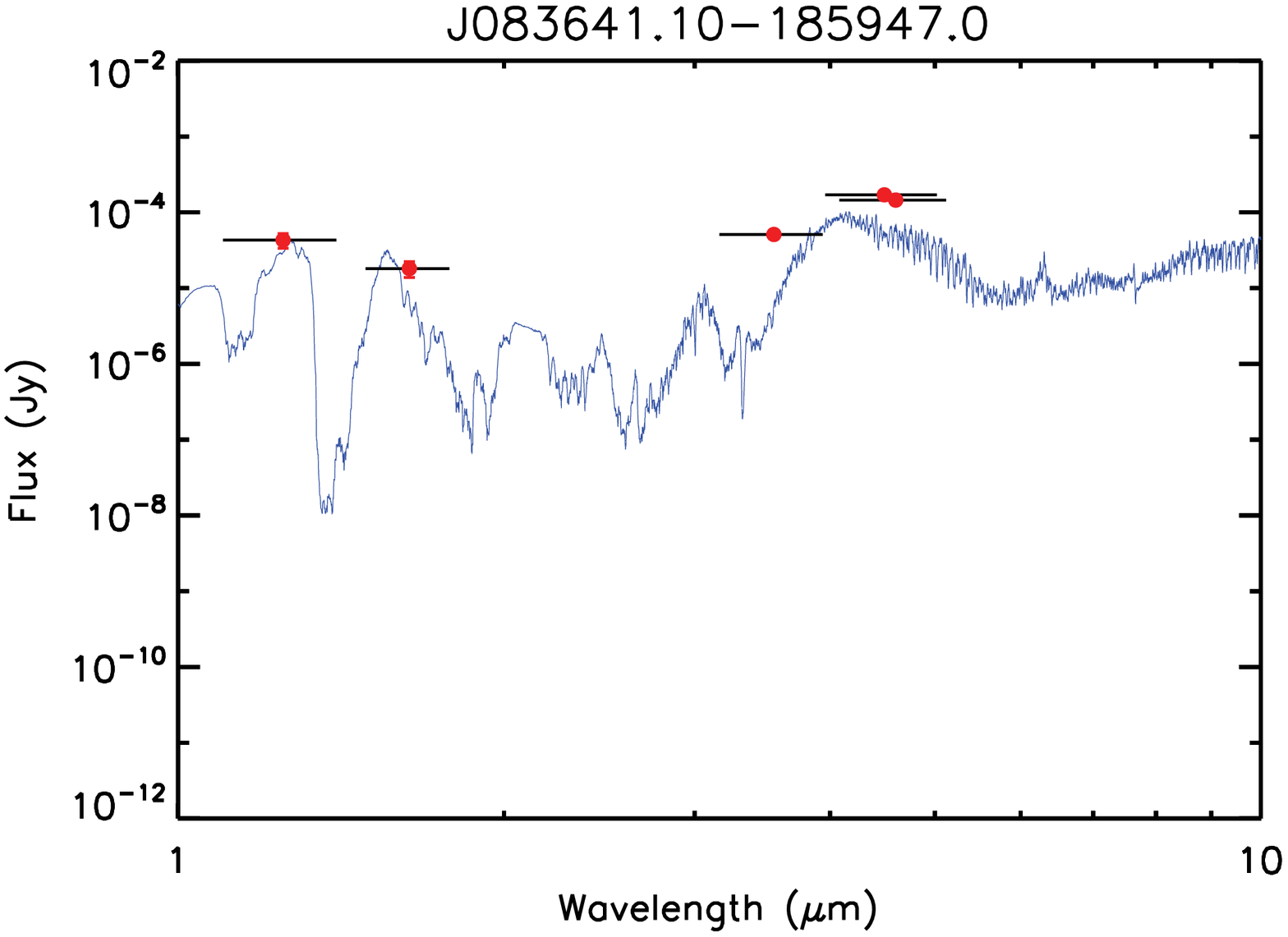}\\%\plottwo{WISE0713.eps}{WISE0836.eps}\\
\plottwo{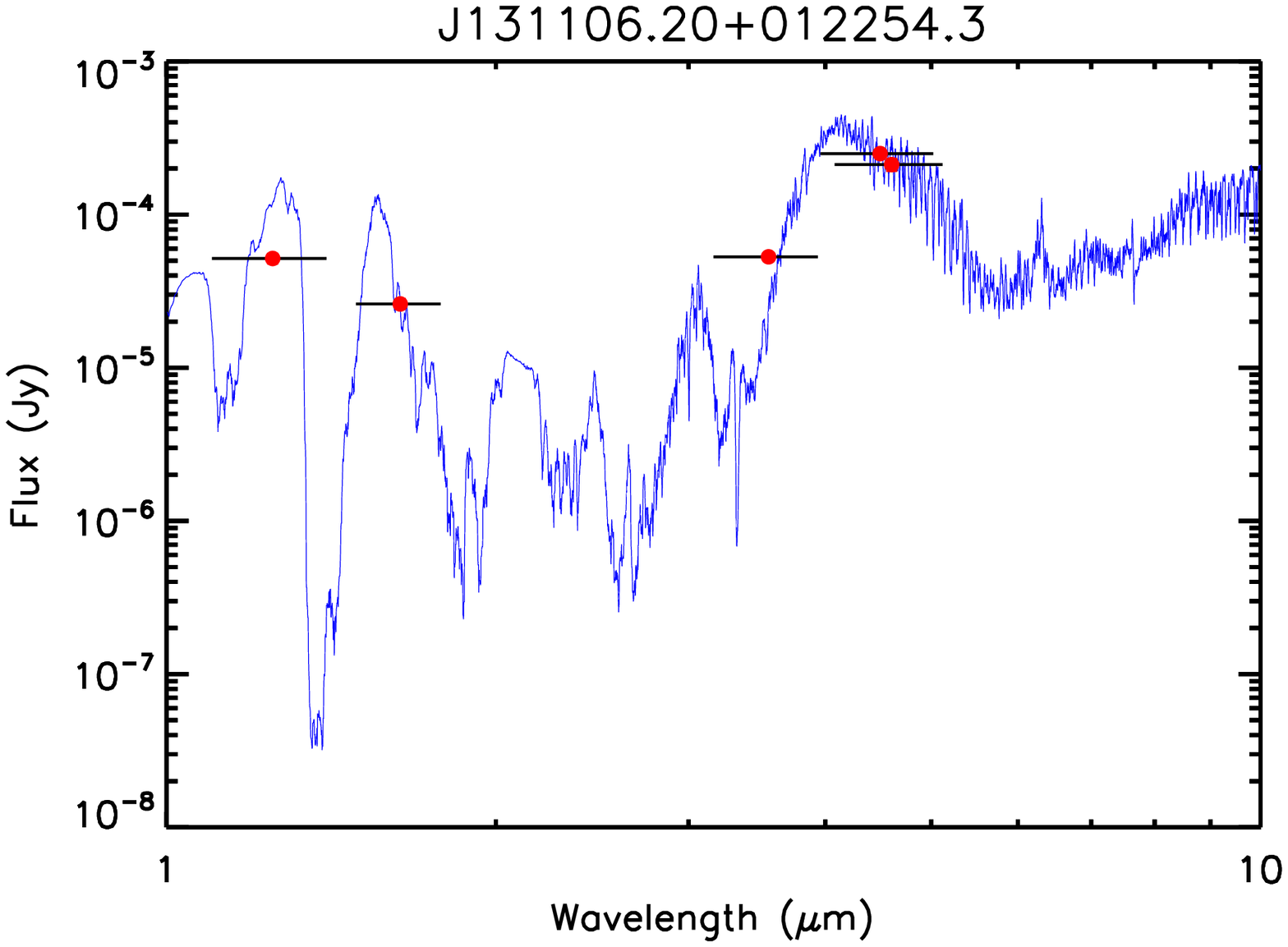}{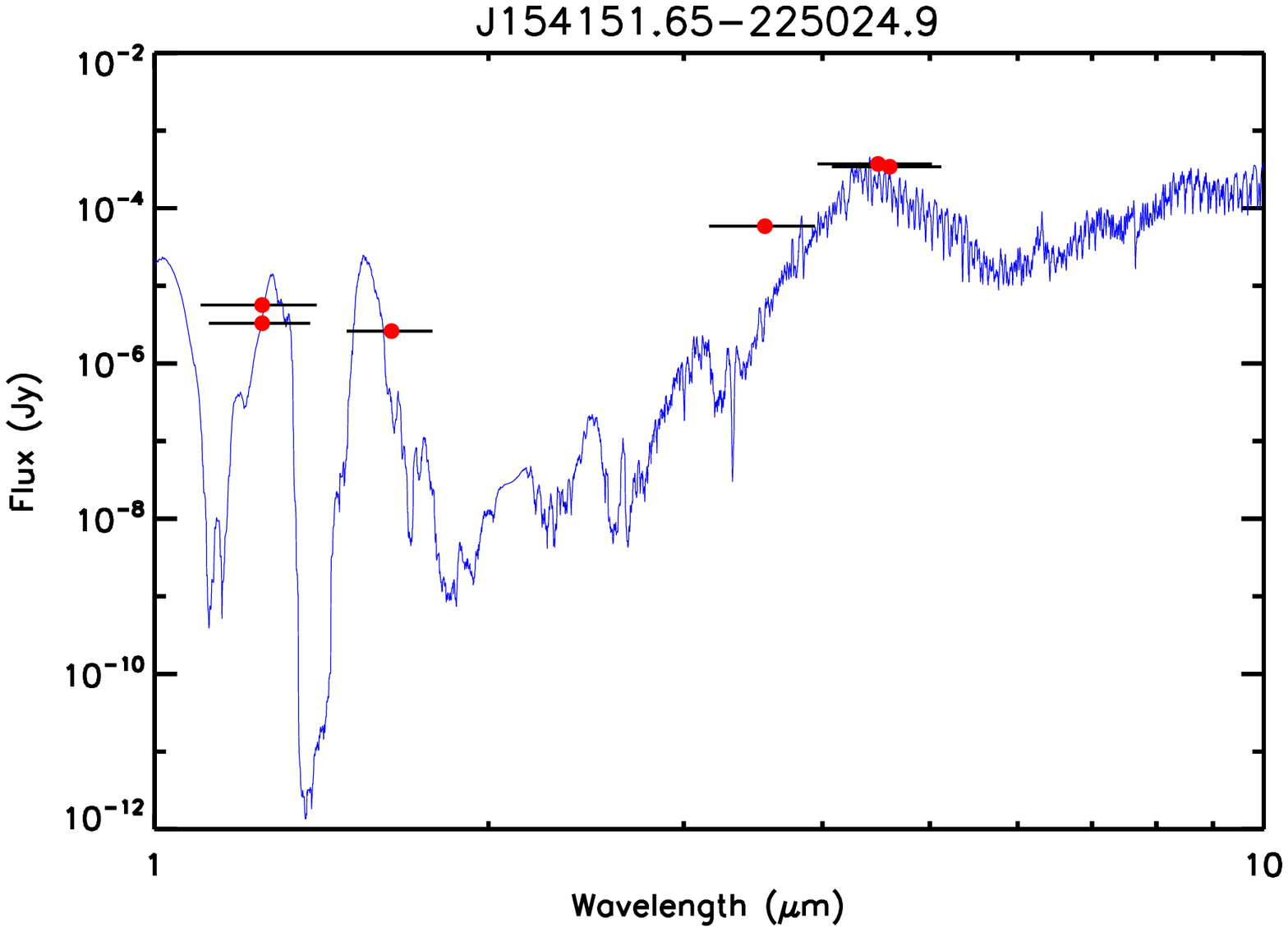}\\%\plottwo{WISE1311.eps}{WISE1541.eps}\\
\end{array}$
\end{center}
\caption{a-h). The result of fitting the photometric colors and absolute absolute 4.5 $\mu$m brightness to the BT-Settl models as described in the text. \label{SampleSpecFits}}
\end{figure*}
\clearpage

\addtocounter{figure}{-1} 
\begin{figure*}
\begin{center}$
\begin{array}{c}
\plottwo{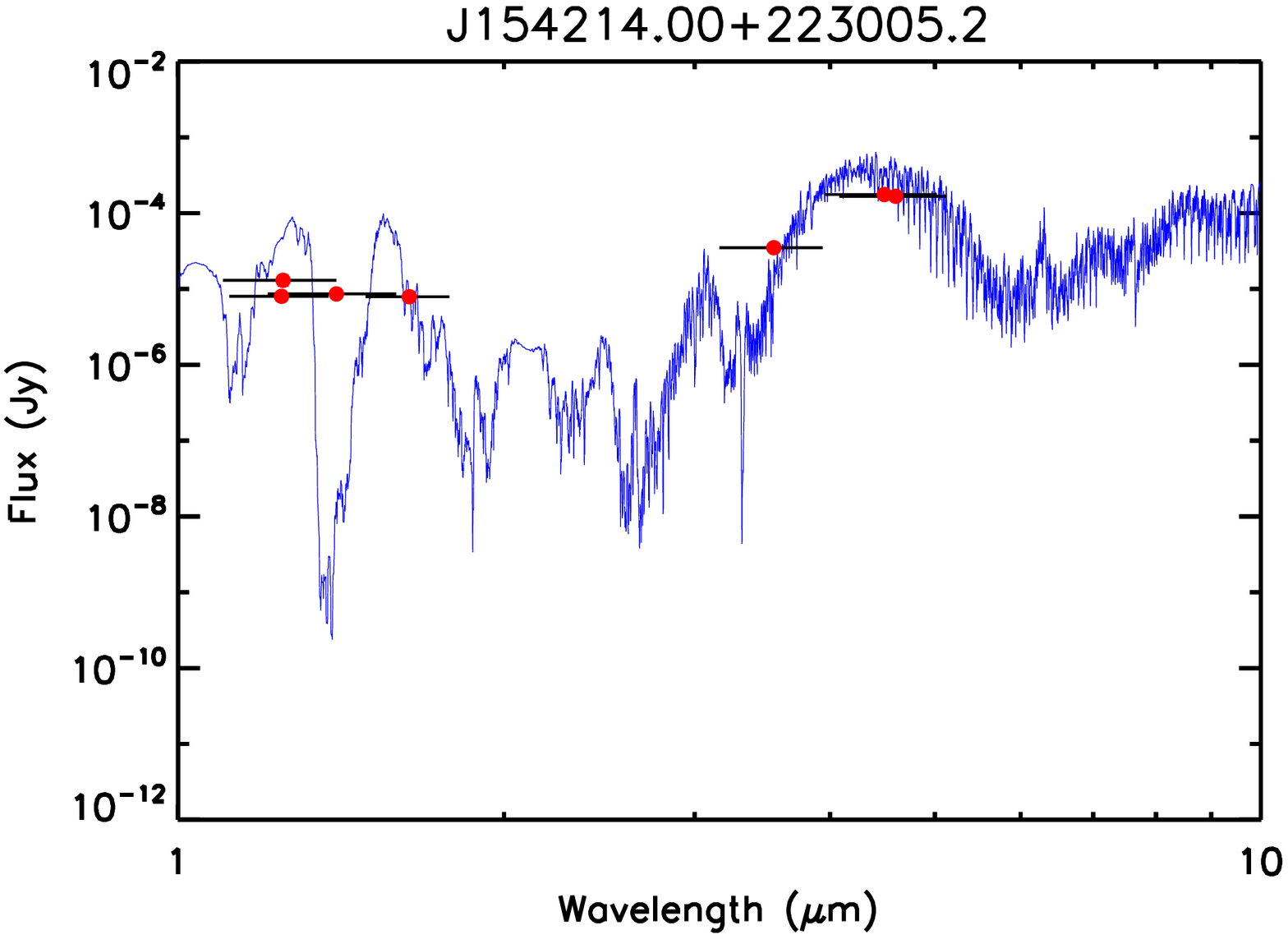}{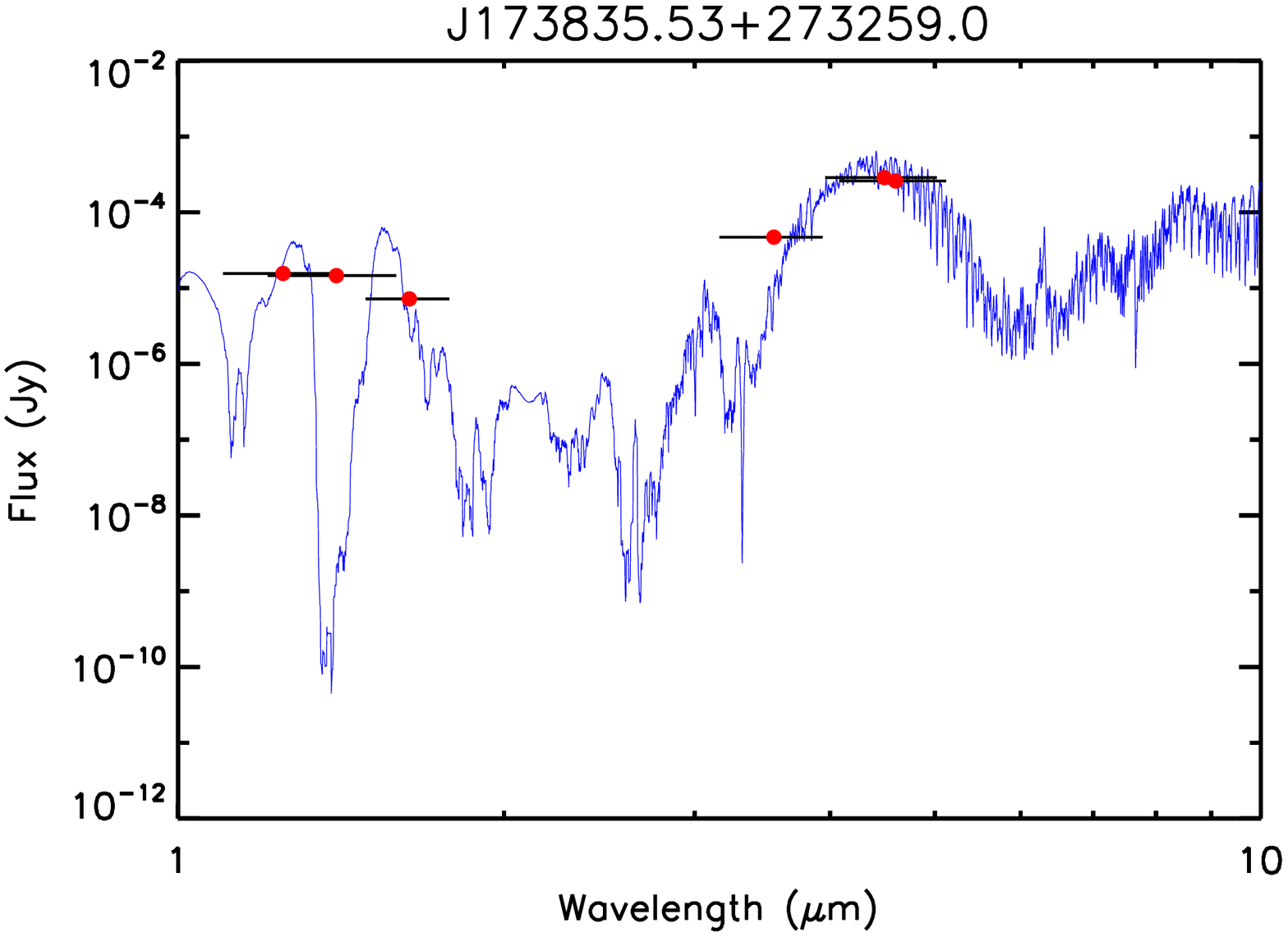}\\%\plottwo{WISE1542.eps}{WISE1738.eps}\\
\plottwo{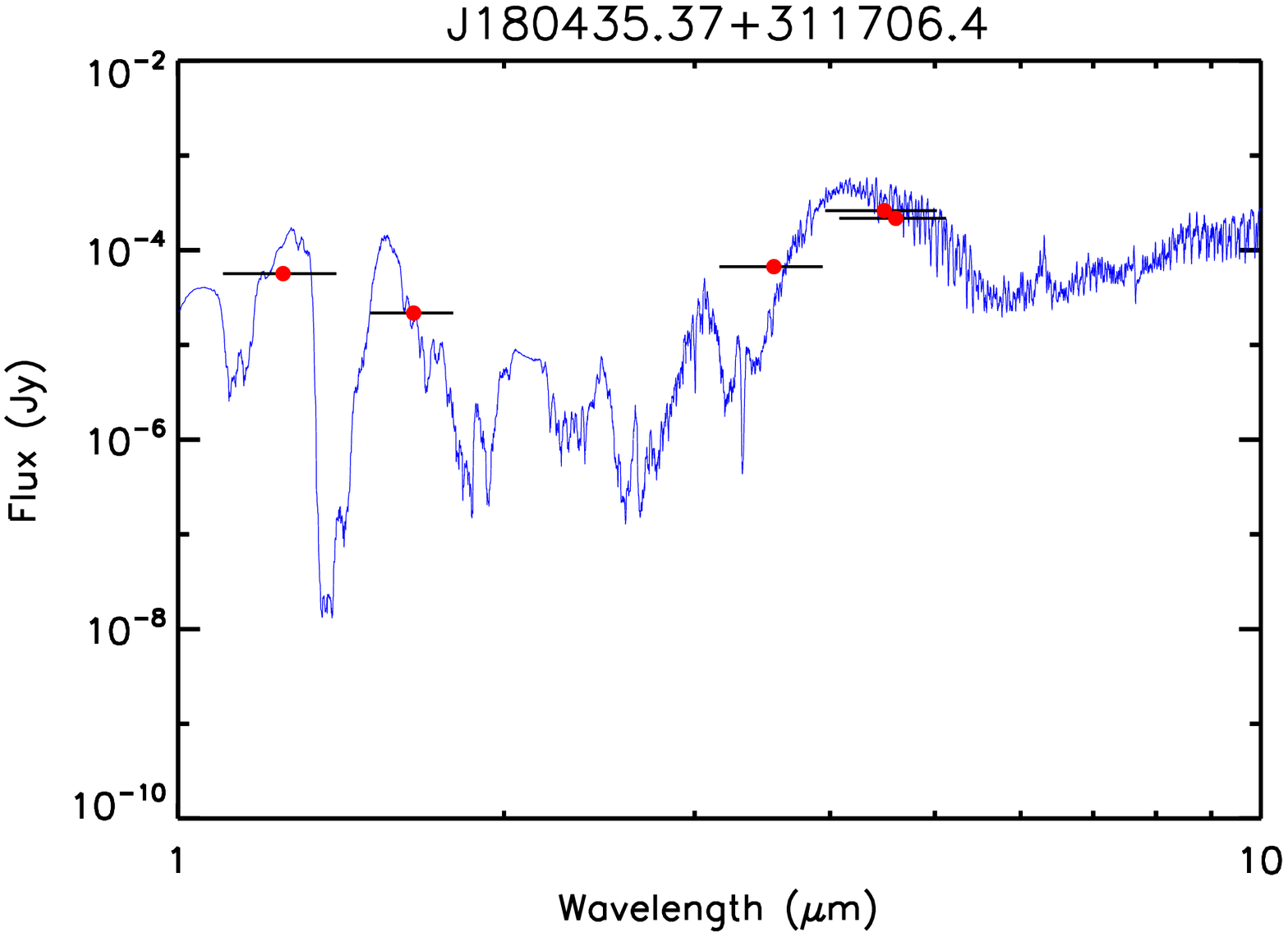}{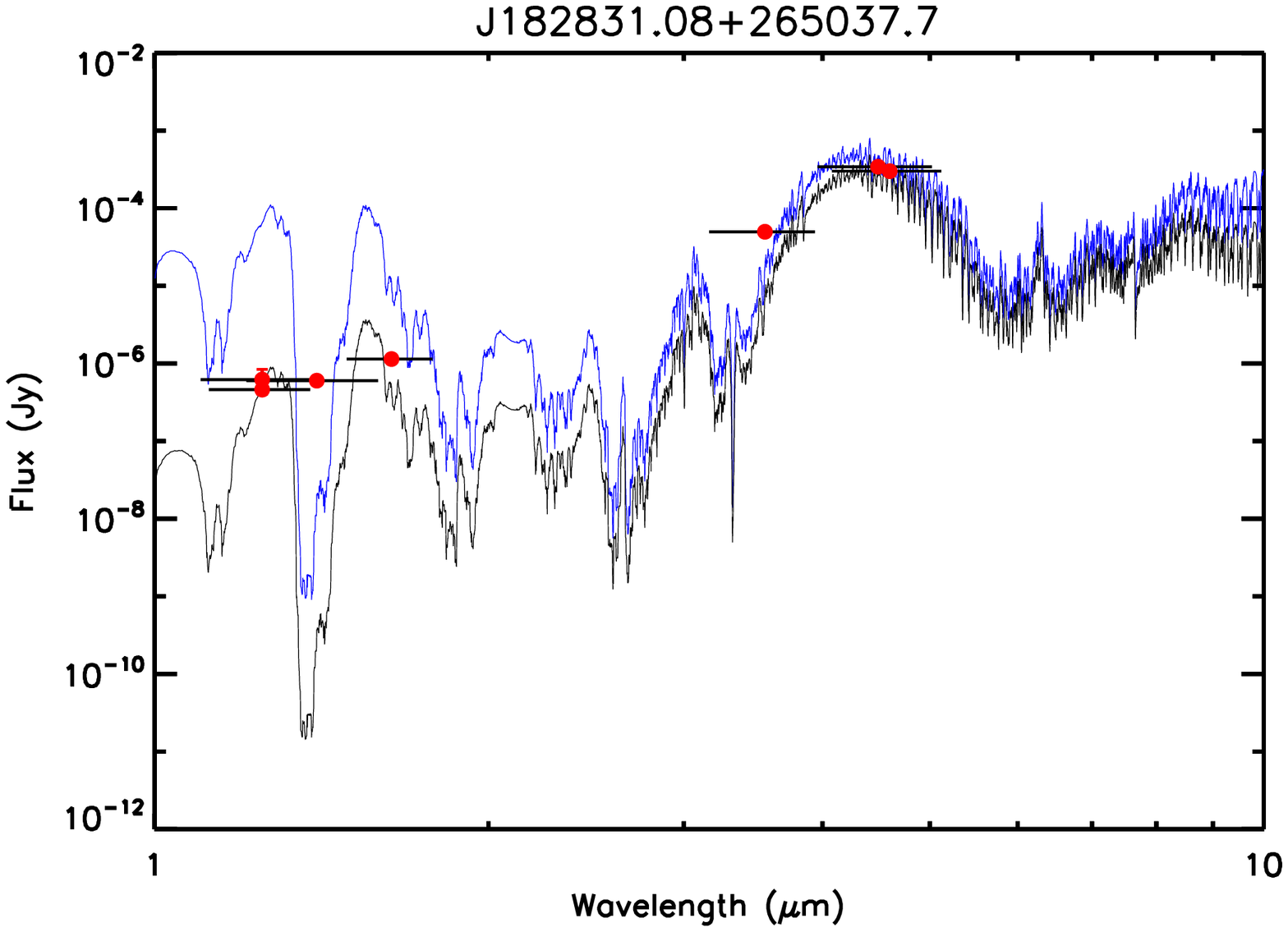}\\%\plottwo{WISE1804.eps}{fig27d.eps}\\
\plottwo{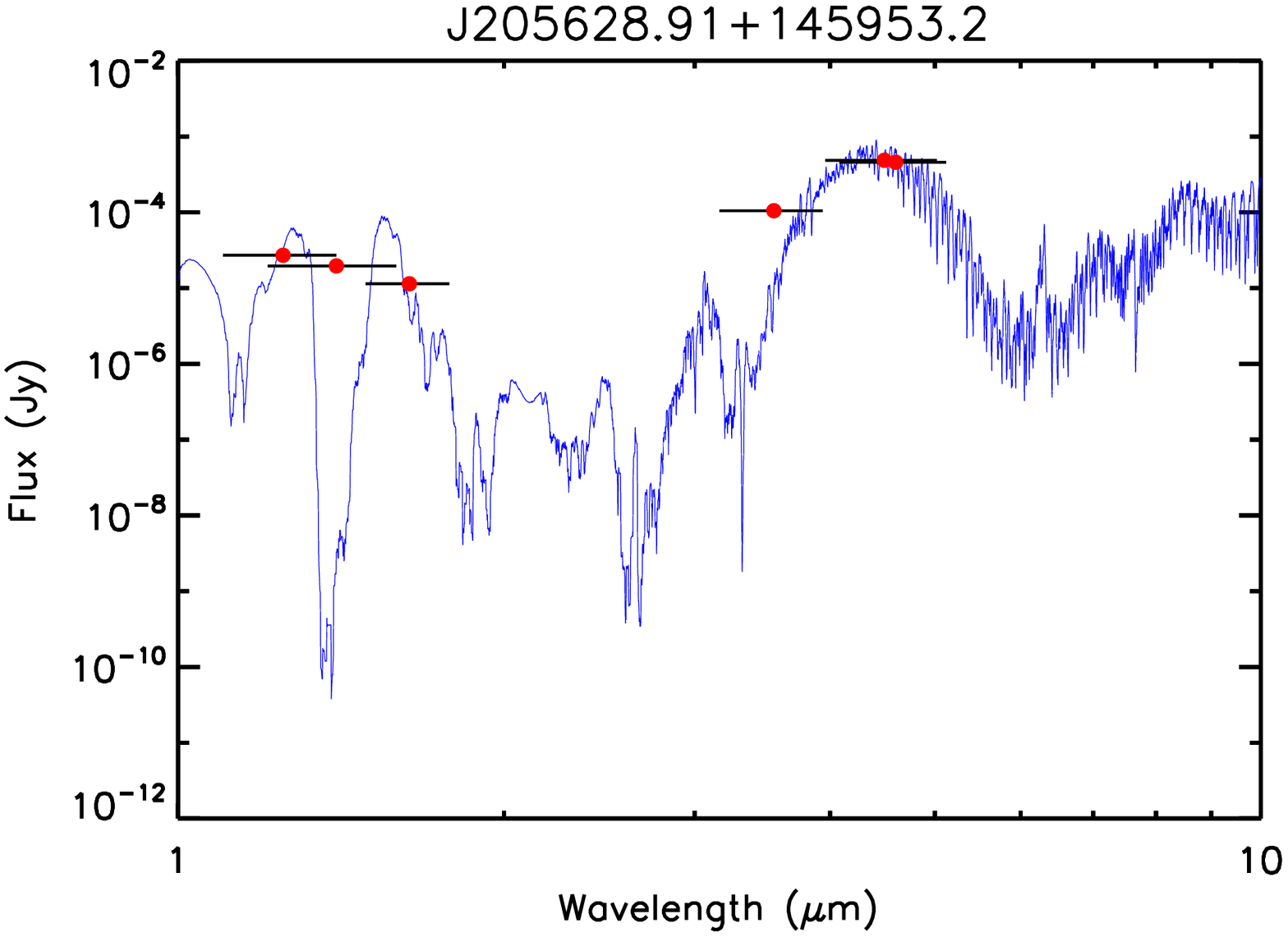}{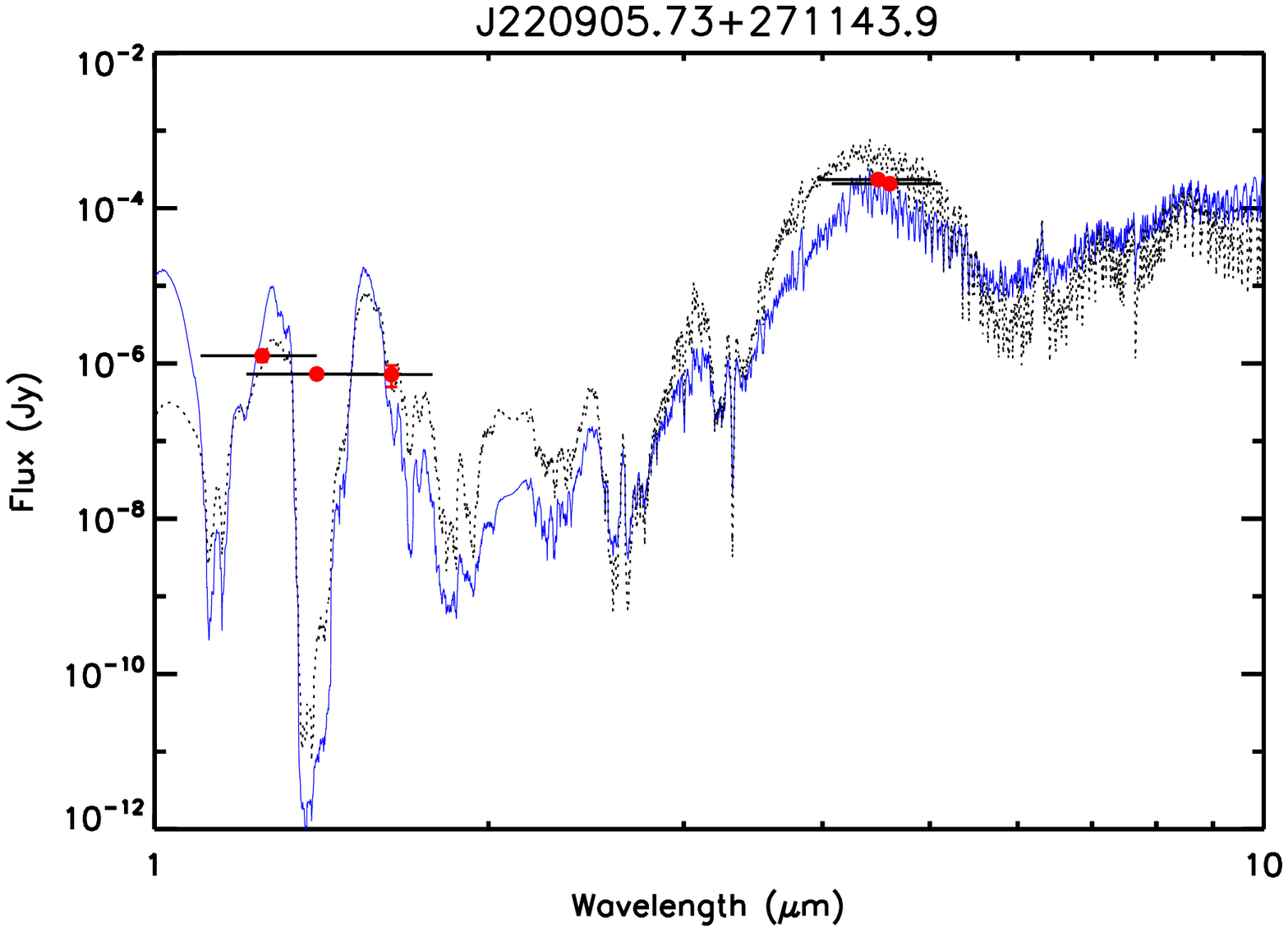}\\% \plottwo{WISE2056.eps}{WISE2209.eps}\\
\includegraphics[width=3in]{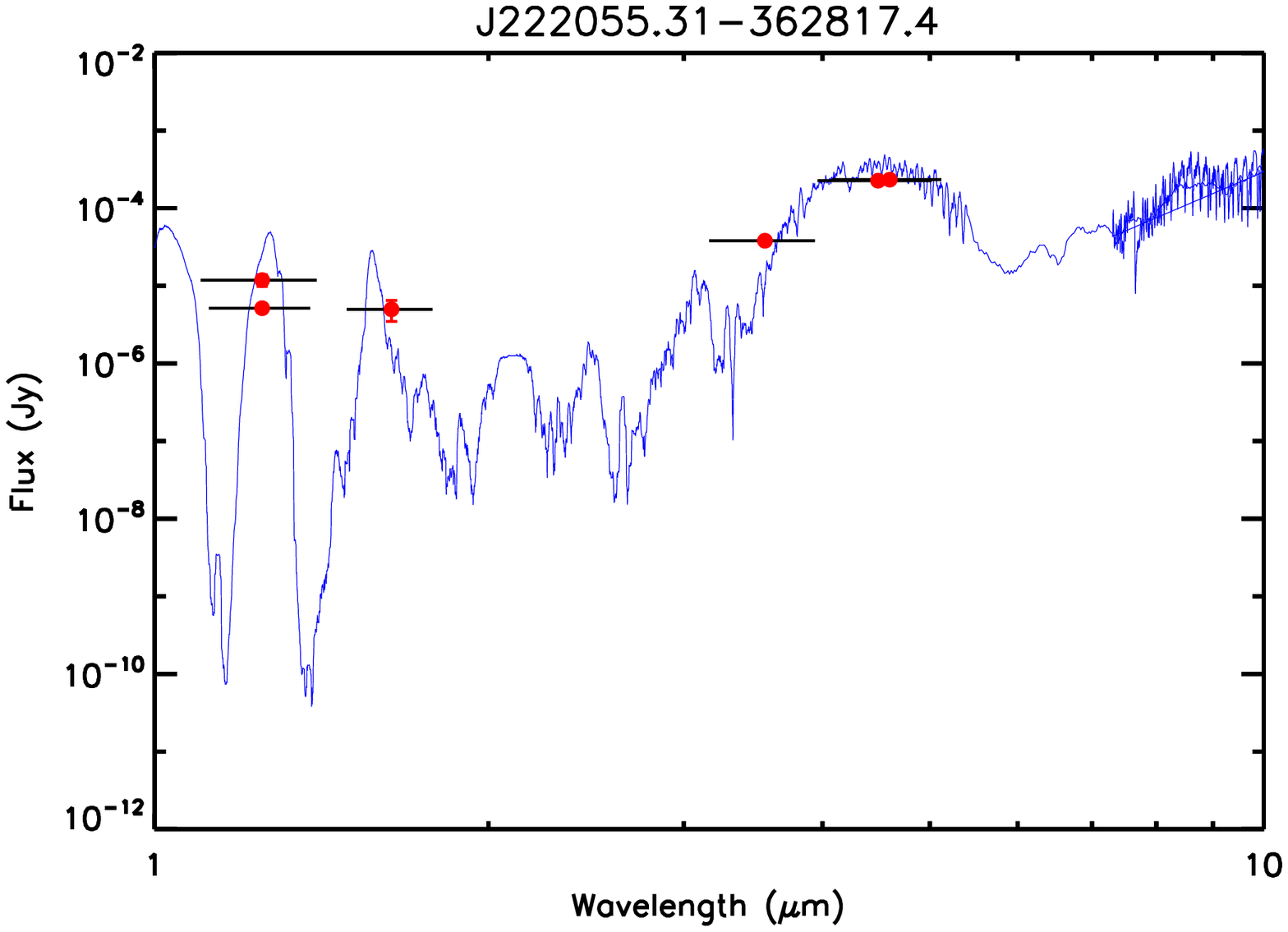}\\
\end{array}$
\end{center}
\caption{i-o). The result of fitting the photometric colors and absolute absolute 4.5 $\mu$m brightness to the BT-Settl models as described in the text. For WISE1828+2650 and WISE 2209+2711, the dotted line shows a model with added interstellar extinction as described in the text. \label{SampleSpecFits2}}
\end{figure*}
\clearpage

\begin{figure*}
\includegraphics[width=6in]{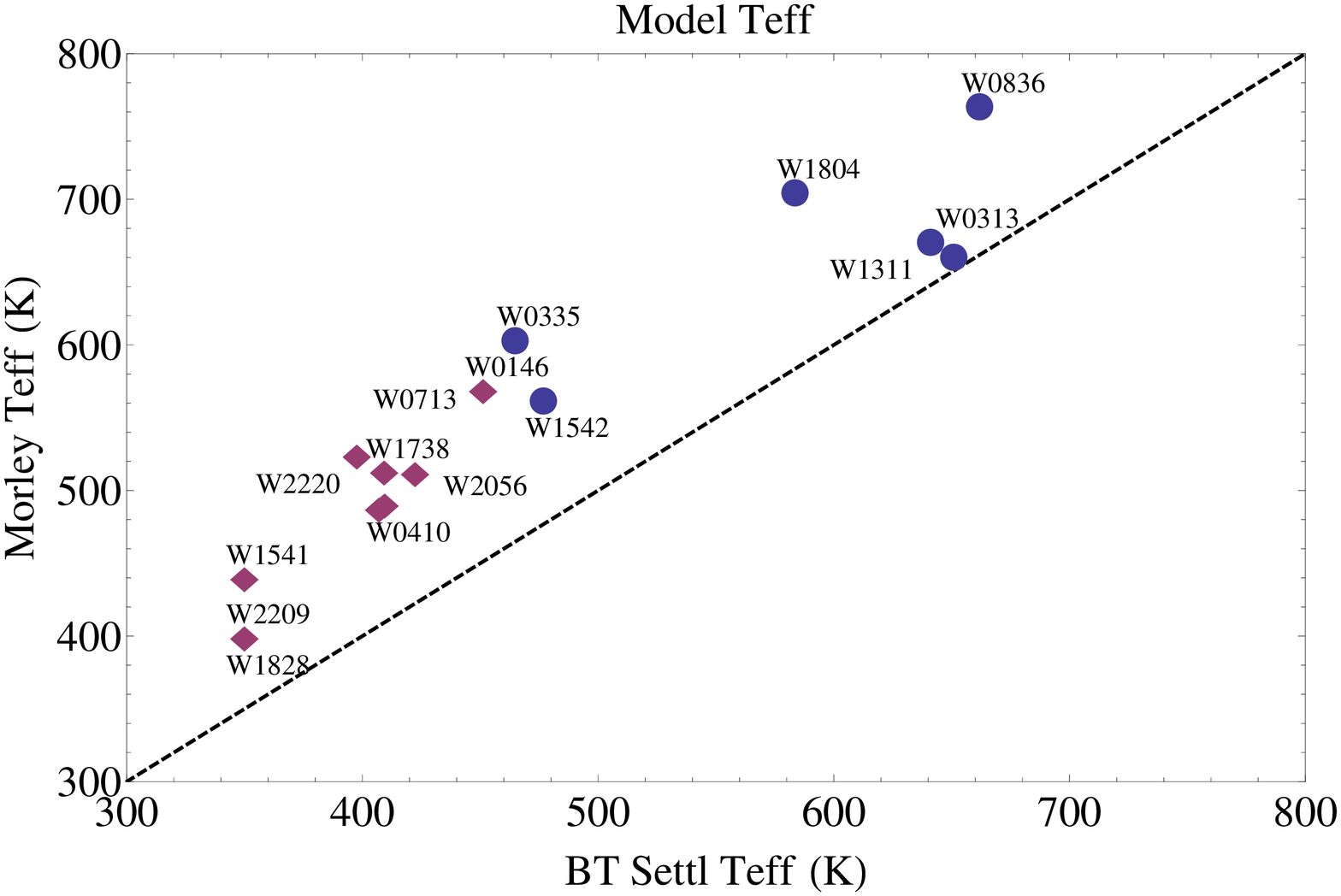}
\caption{left) Comparison of best fitting effective temperatures, T$_{eff}$ for BTSettl and Morley models. The temperatures of the Morley models are $\sim$75 K warmer than the corresponding BT-Settl model. The Y dwarfs are indicated by diamonds and the T dwarfs by circles and are on average $\sim$ 80 K cooler than the T dwarfs. \label{TeffCompare}}
\end{figure*}

\begin{figure*}
\begin{center}$
\begin{array}{c}
\includegraphics[width=3.5in]{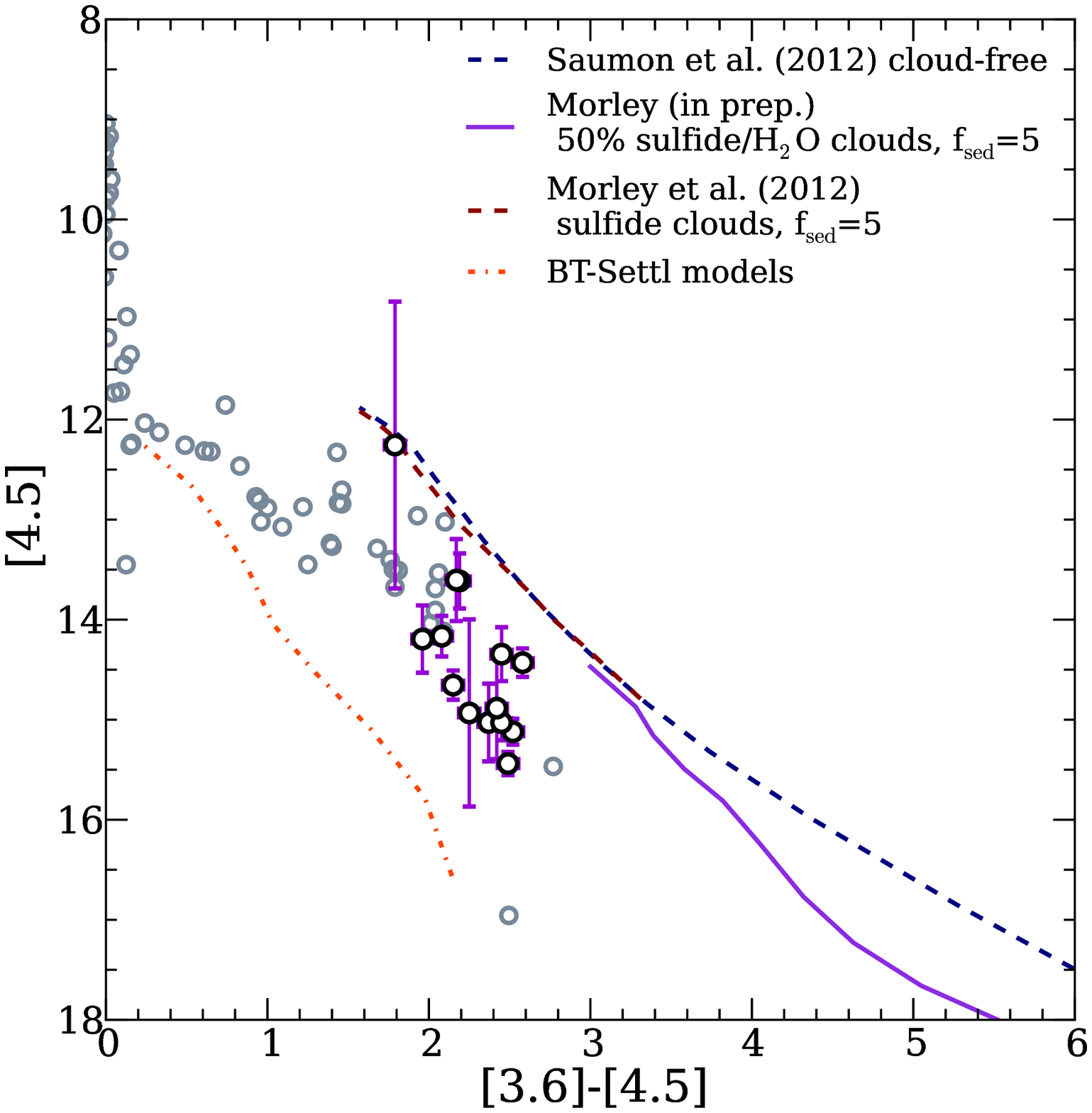}\\
\includegraphics[width=3.5in]{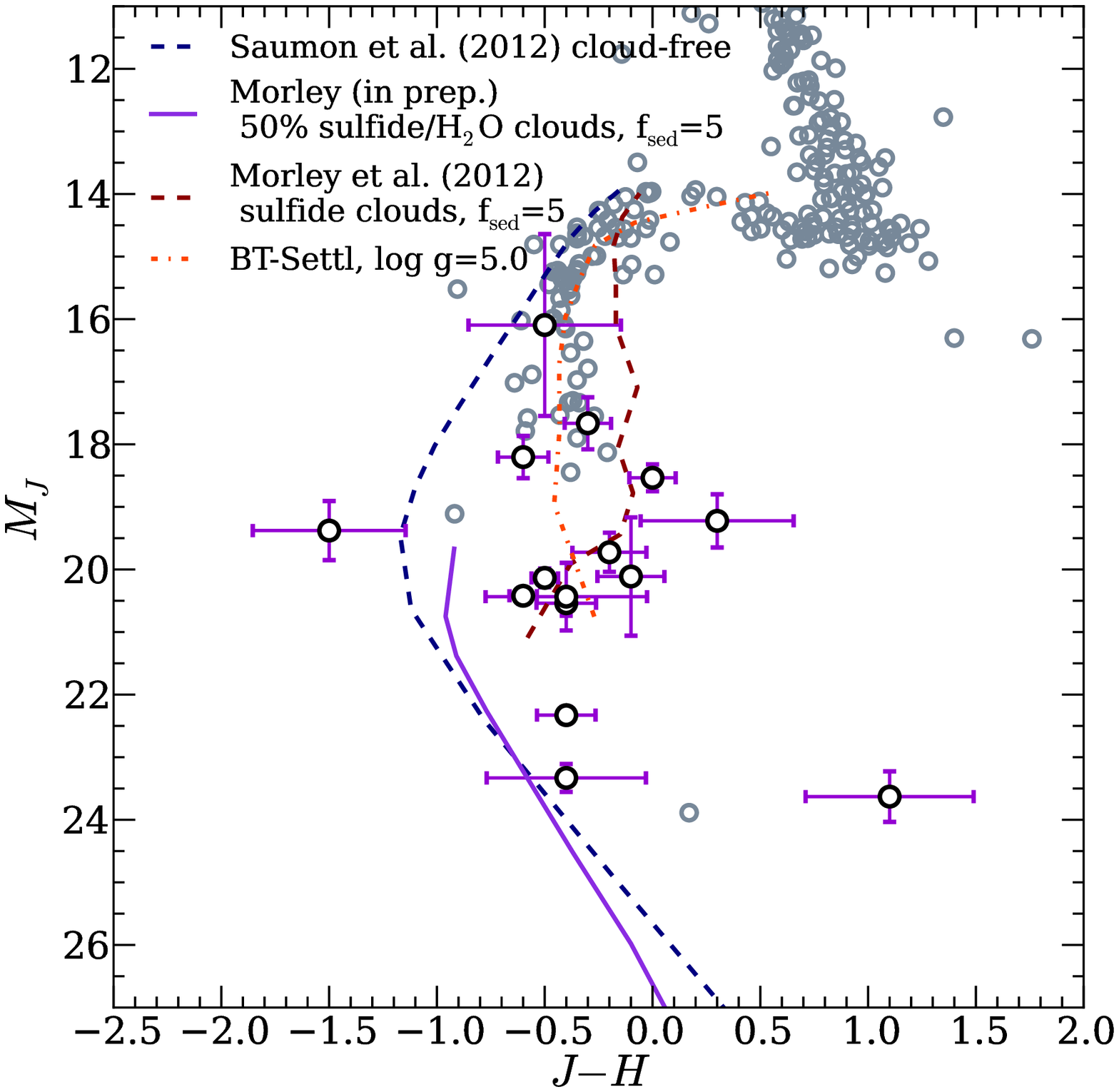}\\
\end{array}$
\end{center}
\caption{top) A Color-Magnitude diagram (CMD) for the two Spitzer bands showing a variety of models, including BT-Settl (orange, dot-dashed) and a variety of Morley models. The late T and Y dwarfs presented in this paper are plotted as well a large number of earlier spectral types taken from the literature. Three varieties of Morley models are shown, one cloud-free, one with sulfide clouds ($f_{sed}$=5, Table ~\ref{MorleyMass}) and one incorporating water clouds (Morley et al in preparation). bottom) A near-IR CMD for the same set of models. The BT-Settl models provide a good fit to the J-H colors and absolute magnitudes for the warmer objects, while the Morley objects do a better job on the colder objects at these wavelengths. WISE 1828+2650 stands out as extremely red in J-H and is poorly fitted in any of the models. \label{MarleyCMD}}
\end{figure*}

\end{document}